\documentclass[12pt]{article}
\pdfoutput=1

\usepackage{putex}
\usepackage{graphicx}
\usepackage{caption}
\usepackage{amsmath}
\usepackage{array}
\usepackage{subcaption}
\usepackage{epstopdf}
\usepackage{enumerate}
\usepackage{cite}
\usepackage{youngtab}
\usepackage{tensor}
\usepackage{slashed}
\usepackage[aligntableaux=center]{ytableau}
\usepackage[utf8]{inputenc}
\usepackage[
      colorlinks=true,
      linkcolor=blue,
      urlcolor=blue,
      filecolor=black,
      citecolor=red,
      ]{hyperref}

\newcommand{\grp}[1]{\mathrm{#1}}

\newcommand{\HH}{\mathbb{H}}

\newcommand{\abs}[1]{\left\lvert #1 \right\rvert}

\newcommand {\be} {\begin {equation}}
\newcommand {\ee} {\end {equation}}

\newcommand {\bes} {\begin {equation*}}
\newcommand {\ees} {\end {equation*}}

\newcommand{\es}[2] {\begin{equation} \label{#1} \begin{split} #2 \end{split} \end{equation}}

\newcommand{\Z}{\mathbb{Z}}

\newcommand{\R}{\mathbb{R}}

\newcommand{\cD}{{\mathcal D}}

\newcommand{\cF}{{\mathcal F}}
\newcommand{\cG}{{\mathcal G}}

\newcommand{\cL}{{\mathcal L}}
\newcommand{\cN}{{\mathcal N}}
\newcommand{\cO}{{\mathcal O}}

\newcommand{\cQ}{{\mathcal Q}}
\newcommand{\cR}{{\mathcal R}}

\newcommand{\cM}{{\mathcal M}}

\newcommand{\cZ}{{\mathcal Z}}

\newcommand{\veps}{\varepsilon}
\newcommand{\vphi}{\varphi}
\newcommand{\tX}{\widetilde{X}}
\newcommand{\tQ}{\widetilde{Q}}
\newcommand{\tq}{\widetilde{q}}
\newcommand{\ta}{\widetilde{a}}

\newcommand{\beq}{\begin{equation}}
\newcommand{\eeq}{\end{equation}}

\newcommand{\ov}{\over}

\def\le{\left}
\def\ri{\right}

\newcommand\al{{\alpha}}
\newcommand\ep{\epsilon}
\newcommand\sig{\sigma}
\newcommand\Sig{\Sigma}
\newcommand\lam{\lambda}
\newcommand\Lam{\Lambda}
\newcommand\om{\omega}

\newcommand\ga{{\ensuremath{{\gamma}}}}
\newcommand\Ga{{\ensuremath{{\Gamma}}}}
\newcommand\de{{\ensuremath{{\delta}}}}
\newcommand\De{{\ensuremath{{\Delta}}}}

\def\p{\partial}

\def\ie{\begin{equation}\begin{aligned}}
\def\fe{\end{aligned}\end{equation}}

\newcommand{\la}{\langle}
\newcommand{\ra}{\rangle}

\newcommand{\m}{\mu}
\newcommand{\n}{\nu}
\newcommand{\pa}{\nabla}
\newcommand{\A}{{\alpha}}
\newcommand{\B}{{\beta}}
\newcommand{\D}{{\delta}}

\newcommand{\CC}{{\Gamma}}

\newcommand{\ve}{{\varepsilon}}

\newcommand{\mR}{{\mathbb R}}

\newcommand{\mf}{\mathfrak }

\numberwithin{equation}{section}

\def\<{\langle}
\def\>{\rangle}

\newenvironment{alignedrows}[1]
{\begin{array}{@{} >{{}}r<{{}} @{}r@{\,} *{#1}{c} @{\,}l@{}}}
	{\end{array}}
\newcommand{\arow}[2]{%
	#1 & ( & #2 & )^T %
}

\begin{document}

\preprint{PUPT-2521}

\institution{PCTS}{Princeton Center for Theoretical Science, Princeton University, Princeton, NJ 08544, USA}
\institution{PU}{Joseph Henry Laboratories, Princeton University, Princeton, NJ 08544, USA}

\title{
A 2d/1d Holographic Duality}

\authors{M\'ark Mezei,\worksat{\PCTS} Silviu S.~Pufu,\worksat{\PU} and Yifan Wang\worksat{\PU}}

\abstract{

We propose $AdS_2$/CFT$_1$ dualities between exactly solvable topological quantum mechanics theories with vector or matrix large $N$ limits (on the boundary) and weakly coupled gauge theories on a fixed $AdS_2$ background (in the bulk).   The boundary theories can be embedded as 1d sectors of 3d ${\cal N} = 4$ superconformal field theories with holographic duals, from which they can be obtained using supersymmetric localization.  We study a few examples of such 1d theories:  theories with vector large $N$ limits that are embedded into 3d theories of many free massless hypermultiplets with $AdS_4$ higher spin duals; and a 1d theory with a matrix large $N$ limit embedded into the 3d ABJM theory at Chern-Simons level $k=1$, which has an $AdS_4$ supergravity dual.  We propose that the $U(N)$ singlet sectors of the 1d vector models are dual to 2d gauge theories on $AdS_2$ whose gauge algebras are finite dimensional and whose full non-linear actions we completely determine in some cases.  The 1d theory embedded into ABJM theory has a $\Z_2$-invariant sector dual to a 2d gauge theory on $AdS_2$ whose gauge algebra is the infinite dimensional algebra of area preserving diffeomorphisms of a two-sphere.  We provide evidence that the 2d gauge theories on $AdS_2$ can be obtained from localizing the $AdS_4$ duals of the 3d SCFTs mentioned above, and thus argue that our 2d/1d dualities can be obtained via supersymmetric localization on both sides of their parent $AdS_4$/CFT$_3$ dualities.  We discuss the boundary terms required by holographic renormalization in the 2d gauge theories on $AdS_2$ and show how they arise from supersymmetric localization.

}
\date{}

\maketitle

\tableofcontents

\section{Introduction}

Almost twenty years after the proposal of the gauge/gravity duality \cite{Maldacena:1997re,Witten:1998qj,Gubser:1998bc}, there has been a large amount of progress in understanding top-down constructions of gauge/gravity dual pairs in various dimensions and with various amounts of supersymmetry.  Notable well-understood examples with maximal supersymmetry come from string theory or M-theory and include the anti-de Sitter / conformal field theory (AdS/CFT) dualities between 4d ${\cal N} =4$ super Yang-Mills theory and string theory on $AdS_5 \times S^5$ \cite{Maldacena:1997re,Witten:1998qj,Gubser:1998bc} and between 3d ABJM theory and M-theory on $AdS_4 \times S^7 / \Z_k$ \cite{Aharony:2008ug}.  A different class of examples include the duality between $O(N)$ vector models and higher spin theories in AdS \cite{Klebanov:2002ja,Giombi:2009wh}.    While in the first class of examples, it is the field theory  that is strongly coupled and hard to study analytically, in the second class of examples it is the higher spin side that is complicated, although there are some cases where the full non-linear equations of motion are known \cite{Vasiliev:1990en,Vasiliev:2003ev} (see \cite{Giombi:2016ejx} for a recent review).

In this work, we propose a 2d/1d holographic duality between solvable one-dimensional field theories and weakly coupled two-dimensional bulk theories.  The field theories are certain topological quantum mechanics (TQM) models\footnote{In this paper, by a topological quantum mechanics we mean a one-dimensional theory, defined on a circle or on a line, described by a convergent Euclidean path integral.  We do not necessarily mean that the Euclidean partition function on $S^1$  computes $\tr e^{-\beta H}$, where $H$ is a hermitian Hamiltonian of a Lorentzian quantum mechanical system.} with a parameter $N$ that can be taken to be large.  We propose that their duals at large $N$ consist of weakly coupled 2d Yang-Mills theory (YM$_2)$ supplemented by higher derivative terms on a fixed $AdS_2$ background---as we will explain in more detail shortly, the 2d metric is not fluctuating in these proposals. We expect that the YM$_2$/TQM dualities we discuss can be embedded into particular instances of the previously-mentioned $AdS_4$/CFT$_3$ duality by applying supersymmetric localization to both sides:  supersymmetric localization in the 3d boundary superconformal field theory (SCFT) gives the TQM, and localization on the gravity side using the corresponding supercharge should give  Yang-Mills theory restricted to an $AdS_2$ slice of the geometry, as we demonstrate in one simple case.  The TQM/YM$_2$ duality we propose here resembles the duality between Chern-Simons theory and Wess-Zumino-Witten models, which can also be embedded into the standard holographic correspondences using localization  \cite{Bonetti:2016nma}.

Our duality can be thought of as a realization of the $AdS_2$/CFT$_1$ correspondence. 
 Indeed, it has been understood for some time  \cite{Maldacena:1998uz,Jensen:2011su,Almheiri:2014cka} that the only truly conformal quantum mechanical systems have vanishing Hamiltonian $H=0$.\footnote{Scaling symmetry implies that the density of states of the quantum mechanical theory is \cite{Jensen:2011su}
  \es{DoS}{
  \rho(E)= A\, \delta(E)+{B\ov E}\,. 
  }
As shown in \cite{Jensen:2011su}, $B>0$ is required for non-topological dynamics, but leads to a non-integrable density of states.  (See also \cite{Iqbal:2011ae}.) This demonstrates that strictly conformal quantum mechanical systems have to have $H=0$.
 \label{footDoS} } Consequently, such a 1d system does not have a local stress tensor operator, so its bulk dual should not contain a fluctuating metric.  Indeed, had there been dynamical (dilaton) gravity in the bulk, the back reaction of  finite energy excitations would be so strong that it would modify the UV asymptotics of the $AdS_2$ region \cite{Maldacena:1998uz}.\footnote{The problem of strong backreaction can be alleviated in cases when the spacetime is $AdS_2\times {\cal M}$, where ${\cal M}$ is non-compact. However, even these cases suffer from the problem explained in footnote \ref{footDoS}.}${}^{,}$\footnote{Another related manifestation of problems with gravity in $AdS_2$/CFT$_1$ is that dilaton and graviton exchange does not produce scale invariant boundary correlators \cite{Almheiri:2014cka}.} Even though the metric is not fluctuating, the duality we propose is still holographic both because it can be embedded in holography with dynamical gravity, as mentioned above, and because, as we will see, the holographic dictionary and computational techniques familiar from higher-dimensional versions of AdS/CFT do apply.  The recent proposed connection between the near-conformal SYK model \cite{Sachdev:1992fk,Kitaev,Polchinski:2016xgd,Jevicki:2016bwu,Maldacena:2016hyu} or tensor models \cite{Gurau:2009tw,Witten:2016iux,Klebanov:2016xxf} in 1d and dilaton gravity in 2d constitute  a different type of duality \cite{Kitaev,Maldacena:2016hyu,Jensen:2016pah,Maldacena:2016upp,Engelsoy:2016xyb}, where the boundary theory is only nearly conformal, and the $AdS_2$ region needs to be cut off near the asymptotic boundary, where the dilaton diverges.

From a bottom-up perspective, the duality we propose can be anticipated as follows.  A TQM is a theory in which correlation functions of local operators $J_a$ are independent of the separations between the insertion points, but could generally depend on the ordering of the insertion points.  Because of the position-independence, one can identify the $J_a$ with conserved currents, and so a TQM is really a theory of conserved currents with a potentially infinite-dimensional non-Abelian global symmetry generated by these currents.   If the TQM has a large $N$ limit that allows an organization of the spectrum of operators into single trace and multi-trace, then its holographic dual should contain one 2d gauge field in $AdS_2$ for every single-trace conserved current $J_a$, as is standard in AdS/CFT constructions.  The bulk action should take the form of Yang-Mills theory plus higher derivative corrections because this is the most general gauge-invariant action one can write down using the bulk gauge fields.  A subtle issue that we will elaborate on is which boundary terms should supplement the bulk action.  The boundary terms we consider are different from those recently considered in the $AdS_2$ literature \cite{Castro:2008ms,Grumiller:2014oha,Grumiller:2015vaa,Jensen:2016pah,Cvetic:2016eiv}, but follow from the fact that, unlike in higher dimensions, a vector field in $AdS_2$ obeys ``alternate'' boundary conditions when the dual operator is a conserved current.  Thus, one must include boundary terms implementing the Legendre transform that needs to be performed when the bulk fields obey alternate boundary conditions \cite{Klebanov:1999tb}.  As we will show, these boundary terms can be also derived using supersymmetric localization of a 4d vector multiplet on $AdS_4$.

We provide two types of examples of YM$_2$/TQM dualities.  In the first type of examples, we have TQM models with vector large $N$ limits, and our dualities can be embedded into the duality between a supersymmetric large $N$ vector model in 3d and a higher spin theory in $AdS_4$.  In the second type of examples, we have TQM models with $N \times N$ matrix degrees of freedom.  We study explicitly an example that can be embedded as a sector of the duality between large $N$ $U(N)_k \times U(N)_{-k}$ ABJM theory at Chern-Simons level $k=1$ and supergravity on $AdS_4 \times S^7$.    One difference between these two types of examples is that in the first type the Yang-Mills coupling scales as $g_\text{YM}^2 L^2 \propto N^{-1}$, while in the second it scales as $g_\text{YM}^2 L^2 \propto N^{-3/2}$ at large $N$.  Another difference is that the Lie algebra of the 2d YM theory is finite-dimensional in the first type of examples, while it is infinite-dimensional in the second.

In every case we study, the embedding of the YM$_2$/TQM duality into the $AdS_4$/CFT$_3$ duality is realized as follows.  We require that the CFT$_3$ is an SCFT with at least ${\cal N} = 4$ supersymmetry.  Recent work \cite{Chester:2014mea, Beem:2016cbd} has shown that any 3d ${\cal N} = 4$ SCFT contains a 1d topological sector representing the cohomology of a certain supercharge.\footnote{There are two such sectors, one associated with the Higgs branch, and one with the Coulomb branch of the 3d SCFT\@.  In this paper we will focus only on the sector associated with the Higgs branch.}  This sector is our TQM\@.    It  is shown in \cite{Dedushenko:2016jxl} that it can be described as a topological gauged quantum mechanics that can be obtained through supersymmetric localization of the SCFT on $S^3$.   This construction applies, in particular, to 3d ${\cal N} = 4$ SCFTs with holographic duals, where one may ask whether a supersymmetric localization computation in the bulk could simplify the bulk theory (which is either described by supergravity or by a higher-spin theory) to precisely the 2d holographic duals of the topological gauged quantum mechanics.   A similar connection was developed in \cite{Bonetti:2016nma} between 2d chiral algebras arising as a sector of 4d ${\cal N} = 4$ super Yang-Mills theory (SYM) and Chern-Simons theory on $AdS_3$, which is believed to arise from localizing supergravity on $AdS_5$. 

Performing supersymmetric localization in $AdS_4$ appears to be very complicated (and in the higher spin case impossible without knowing the action of the higher spin theory), and we do not pursue it here.  However, in the examples mentioned above, the $AdS_4$ theory contains an ${\cal N} = 4$ SYM theory as a consistent truncation, and as a first step towards localizing the full $AdS_4$ theory we begin by localizing this ${\cal N} = 4$ SYM theory on a fixed $AdS_4$ background.  From the point of view of the $AdS_4$/CFT$_3$ duality, the 4d ${\cal N} = 4$ SYM theory in the bulk is dual to half BPS superconformal multiplets  that contain flavor currents in the boundary 3d SCFT\@.   As we will show, the localization of the 4d SYM theory on $AdS_4$ yields precisely a Yang-Mills theory on $AdS_2$ that is part of the full YM$_2$ theory dual to our TQM\@.     This supersymmetric localization computation is related to that performed by Pestun in \cite{Pestun:2009nn}, where  it was shown that an ${\cal N} = 4$ vector multiplet on $S^4$ localizes to 2d Yang-Mills theory on $S^2$.  (See also \cite{Giombi:2009ds,Giombi:2009ek}.)   Since $\cN=4$ SYM is conformal, and since $AdS_4$ and $S^4$ are related by a Weyl transformation, one can argue that Pestun's computation implies that the $AdS_4$ Yang-Mills theory localizes to 2d Yang-Mills theory on $AdS_2$.  One significant difference between $S^4$ and $AdS_4$, however, is that while $S^4$ is compact, $AdS_4$ has an asymptotic boundary, and thus in the localization computation on $AdS_4$ one should also keep track of boundary terms.  Quite nicely, the $AdS_2$ boundary terms obtained through this procedure match precisely what one would have expected based on holographic renormalization and the alternate quantization of a 2d gauge field dual to a conserved current. 

Let us now discuss in more detail specific examples.

\paragraph{Dualities with a vector large $N$ limit.}

The first example starts with the 4d/3d duality between the $\mathfrak{u}(N)$ singlet sector of $N$ free hypermultiplets and the ${\cal N} = 4$ supersymmetric version of Vasiliev theory constructed in \cite{Vasiliev:1992av,Vasiliev:1995dn,Vasiliev:1999ba,Engquist:2002vr,Engquist:2002gy,Chang:2012kt}.  By $\mathfrak{u}(N)$ singlet sector we mean that we consider only operators that are invariant under a $\mathfrak{u}(N)$ subalgebra of the flavor symmetry algebra $\mathfrak{usp}(2N)$ of $N$ free hypermultiplets.\footnote{Perhaps more rigorously, one can consider \cite{Chang:2012kt} a $\mathfrak{u}(N)_k \times \mathfrak{u}(1)_{-k}$ theory with $N$ bifundamental hypermultiplets, and take the Chern-Simons level $k \to \infty$.  The Chern-Simons interactions with large $k$ essentially implement the $\mathfrak{u}(N)$ singlet constraint. \label{CSFootnote} } Within $\mathfrak{usp}(2N)$, this $\mathfrak{u}(N)$ algebra commutes only with the diagonal $\mathfrak{u}(1)$ in $\mathfrak{u}(N)$.  This $\mathfrak{u}(1)$ acts as a global flavor symmetry of the $\mathfrak{u}(N)$ singlet sector.  In AdS/CFT global symmetries on the boundary are gauged in the bulk, and indeed,  the ${\cal N} = 4$ higher spin theory contains, among many other fields, an ${\cal N} = 4$ $\mathfrak{u}(1)$ vector multiplet in $AdS_4$.

The topological quantum mechanics that can be embedded in the theory of $N$ free hypermultiplets can be described in terms of complex fields $Q_{I}$ and $\tQ^I$, where $I = 1, \ldots, N$ and lower/upper indices are $\mathfrak{u}(N)$ fundamental/anti-fundamental indices.  When this theory is defined on a circle, the fields $Q_{I}$ and $\tQ^I$ are anti-periodic, and the action is proportional to
 \es{TopSimple}{
    \int \, \tQ^I\, d Q_{I}  \,.
 }
While the $\mathfrak{u}(N)$ singlet sector of the 3d SCFT consists of many single trace operators as well as multi-trace operators, the $\mathfrak{u}(N)$ singlet sector of the 1d topological theory\footnote{One can also consider the analogous construction to that in footnote~\ref{CSFootnote}.  In the 1d theory we can consider  coupling the $Q_I$ to a $\mathfrak{u}(N)$ gauge field ${\cal A}_I{}^J$ and consider the total action
\es{TopMoreComplicated}{
 \left( \int \, \tQ^I\, ( d Q_{I} + {\cal A}_I{}^J  Q_J )  \right) + \kappa  \left( \int {\cal A}_I{}^J \right)
  \left( \int {\cal A}_J{}^I \right)  
}
with a coefficient $\kappa$ related to $k$.  Taking $\kappa$ to infinity essentially imposes the $\mathfrak{u}(N)$ singlet constraint.} \eqref{TopSimple} is much simpler:  its single trace sector consists of only one operator,
 \es{SingleTrace}{
  \tQ^{I} Q_{I} \,.
 }
Additionally, the multi-trace operators in this 1d theory are simply just powers of $\tQ^{I} Q_{I}$.   Our holographic duality proposal in this case is that at large $N$ the $\mathfrak{u}(N)$ singlet sector of the theory \eqref{TopSimple} is dual to Maxwell theory in $AdS_2$, supplemented by higher derivative terms:
 \es{SchematicYM}{
  S = \frac{1}{2e^2} \int d^2 x \sqrt{g} \left[ F^2 + w_4 F^4 + w_6 F^6 + \ldots  \right] \,,
 }
where we defined $F \equiv \frac 12 F_{\mu\nu} \epsilon^{\mu\nu}$, $e$ is the gauge coupling constant, and $w_n$ are coefficients to be determined. We will determine all the coefficients $w_n$ in Section~\ref{VECTORABELIAN}.

This proposal can be generalized to a 3d theory of $N M$ free massless hypermultiplets, which is dual to the supersymmetric Vasiliev theory mentioned above with additional $U(M)$ Chan-Paton factors.  Thus the Vasiliev theory contains, among many other fields, a $\mathfrak{u}(M)$ ${\cal N} = 4$ vector multiplet in $AdS_4$.  The topological theory now contains fields $Q_{Ii}$ and $\tQ^{Ii}$, where   $I = 1, \ldots, N$ and $i = 1, \ldots, M$.  The 1d action is the same as \eqref{TopSimple} with the index $I$ replaced by $Ii$.  The $\mathfrak{u}(N)$ singlet sector consists of the single trace operators $\tQ^{Ii} Q_{Ij}$, which transform in the adjoint representation of $\mathfrak{u}(M)$.  Thus, our proposal in this case is that the $\mathfrak{u}(N)$ singlet sector of the topological theory is dual to $\mathfrak{u}(M)$ Yang-Mills theory on $AdS_2$.

\paragraph{Dualities with a matrix large $N$ limit.}

As mentioned above, we also consider YM$_2$/TQM dualities with a matrix large $N$ limit and $AdS_4$/CFT$_3$ parents arising from M-theory, and in particular we focus on an example coming from the $U(N)_k \times U(N)_{-k}$ ABJM theory \cite{Aharony:2008ug} at Chern-Simons level $k=1$.  In the deep infrared, this theory is believed to flow to a 3d interacting SCFT with ${\cal N} = 8$ supersymmetry, which, at large $N$, is dual to 11d supergravity on $AdS_4 \times S^7$.  The same IR fixed point can be reached from a different UV theory studied by Bashkirov and Kapustin \cite{Bashkirov:2010kz}, which is the UV description we will focus on.  In ${\cal N} = 4$ notation, this theory has a $U(N)$ vector multiplet coupled to an adjoint hypermultiplet and a fundamental hypermultiplet.  The TQM can be written down using the results of \cite{Dedushenko:2016jxl}.  It contains a $\mathfrak{u}(N)$ gauge field ${\cal A}$ coupled to a $\mathfrak{u}(N)$ fundamental field $Q$, an anti-fundamental $\tQ$, and two adjoint fields $X$ and $\tX$.  The action is
 \es{Top}{
   \int \left(\tQ (d + {\cal A}) Q + \tr ( \tX dX + {\cal A} [X, \tX] )\right)   \,,
 }
where the matter fields are anti-periodic when the theory is defined on a circle.  This theory has a $\Z_2$ global symmetry that acts by $X \to -X$ and $\tX \to -\tX$.

With no operator insertions, the partition function of \eqref{Top} equals the partition function of the 3d SCFT placed on a round $S^3$.\footnote{The fact that the $S^1$ partition function equals the 3d $S^3$ partition function of a unitary SCFT implies that $Z_{S^1} = Z_{S^3} < 1$, because in 3d $Z_{S^3} = e^{-F_{S^3}}$ with $F_{S^3}>0$ after regularization.  (Provided that there exists an RG flow from the 3d SCFT to the empty theory, one can derive that $F_{S^3}>0$ using the $F$-theorem \cite{Jafferis:2011zi,Klebanov:2011gs,Myers:2010xs,Myers:2010tj} proven in \cite{Casini:2012ei}---see \cite{Pufu:2016zxm} for a review.)  This might seem confusing in light of the $g$-theorem \cite{Affleck:1991tk,Affleck:1992ng} (or a generalized $F$-theorem \cite{Giombi:2014xxa}) , which states that for a CFT$_1$ where $Z_{S^1}$ measures the ground state degeneracy, this quantity decreases under RG flow and hence should obey  $Z_{S^1} > 1$ if there exists an RG flow connecting the CFT$_1$ to a trivial theory.  We suspect that the resolution of this puzzle is that because of the anti-periodicity of the scalars of our theories on $S^1$, the $S^1$ partition function cannot be interpreted as $\tr e^{-\beta H}$ for a quantum system with a Hermitian Hamiltonian $H$.}  Indeed, after gauge fixing and integrating out the matter fields, as explained in \cite{Dedushenko:2016jxl}, Eq.~\eqref{Top} reduces to the $N$-dimensional integral
 \es{MatrixModel}{
  Z = \frac{1}{N!} \int \prod_{i=1}^N d\lambda_i  \frac{ \prod_{i<j} (2 \sinh^2 (\pi (\lambda_i - \lambda_j) ))}
   {\prod_{i, j} \cosh (\pi (\lambda_i - \lambda_j)) \prod_i \cosh(\pi \lambda_i) } \,,
 }
which can also be obtained by calculating the $S^3$ partition function from the supersymmetric localization results of Kapustin, Willett, and Yaakov \cite{Kapustin:2009kz}.  This integral can be viewed as a matrix model where the $\lambda_i$ are the eigenvalues of an $N\times N$ matrix.  At large $N$, this matrix model can be solved using a variety of techniques developed in \cite{Herzog:2010hf, Jafferis:2011zi, Santamaria:2010dm,Marino:2011eh, Mezei:2013gqa, Grassi:2014vwa}.   A special property of this matrix model is that there are no long-range interaction forces between the eigenvalues $\lambda_i$  \cite{Herzog:2010hf}.  This property leads to an unconventional scaling with $N$ of the $S^3$ free energy $F \equiv - \log Z$, namely $F \propto N^{3/2}$ at large $N$, in agreement with expectations from the $AdS_4 \times S^7$ dual \cite{Drukker:2010nc}.

The topological theory \eqref{Top} contains gauge-invariant operators that can be written in terms of $X$ and $\tilde X$.  These are the single trace operators  of the form
 \es{singleTr}{
  \tr (X^{p_1} \tX^{q_1}  X^{p_2} \tX^{q_2} \cdots ) \,,
 }
as well as multi-trace operators that are suppressed at large $N$.  For simplicity, we focus on the $\Z_2$-invariant sector that contains only operators of the form \eqref{singleTr} with an even total number of $X$ and $\tX$.  All these operators have vanishing scaling dimension in the 1d topological theory and can be thought of as conserved currents. They enhance the manifest $\mathfrak{su}(2)$ symmetry algebra of the theory to an infinite dimensional symmetry. The correlation functions of the single trace operators \eqref{singleTr} can be computed from \eqref{Top} by replacing these operators with appropriate expressions in terms of $\lambda_i$, as we will show in detail.  As we will see, at large $N$, the correlation functions of the single trace operators are dominated by planar diagrams, as is common in matrix large $N$ theories.  However, because of the local nature of the interactions between the eigenvalues in \eqref{MatrixModel}, all these correlation functions have an unconventional scaling in $N$.  This unconventional scaling is consistent with the expectations coming from the $AdS_4 \times S^7$ supergravity dual.

Our proposal is then that at large $N$, the holographic dual of \eqref{Top} is a 2d Yang-Mills theory whose gauge algebra has one generator corresponding to each of the operators \eqref{singleTr}. As we will explain in Section~\ref{MATRIX}, from computing the 2-, 3-, and 4-point functions of these operators at large $N$, we identify the structure constants of the gauge algebra as well as all 3-derivative and a specific 4-derivative term appearing in the 2d action.  As we will see, the gauge algebra corresponding to the $\Z_2$-invariant sector of \eqref{Top} is the algebra $\text{SDiff}(S^2)$ of area-preserving diffeomorphisms of a two-sphere.  

\paragraph{}

The rest of this paper is organized as follows.  In Sections~\ref{ABELIAN} and \ref{NONABELIAN} we describe the bulk side of our TQM/YM$_2$ dualities by first focusing on the Abelian case in Section~\ref{ABELIAN} and then on the non-Abelian case in Section~\ref{NONABELIAN}.   In particular, we use the holographic dictionary to calculate 2-, 3-, and 4-point functions in the boundary theory dual to YM$_2$ theory in $AdS_2$.  Sections~\ref{TOPOLOGICAL}--\ref{MATRIX} are devoted to the boundary side of the TQM/YM$_2$ duality, and can be read independently from the rest of the paper.  In particular, we start in Section~\ref{TOPOLOGICAL} by reviewing the embeddings of the TQM models into 3d ${\cal N} = 4$ SCFTs, and then we discuss their infinite dimensional symmetry that arises by interpreting all operators in the TQM as conserved currents.  Sections~\ref{VECTOR} and \ref{MATRIX} contain specific examples of the YM$_2$/TQM dualities:  in Section~\ref{VECTOR} we present examples with a vector large $N$ limit, and in Section~\ref{MATRIX} we present an example with a matrix large $N$ limit.   Section~\ref{LOCALIZATION} then contains a description of how supersymmetric localization of $
\cN=4$ SYM on $AdS_4$ yields Yang-Mills theory on $AdS_2$, thus providing the beginning of a derivation of the dualities mentioned above.  We end with a summary of our results as well as a list of new directions in Section~\ref{CONCLUSIONS}.  In Appendix~\ref{WITTENDIAG} we reproduce many of the bulk results by computing Witten diagrams. Several technical details of the localization computation of Section~\ref{LOCALIZATION} are delegated to Appendix~\ref{LOCAPP}.

\section{Abelian gauge theory in $AdS_2$}
\label{ABELIAN}

\subsection{Quadratic bulk action}

Let us begin with the study of an Abelian gauge theory in $AdS_2$ (or in its Euclidean analog $\HH^2$), with $A_\mu$ being the 2d gauge field and $F_{\mu\nu} = \partial_\mu A_\nu - \partial_\nu A_\mu$ being its field strength.  In Euclidean signature, the action takes the form
  \es{MaxwellAction}{
  S_\text{bulk} = \frac{1}{4 e^2} \int d^2 x \, \sqrt{g} F_{\mu\nu} F^{\mu\nu} + \cdots \,.
 }
where the ellipsis stands for potential higher derivative terms that we suppress for now.   Without these higher derivative terms, \eqref{MaxwellAction} is just the Maxwell action.  The gauge field $A_\mu$ is dual to a conserved current $j$ in the boundary theory whose correlation functions we will now compute using AdS/CFT techniques.

As is common in AdS, the action \eqref{MaxwellAction} is not finite on-shell and needs to be supplemented by boundary terms, a procedure referred to as holographic renormalization.  Although various boundary terms supplementing the Maxwell action \eqref{MaxwellAction} have been considered in the literature \cite{Castro:2008ms,Grumiller:2014oha,Grumiller:2015vaa,Jensen:2016pah,Cvetic:2016eiv}, we will argue that for our purposes we should consider:
 \es{SBdy}{
  S_\text{bdy} = \frac{1}{2 e^2} \int_\partial dx\, \sqrt{\gamma}\, \left(A^i A_i - 2 A^i F_{\mu i} n^\mu \right)
   + \ldots \,,
 } 
where the ellipsis again stands for higher derivative terms that we drop for now, $\gamma_{ij}$  (a $1\times1$ matrix) is the induced metric on the boundary, $\gamma$ is its determinant, and $n^\mu$ is the unit normal vector to the boundary obeying $n^\mu n_\mu = 1$.  The boundary term \eqref{SBdy} appears to be new.

In order to argue that \eqref{SBdy} is the correct boundary term, let us first solve the theory classically.  To do so, let us commit to the specific form of the hyperbolic space metric
 \es{Back}{
  ds^2 = dr^2 + \sinh^2 r\, d\vphi^2 \,,
 }
where we set the curvature radius of $\HH^2$ to $L=1$ as a choice of units.  In the absence of sources, the equations of motion $\nabla^\mu F_{\mu\nu} = 0$ reduce to
 \es{eoms}{
  \partial_\vphi (\sinh^{-1} r\, F_{r\vphi} ) = \partial_r (\sinh^{-1} r\, F_{r\vphi}) = 0 \,.
 }
Their exact solution is $F_{r\vphi} = Q \sinh r$, where $Q$ is an integration constant.  As a boundary condition on the gauge field, we impose that the radial component $A_r$ decays near the boundary, $\lim_{r\to\infty} A_r=0$.  The most general gauge field configuration obeying this boundary condition that gives the profile $F_{r\vphi}= Q \sinh r$ is  
 \es{AsympExpansion}{
  A_\vphi &= Q \cosh r + a(\vphi) + \partial_\vphi \Lambda(r, \vphi)  \,, \qquad A_r = \partial_r \Lambda(r, \vphi) \,,
 }
where $\Lambda(r, \vphi)$ is a gauge parameter that decays at infinity and $a(\vphi)$ is so far an arbitrary function. Actually, $Q$ and $a(\vphi)$ are not independent, because, in general, \eqref{AsympExpansion} may reproduce our expression for $F_{r\vphi}$ only up to a delta function contribution supported at $r=0$.  By the Stokes theorem, requiring that there should be no such delta function is equivalent to requiring $\oint A = 0$ along a small circle around $r=0$.  This condition implies
 \es{GotQ}{
  Q = -\frac{1}{2 \pi} \int_{-\pi}^{\pi} d\vphi\, a(\vphi) \,.
 } 
The expression \eqref{AsympExpansion} with the condition \eqref{GotQ} thus represents the most general classical solution of our theory.  Note that the expression \eqref{AsympExpansion} is of course not well-defined at $r=0$ unless $\Lambda$ is chosen appropriately in a patch containing this point.

According to the AdS/CFT dictionary, the bulk gauge field $A$ can either be dual to a gauge field in the boundary theory or to a conserved current, depending on the boundary conditions it obeys \cite{Witten:2003ya,Marolf:2006nd}.  We are interested in the case in which the boundary theory contains a conserved current.  This conserved current is a dimension $0$ operator that couples to a dimension $1$ source, a background gauge field, which should be identified with the function $a(\vphi)$ in the asymptotic expansion \eqref{AsympExpansion}.  Because $a(\vphi)$ is a source for the boundary operator, the Euler-Lagrange problem in the bulk should be well-posed provided that $\delta a(\vphi) = 0$.  In analyzing whether a variational problem is well-posed or not, one should only be concerned with the solution of the equations of motion close to the boundary and should not impose regularity away from it.  Thus, in studying the Euler-Lagrange problem we treat $\delta a(\vphi)$ and $\delta Q$ as independent and not related through \eqref{GotQ}.  Requiring that the Euler-Lagrange problem is well-posed when $\delta a = 0$ but for arbitrary $a$, $Q$, and $\delta Q$ is reminiscent of the alternate quantization of Klebanov-Witten \cite{Klebanov:1999tb}, and the procedure for determining the needed boundary terms is similar and consists of two steps.  The first step is to add the counterterm
 \es{Counter1}{
  S_\text{bdy(1)} = -\frac{1}{2e^2} \int d\vphi \, \sinh^{-1} r\,  (A_\vphi)^2
 }
in order to make the on-shell action finite on any solution to the equations of motion \eqref{eoms}.  This boundary term can be converted into the bulk term $-\frac{1}{2e^2} \int d\vphi\, dr\, \partial_r \left[ \sinh^{-1} r\,  (A_\vphi)^2 \right]$, which, when considered together with \eqref{MaxwellAction}, implies that the canonical momentum conjugate to $A_\vphi$ is
 \es{CanMom}{
  \pi^\vphi(r, \vphi) = \frac{1}{e^2} \sinh^{-1}r (F_{r\vphi} - A_\vphi ) \,.
 }
When expanded at large $r$, we see that $\pi^\vphi(r, \vphi) \approx -\frac{2}{e^2} e^{-r} a(\vphi)$, so the leading boundary behavior of $\pi^\vphi(r, \vphi)$ is what we want to identify as the source for the boundary operator.

Under an Euler-Lagrange variation, so far we have
 \es{var}{
  \delta (S_\text{bulk} + S_\text{bdy(1)} ) = \int d\vphi\,  (\delta A_\vphi) \pi^\vphi \bigg|_{r=r_0}
   = -\frac{1}{e^2} \int d\vphi\, a(\vphi)\, \delta Q \,,
 }
which does not vanish under the assumption that $\delta a(\vphi) = 0$.  (It would have vanished under the assumption $\delta Q = 0$.)  To obtain a well-defined variational principle under the assumption $\delta a(\vphi) = 0$, we should perform a Legendre transform of $S_\text{bulk} + S_\text{bdy(1)}$ by adding the additional counterterm
 \es{Counter2}{
  S_\text{bdy(2)} = - \int d\vphi \, A_\vphi \pi^\vphi \,,
 }
because then 
 \es{var2}{
  \delta (S_\text{bulk} + S_\text{bdy(1)}  + S_\text{bdy(2)} ) = - \int d\vphi\, e^{-r} A_\vphi \, \delta \pi^\vphi \bigg|_{r=r_0} 
   = \frac{1}{e^2} \int d\vphi\, Q\, \delta a(\vphi) \,.
 }
The expression \eqref{var2}  indeed vanishes when $\delta a(\vphi) = 0$.  We conclude that the complete boundary counterterm we need is $S_\text{bdy(1)}  + S_\text{bdy(2)}$.  The covariant expression for it is given in \eqref{SBdy}.

As discussed in the Introduction, we propose that the on-shell action of the bulk theory for fixed $a(\vphi)$ corresponds to the partition function of the boundary theory of currents coupled to a background gauge field $a(\vphi)$. An important check of the proposed action is that it is gauge invariant under gauge transformations that go to a constant at the boundary. This is required, as these large gauge transformations correspond on the field theory side to gauge transformations of the background gauge field $a(\vphi)$. To check gauge invariance, note that $S_\text{bulk}$ is gauge invariant on its own, while the boundary terms change under \eqref{AsympExpansion} as:
 \es{Counter1b}{
  \De_\Lam\,  S_\text{bdy(1)} &\approx -\frac{1}{e^2} \int d\vphi \, Q  \partial_\vphi \Lambda(\infty, \vphi) \,, \\
  \De_\Lam\,  S_\text{bdy(2)} &\approx \frac{1}{e^2} \int d\vphi \, Q  \partial_\vphi \Lambda(\infty, \vphi)\,.
 }
Adding the two together, we find that the boundary term \eqref{SBdy} is gauge invariant, as desired.

\subsection{Boundary two-point function}

On-shell the Maxwell action is a total derivative, so it evaluates to a boundary term
 \es{MaxOnShell}{
  S_\text{bulk}^\text{on-shell} =  \frac{1}{2 e^2} \int dx \, \sqrt{\gamma} A^i F_{\mu i} n^\mu  \,,
 }
and the combined bulk and boundary action \eqref{MaxwellAction},~\eqref{SBdy} reduces to 
 \es{STotal}{
  S^\text{on-shell} = \frac{1}{2 e^2} \int_\partial dx\, \sqrt{\gamma}\, \left(A^i A_i - A^i F_{\mu i} n^\mu \right)
   =  -\lim_{r \to \infty} \frac{1}{2 e^2} \int d\vphi \, \sinh^{-1}r\, A_\vphi (F_{r\vphi} - A_\vphi) \,.  
 }
Using \eqref{AsympExpansion} together with the relation \eqref{GotQ} between $Q$ and $a(\vphi)$ as well as the fact that $\Lambda$ vanishes at large $r$, we can write the expression in \eqref{STotal} as
 \es{STotalAgain}{
  S^\text{on-shell}[a] = - \frac{1}{4 \pi e^2} \int d\vphi\, d\vphi' \, a(\vphi) a(\vphi') \,.
 }

According to the GKPW holographic dictionary \cite{Gubser:1998bc,Witten:1998qj},
\es{GKPWMain}{
e^{-S^\text{on-shell}[a]}=\langle e^{-\int a(\vphi) j(\vphi)}\rangle_\text{CFT}\,.
} 
Taking two functional derivatives of $-S^\text{on-shell}[a]$ with respect to $a$ thus yields the two-point function
 \es{jjCorr}{
  \langle j(\vphi_1) j(\vphi_2) \rangle = \frac{1}{2 \pi e^2} \,.
 }
Note that this two-point function is positive-definite as long as the bulk action has the same property.

\subsection{Non-linear bulk action and boundary $n$-point function}

Let us now generalize the previous discussion to a full non-linear bulk action and calculate $n$-point correlators of the dual operator $j$.  To get started, note that because the dual boundary theory is topological, the bulk theory has to have the reparametrization of the asymptotic boundary as a symmetry.  The Maxwell action \eqref{MaxwellAction} is invariant under area preserving diffeomorphisms, which indeed contains the boundary reparametrization symmetry. When writing down a non-linear bulk action, we want to preserve this symmetry.
The more general bulk action takes the form
 \es{BulkNonLinear}{
  S_\text{bulk} = \frac{1}{2 e^2} \int d^2 x \, \sqrt{g}\, W(F) \,,
 }
where $F\equiv \frac 12 \epsilon^{\mu\nu} F_{\mu\nu}$.  In this expression $\epsilon^{r\vphi} = - \epsilon^{\vphi r} = \frac{1}{\sqrt{g}}$, so with the metric \eqref{Back} we have $F = \sinh^{-1}r \,F_{r \vphi}$.  The function $W(F)$ is such that at small $F$ we have $W(F) = F^2 + \ldots$ so that it reproduces the action  \eqref{MaxwellAction} in the quadratic approximation.    Due to the fact that gauge fields do not propagate in 2d, the non-linear theory \eqref{BulkNonLinear} can also be solved exactly classically.   The solution of the equations of motion is still given by \eqref{AsympExpansion}.

With inspiration from the quadratic boundary term \eqref{SBdy}, we supplement the bulk action \eqref{BulkNonLinear} by the boundary term 
 \es{SBdyGeneral}{
 	S_\text{bdy} = \frac{1}{2 e^2} \int_\partial dx\, \sqrt{\gamma}\, \left(W_1(\epsilon^{\m\n}n_\m A_\n ) + W_2(F) A^i F_{\mu i} n^\mu \right) \,,
 }
 
for some functions $W_1$ and $W_2$ to be determined in terms of $W$.   In the coordinates \eqref{Back}, the total action takes the form
 \es{STotalGen}{
  S = \frac{1}{2 e^2} \int d\vphi \left[ \sinh r_0 W_1(A_\vphi \sinh^{-1} r_0)
    + W_2(F) F A_\vphi  +  \int_{0}^{r_0} dr\, \sinh r\, W[F] 
    \right] \,,
 }
where we set the boundary at a large value $r = r_0$.

The functions $W_1$ and $W_2$ are determined from the requirement that the combined on-shell action is well-defined and that the variational problem is well-posed.  Recalling $F = F_{r\vphi} \sinh^{-1} r$, using  \eqref{AsympExpansion}, and expanding at large $r_0$, we find the on-shell action
 \es{SOnShellGen}{
  S_\text{on-shell}
   &=  \frac{2 \pi e^{r_0}}{4 e^2} \left[   W_1(Q ) 
    + Q^2 W_2(Q)  +  W(Q) 
    \right]  - \frac{2\pi }{2 e^2}  \left[Q W_1'(Q) 
    + Q^2 W_2(Q)     + W(Q) \
    \right] \,,
 }
where $Q$ is given by \eqref{GotQ}.  The absence of UV divergences in the on-shell action requires
 \es{DivAbsence}{
   W_1(Q ) 
    + Q^2 W_2(Q)  +  W(Q)  = 0 \,.
 }
Imposing this relation, we find that the Euler-Lagrange variation results in the boundary term 
 \es{BdyTermEL}{
   \delta S &= \frac{1}{2 e^2} \int d\vphi \Biggl[  \biggl( W_1''(Q)  +  W_2(Q) + Q W_2'(Q) \biggr)  a \delta Q + \biggl(  W_1'(Q) + Q W_2(Q)  + W'(Q) \biggr)  \delta a 
    \Biggr] \,,
 }
so in order for the Euler-Lagrange variation to imply the equations of motion when $\delta a= 0$ we must have that the term proportional to $\delta Q$ vanishes.  Thus, 
 \es{ELCondition}{
  W_1''(Q)  +  W_2(Q) + Q W_2'(Q) = 0 \,.
 }
This equation is obeyed provided that
 \es{W1pEq}{
  W_1'(Q) + Q W_2(Q) = 0 \,.
 }
Eqs.~\eqref{DivAbsence} and \eqref{W1pEq} can be used to determine $W_1$ and $W_2$ in terms of $W$.\footnote{It is useful to have the solution explicitly. If we write
\es{WTaylor}{
W(Q)\equiv\sum_{n=2}^\infty w_n Q^n\,,
}
then $W_1$ and $W_2$ are given by:
\es{W12Taylor}{
W_1(Q)&=\sum_{n=2}^\infty {w_n\ov n-1} \,Q^n\,, \qquad W_2(Q)=-\sum_{n=2}^\infty {n\, w_n\ov n-1}\, Q^{n-2}\,.
}
}  

Using \eqref{W1pEq}, we see that the full non-linear on-shell action is 
 \es{OnShellGeneral}{
  S_\text{on-shell}[a] = - \frac{2\pi }{2 e^2}  W\left(-\frac{\int d\vphi\, a(\vphi)}{2 \pi} \right) \,.
 }
Since the on-shell action is minus the generating functional of connected correlators of the dual operator $-j(\vphi)$ \eqref{GKPWMain}, we can easily compute $n$-point functions of $j$ by taking functional derivatives of \eqref{OnShellGeneral}:
 \es{jCorr}{
  \langle j(\vphi_1) \cdots j(\vphi_n) \rangle 
   =  \frac{1}{2e^2} \frac{1}{(2\pi)^{n-1}} W^{(n)}(0) \,,
 }
where $W^{(n)}(0)$ is the $n$th derivative of the function $W$ evaluated at $F = 0$.  We will use this equation later when we study a boundary theory for which we can calculate all correlation functions in \eqref{jCorr}, and thus obtain the full non-linear bulk dual.

\section{Non-Abelian gauge theory in $AdS_2$}
\label{NONABELIAN}

\subsection{Yang-Mills theory}

Let us now extend the gauge theory presented in the previous section to the non-Abelian case.  For concreteness and simplicity, we will focus on the case where the gauge algebra is a simple Lie algebra $\mathfrak{g} = \text{Lie}(G)$, with $G$ being a Lie group, but our discussion can be generalized to other gauge algebras, and the final results do not make use of the existence or global structure of $G$.  The simplest bulk action is that of Yang-Mills theory,
 \es{YMAction}{
  S_\text{bulk} =  \frac{1}{2 g_\text{YM}^2} \int d^2 x \, \sqrt{g}\,  \tr (F_{\mu\nu} F^{\mu\nu} ) \,,
 }
the trace being taken in the fundamental representation.  
As was the case in the Abelian case, this action must be supplemented by appropriate boundary terms:
 \es{YMBdry}{
  S_\text{bdy} = \frac{1}{g_\text{YM}^2} \int_\partial dx\, \sqrt{\gamma}\, \tr \left(A^i A_i - 2 A^i F_{\mu i} n^\mu \right) \,.
 }
Here, $A = A^a T^a$, $F = F^a T^a$, etc., where $T^a$ are the Hermitian generators of the gauge algebra.  We normalize them such that $\tr T^a T^b = \frac 12 \delta^{ab}$, and we define the structure constants via $[T^a, T^b] = i f^{abc} T^c$. The action is gauge invariant under gauge transformations that go to a constant at the boundary, as can be checked by explicit computation.\footnote{
As in the Abelian case, $S_\text{bulk}$ is gauge invariant on its own. Under a gauge transformation $U(r, \vphi)$ that does not blow up near the boundary we have
 \es{Counter2b}{
 \De_U\,A &= U\le(i d +A\ri) U^{-1}-A\\
  \De_U\, \le[ \sqrt{\gamma}\, \tr \left(A^i A_i \ri)\ri] &\approx {2 \sinh^{-1} r} \, \tr \left(A_\vphi \,U (i\p_\vphi U^{-1})   \ri)\\
   \De_U\, \le[  \sqrt{\gamma}\, \tr \left(- 2 A^i F_{\mu i} n^\mu  \ri) \ri] &\approx -{2 \sinh^{-1} r} \, \tr \left(F_{r\vphi} \, U(i\p_\vphi U^{-1})   \ri)\,.
 }
With the assumption that $A_r$ decays near the boundary, the asymptotic analysis of the field equations implies that  $F_{r\vphi}\approx A_\vphi$. Using this, and adding the two lines of \eqref{Counter2} together, we get that the boundary term \eqref{YMBdry} is gauge invariant.} 

To construct the most general solution of the bulk equations of motion following from \eqref{YMAction}, we start, in the coordinate system \eqref{Back} with the solution
 \es{SimpleSol}{
  F_{r\vphi} &= Q \sinh r \,, \qquad
  A_{\vphi} = Q (\cosh r  - 1) \,, \qquad A_r = 0 \,,
 }
for constant Lie-algebra-valued $Q$.  The most general solution can be obtained from \eqref{SimpleSol} through a large gauge transformation with gauge parameter $U$, 
 \es{GeneralSol}{
  F_{r \vphi} &= U Q U^{-1} \sinh r \,, \\
  A_\vphi &= U Q U^{-1} (\cosh r  - 1) + i U \partial_\vphi U^{-1} \,, \\
  A_r &=  i U \partial_r U^{-1} \,.
 }
The Lie-group-valued $U(r, \vphi)$ is an arbitrary function of $r$ and $\vphi$, which has a finite limit  $U(\infty, \vphi) \equiv u(\vphi)$ as $r \to \infty$, corresponding to large gauge transformations.

To compute correlation functions of the non-Abelian conserved currents in the boundary theory, first note that under an Euler-Lagrange variation, we have
 \es{ELNonAb}{
  \delta (S_\text{bulk} + S_\text{bdy} ) = \frac{2}{g_\text{YM}^2} \int d\vphi \, \sinh^{-1} r\, \tr \left( 
   A_\vphi \delta (A_\vphi - F_{r \vphi} ) \right) \,,
 }
so the action $S_\text{bulk} + S_\text{bdy}$ implies the equations of motion provided that the boundary limit of $A_\vphi - F_{r \vphi} $ is held fixed.  We identify this boundary limit as the source $a(\vphi)$ of the non-Abelian conserved current in the boundary theory:
 \es{Source}{
  a(\vphi) = A_\vphi - F_{r \vphi} \biggr|_{r \to \infty}
   = i u \partial_\vphi u^{-1} - u Q u^{-1} \,.
 }

In the classical approximation, the on-shell action is minus the generating functional for connected correlators of $-j(\vphi)$.  It is easy to find this on-shell action by evaluating $S_\text{bulk} + S_\text{bdy}$ on the solution \eqref{GeneralSol}:
 \es{OnShellYM}{
  S_\text{on-shell}[a] = -\frac{2\pi}{g_\text{YM}^2} \tr Q^2 \,.
 }
Thus, our remaining task is to write $Q$ in terms of $a$, and then take functional derivatives of \eqref{OnShellYM} with respect to $a$ in order to calculate connected correlators in the boundary theory.  To this end, Eq.~\eqref{Source} can be recast as
 \es{uEq}{
  i \partial_\vphi u = - a u - u Q \,.
 }
The solution is
 \es{uSol}{
  u(\vphi) = P \exp \left[ i \int_0^\vphi d\vphi\, a(\vphi) \right] \, u_0 \, e^{i \vphi Q} \,,
 }
where $P$ stands for path ordering, and $u_0$ is a constant.  The gauge parameter $u(\vphi)$ being well-defined implies $u(0) = u(2 \pi)$, which further implies
 \es{QFormula}{
  Q = \frac{i}{2 \pi} u_0^{-1} \log \left( P \exp \left[ i \int_0^{2 \pi} d\vphi\, a(\vphi) \right]  \right)  u_0\,.
 }
This expression, together with \eqref{OnShellYM}, provides quite an explicit solution of the classical YM theory.\footnote{Note that the constant matrix $u_0$ drops out from the expression for the on-shell action.}  One can further extract boundary correlators using \eqref{GKPWMain}
 \es{Correl}{
  \langle j^{a_1}(\vphi_1) \cdots j^{a_n}(\vphi_n)  \rangle
   = (-1)^n \frac{\delta (- S_\text{on-shell}[a])}{\delta a^{a_1}(\vphi_1) \cdots \delta a^{a_n}(\vphi_n)  }\Bigg\vert_{a=0}\,.
 }

We remark that there is a quick argument that allows us to omit the steps \eqref{GeneralSol}--\eqref{QFormula} and to immediately go from \eqref{SimpleSol} to writing down the final result for the on-shell action obtained by plugging \eqref{QFormula} into \eqref{OnShellYM}. First, we note that action is invariant under large gauge transformations, hence the only object that it can depend on is the holonomy of $a(\vphi)$, $\Ga=P \exp \left[ i \int_0^{2 \pi} d\vphi\, a(\vphi) \right]$. The functional dependence on the holonomy can be determined by evaluating the on-shell action on the simple solution \eqref{SimpleSol}, and noting that for \eqref{SimpleSol} the holonomy is given by $\Ga=\exp\left[ -2\pi i\, Q  \right]$.  One then immediately obtains \eqref{OnShellYM} with $Q$ defined by \eqref{QFormula} with $u_0 = 1$.  (As mentioned before, $u_0$ drops out from plugging \eqref{QFormula} into \eqref{OnShellYM}.)

A comment on path ordering of points on a circle is in order.  The path ordering $P$ on the circle can be defined  by considering a reference point $\vphi_0$, and then taking the values of the insertion points $\vphi_i$ to be in the interval $[\vphi_0, \vphi_0 + 2 \pi)$.  On this interval, one may use the usual definition of path ordering on a line.  (Above, we implicitly used this definition with $\vphi_0 = 0$.)  One can check quite explicitly that the final answers for the correlators do not depend on the precise value of $\vphi_0$.  More conceptually, the holonomy $\Gamma$ does depend on the reference point, but only up to a gauge transformation, so the dependence on $\vphi_0$ drops out from gauge-invariant quantities like the on-shell action \eqref{OnShellYM}.

\subsection{Examples of correlation functions}
\label{CORRELATORS}

Let us now provide explicit formulas for 2-, 3-, and 4-point functions.  We will compare these formulas with the field theory formulas in Sections~\ref{VECTOR} and~\ref{MATRIX}.  To obtain explicit formulas, we can obtain $u_0 Q u_0^{-1}$ to the first few orders in $a$ using the Magnus expansion:
 \es{QExpansion}{
  (u_0 Q u_0^{-1})^a &= -\frac{1}{2\pi} \int_0^{2\pi} d\vphi_1 \, a^a(\vphi_1)
   + \frac{f^{abc}}{4 \pi} \int_0^{2 \pi} d \vphi_1 \, d \vphi_2 \,a^b(\vphi_1) a^c(\vphi_2) \theta( \vphi_{12}) \\
   &{}-\frac{f^{abc} f^{cde}}{12 \pi}  
    \int_0^{2 \pi} d \vphi_1 \, d \vphi_2 \, d \vphi_3 \,
     a^b(\vphi_1) a^d(\vphi_2) a^e(\vphi_3) \left[ \theta (\vphi_{12}) \theta( \vphi_{23}) + \theta (\vphi_{32}) \theta(\vphi_{21}) \right] + \cdots
 }

Then the on-shell action becomes:
 \es{SOnShell}{
  S_\text{on-shell}[a] &= -\frac{2 \pi}{2 g_\text{YM}^2}
   \Biggl[\frac{1}{4 \pi^2} \int_0^{2 \pi} d \vphi_1\, d\vphi_2\, a^a(\vphi_1) a^a(\vphi_2) \\
   &{}-\frac{f^{abc}}{4 \pi^2}  \int_0^{2 \pi} d \vphi_1\, d\vphi_2\, d\vphi_3 \,
    a^a(\vphi_1) a^b(\vphi_2) a^c (\vphi_3) \theta(\vphi_{23}) \\
   &{}+\frac{f^{abc} f^{ade}}{4 \pi^2}  
    \int_0^{2 \pi} d \vphi_1\, d\vphi_2\, d\vphi_3 \, d\vphi_4\, 
     a^b(\vphi_1) a^c(\vphi_2) a^d(\vphi_3) a^e (\vphi_4)
     \\ &\qquad\qquad\qquad\qquad{}\times \left( \frac{\theta(\vphi_{12}) \theta(\vphi_{34})}{4} + \frac{ \theta (\vphi_{23}) \theta( \vphi_{34}) + \theta (\vphi_{43}) \theta(\vphi_{32}) }{3} \right)
   \Biggr] + \cdots \,.
 }

Using \eqref{Correl} and \eqref{SOnShell}, the two-point function can then be written as
 \es{TwoPointYM}{
  \langle j^a(\vphi_1) j^b(\vphi_2) \rangle = \frac{\delta^{ab}}{2 \pi g_\text{YM}^2} \,.
 } 
For the three-point function, a bit of algebra gives
  \es{ThreePointYM}{
  \langle j^a(\vphi_1) j^b(\vphi_2) j^c(\vphi_3) \rangle &= 
    -\frac{f^{abc}}{4 \pi g_\text{YM}^2}  \sgn (\vphi_{12} \vphi_{23} \vphi_{31})  \,.
 } 
For the four-point function, also a bit of algebra as well as the use of the Jacobi identity results in the expression
 \es{FourPointYMFinal}{
  \langle j^a(\vphi_1) j^b(\vphi_2) j^c(\vphi_3) j^d(\vphi_4) \rangle
   &= \frac{1}{24 \pi g_\text{YM}^2}  
 \Biggl[ 
    f^{abe} f^{cde}  \left[\sgn (\vphi_{12} \vphi_{24} \vphi_{43} \vphi_{31} )
     - \sgn(\vphi_{12} \vphi_{23} \vphi_{34} \vphi_{41})
       \right] \\
   &{}+f^{ace} f^{bde} \left[\sgn (\vphi_{12} \vphi_{24} \vphi_{43} \vphi_{31} )
   - \sgn(\vphi_{13} \vphi_{32} \vphi_{24} \vphi_{41})
        \right] \\
  &{}+f^{ade} f^{cbe}  \left[ \sgn(\vphi_{13} \vphi_{32} \vphi_{24} \vphi_{41})
   - \sgn(\vphi_{12} \vphi_{23} \vphi_{34} \vphi_{41})
      \right]
   \Biggr] \,.
 }
 The simplicity of YM$_2$ on the $AdS_2$ background allowed for a very explicit formula for the on-shell action given in \eqref{OnShellYM} and \eqref{QFormula}. In AdS/CFT, correlation functions are usually computed using Witten diagrams. In Appendix~\ref{WITTENDIAG} we demonstrate how Witten diagrams reproduce the 2- and 3-point functions, and make some steps towards computing the the 4-point function \eqref{FourPointYMFinal}.

\subsection{Higher-derivative generalization}
\label{NONABELIANHIGHER}

More generally, we can consider the non-Abelian counterpart of the non-linear bulk action \eqref{BulkNonLinear}
 \es{SbulkNonAbNonLinear}{
  S_\text{bulk} = \frac{1}{g_\text{YM}^2} \int d^2x \, \sqrt{g} \le[\frac12 F^a F^a+ {2\pi \,d_3^{abc}\ov  3!} F^aF^b F^c+\dots+ {(2\pi)^{n-2} \, d_n^{a_1\dots a_n} \ov   n!}F^{a_1}\dots F^{a_n}+\dots\ri] \,,
 }
where $F^a = \frac{1}{2} \epsilon^{\mu\nu} F^a_{\mu\nu}$, and where $d_n^{a_1\dots a_n}$ are totally symmetric tensors required by  gauge invariance to also be invariant tensors of the Lie algebra. The boundary terms can be written down in analogy with the non-linear Abelian case
\es{SBdyNonAbNonLinear}{
  S_\text{bdy} =& \frac{1}{g_\text{YM}^2} 
   \int d\vphi \  A^a_\vphi\left[\frac12\le({A^a_\vphi\ov \sinh r_0}-2 F^a\ri)+ {2 \pi \, d_3^{abc}\ov  3!} \le(\frac12 \, {A^b_\vphi A^c_\vphi\ov \sinh^2 r_0}-\frac32 \, F^b F^c \ri)+\dots\ri.\\
   &\le.+{(2\pi)^{n-2} \, d_n^{a a_2\dots a_n}\ov  n!} \le({1\ov n-1} \, {A^{a_2}_\vphi \dots A^{a_n}_\vphi\ov \sinh^{n-1} r_0}-{n\ov n-1} \, F^{a_2} \dots F^{a_n} \ri)+\dots\ri] \,,
 }
 where the numerical coefficients are identical to the ones appearing in \eqref{W12Taylor}. It can be shown in the same way as in the two-derivative case (going order by order in the fields) that the action is invariant under large gauge transformations,\footnote{To see this it is important to use the defining identity of an invariant tensor
 \es{InvTensor}{
0=f^{ b c (a}\, d_n^{a_1 \dots a_{n-1}) c} \,.
 }} and the variational problem is well-posed with the boundary condition that $A_\vphi- F_{r\vphi}$ is held fixed at the boundary. As a consequence of a having a theory with no propagating degrees of freedom, the equations of motion for the non-linear theory are still solved by the expression \eqref{GeneralSol} we found for the two-derivative theory. Hence, the on-shell action evaluates to
  \es{SOnShellNonAb}{
  S_\text{on-shell}[a] = -\frac{1}{  2\pi g_\text{YM}^2} \le[{ (2\pi)^2\ov2}  Q^a Q^a+ {(2 \pi)^3\,d_3^{abc}\ov 3! } Q^a Q^b Q^c+\cdots + {(2\pi)^{n} \, d_n^{a_1\dots a_n} \ov n!}Q^{a_1} \cdots Q^{a_n}+\cdots\ri]\,,
 }
 with $Q$ being given by \eqref{QFormula} in terms of the boundary source $a$.  One can then explicitly compute correlation functions as in the previous subsection.

The contribution of the $Q^a Q^a$ term in \eqref{SOnShellNonAb} to 2-, 3-, and 4-point functions was worked out for a general gauge group in Section~\ref{CORRELATORS}. The cubic (quartic) term only starts to contribute to 3-point (4-point) functions, while the higher order terms in \eqref{SbulkNonAbNonLinear} and \eqref{SOnShellNonAb} contribute only to higher point functions.
The cubic term contributes to the 3-point function a space independent term:
\es{Contrib3pt}{
  \langle j^a(\vphi_1) j^b(\vphi_2) j^c(\vphi_3) \rangle &=
    -\frac{f^{abc}}{4 \pi g_\text{YM}^2}  \sgn (\vphi_{12} \vphi_{23} \vphi_{31}) +{d_3^{abc} \ov 2 \pi g_\text{YM}^2}\,,
}
where we also included the contribution of the Yang-Mills term \eqref{ThreePointYM}. Note that the contribution of the Yang-Mills term is totally antisymmetric (and dependent on the ordering of the currents), while the higher derivative term adds a term symmetric in the adjoint indices. The contribution to the 4-point function is:
\es{Contrib4pt}{
& \langle j^a(\vphi_1) j^b(\vphi_2) j^c(\vphi_3) j^d(\vphi_4) \rangle \\
 &=\text{Eq.~\eqref{FourPointYMFinal}}+{1  \ov 2 \pi g_\text{YM}^2}\, \le[d_4^{abcd}+ {1\ov 2} \le(d_3^{cde}f^{eab}\sgn(\vphi_{12})+d_3^{bde}f^{eac}\sgn(\vphi_{13})+d_3^{bce}f^{ead}\sgn(\vphi_{14})\ri.\ri.\\
& \le.\le.+d_3^{ade}f^{ebc}\sgn(\vphi_{23})+d_3^{ace}f^{ebd}\sgn(\vphi_{24})+d_3^{abe}f^{ecd}\sgn(\vphi_{34})\ri)\ri]\,,
}
where again there is a difference in symmetry between the new contributions and the Yang-Mills result
\eqref{FourPointYMFinal}. We can isolate $d_n^{a_1\dots a_n}$ from correlation functions by considering the totally symmetric part:
\es{Contribnpt}{
\langle j^{(a_1}(\vphi_1)\dots j^{a_n)}(\vphi_n) \rangle={1 \ov  2\pi g_\text{YM}^2}\, d_n^{a_1\dots a_n}\,.
}

There is an interesting puzzle that arises. We observe that from \eqref{SOnShellNonAb} we can only ever get contributions to the $k$-point functions that are linear in the tensors $d_n^{a_1\dots a_n}$ with $n\leq k$. However, if we thought about the $d_n^{a_1\dots a_n}$ as interaction vertices in the theory, there should be Witten diagrams that contain multiple $d_n^{a_1\dots a_n}$ insertions. Eq.~\eqref{SOnShellNonAb} is of course correct, and we explain why in  Appendix~\ref{WITTENDIAG}.

\subsection{Example:  $\mathfrak{su}(2)$ theory}
\label{SU2EXAMPLE}

As an example that will be useful later, let us consider an $\mathfrak{su}(2)$ gauge theory. All totally symmetric invariant tensors of the $\mathfrak{su}(2)$ Lie algebra are generated by $\de^{ab}$.  For example:\footnote{Another familiar way of saying this is that  in the fundamental representation of  $\mathfrak{su}(2)$, $F = \frac 12 F^a \sigma_a$ , we have $\tr F^4 = \frac 12 (\tr F^2)^2$.}
\es{d4}{
d_3^{abc}&=0 \,, \\
d_4^{abcd}&=w_4\, \de^{(ab}\de^{cd)}={w_4\ov 3}\, \le( \delta^{ab} \delta^{cd} 
   + \delta^{ac} \delta^{bd} + \delta^{ad} \delta^{bc}\ri)\,.
}
It then follows that the most general non-linear $\mathfrak{su}(2)$  theory can be rewritten in the convenient form
 \es{SbulkNonAbNonLinearSU2}{
  S_\text{bulk} &= \frac{1}{g_\text{YM}^2} \int d^2x \, \sqrt{g} \tr W(F) \,, \\
  S_\text{bdy} &= \frac{1}{g_\text{YM}^2} 
   \int_\partial dx \sqrt{\gamma} \tr \left( 
    W_1 (\epsilon^{\m\n}n_\m A_\n) + W_2(F) A^i F_{\mu i} n^\mu \right)  \,,
 }
with $W(F)=F^2+{4\pi^2 w_4\ov 3}\, F^4+\dots$, and the functions $W_1$ and $W_2$ determined by Eqs.~\eqref{DivAbsence} and \eqref{W1pEq} as in the Abelian case.

Using the general formulas \eqref{Contrib3pt} and \eqref{Contrib4pt}, we see that only the 4-point function gets modified from its Yang-Mills value by 
 \es{Contrib}{
     \frac{w_4}{6 \pi g_\text{YM}^2} \left( \delta^{ab} \delta^{cd} 
   + \delta^{ac} \delta^{bd} + \delta^{ad} \delta^{bc} \right) \,.
 } 
Using that the structure constants are $f^{abc} = \epsilon^{abc}$, the 2-, 3-, and 4- point functions are thus
 \es{su2Correl}{
  \langle j^a(\vphi_1) j^b(\vphi_2) \rangle &= \frac{\delta^{ab}}{2 \pi g_\text{YM}^2} \,, \\
  \langle j^a(\vphi_1) j^b(\vphi_2) j^c(\vphi_3) \rangle &= -\frac{\epsilon^{abc}}{4 \pi g_\text{YM}^2}  \sgn (\vphi_{12} \vphi_{23} \vphi_{31}) \,, \\
  \langle j^a(\vphi_1) j^b(\vphi_2) j^c(\vphi_3) j^d(\vphi_4) \rangle &= \frac{ 1+{4 w_4}}{24 \pi g_\text{YM}^2}  \\
   &{}\times 
 \Biggl[ \left(   \delta^{ab} \delta^{cd} + \delta^{ad} \delta^{bc}  - \frac{ 2- 4w_4 }{1 + 4w_4}\delta^{ac} \delta^{bd}  \right)
      \sgn(\vphi_{12} \vphi_{23} \vphi_{34} \vphi_{41})
       \\  
    &{}+\left(\delta^{ab} \delta^{cd} + \delta^{ac} \delta^{bd} - \frac{ 2- 4w_4}{1 + 4w_4 } \delta^{ad} \delta^{bc} )  \sgn (\vphi_{12} \vphi_{24} \vphi_{43} \vphi_{31} \right)
       \\
  &{}+\left(\delta^{ac} \delta^{bd} +\delta^{ad} \delta^{bc} - \frac{ 2- 4w_4 }{1 + 4w_4 } \delta^{ab} \delta^{cd} \right)  
   \sgn(\vphi_{13} \vphi_{32} \vphi_{24} \vphi_{41})
   \Biggr]   \,,
 }
where we combined \eqref{Contrib} with \eqref{TwoPointYM}, \eqref{ThreePointYM}, and \eqref{FourPointYMFinal} and used that $\sgn(\vphi_{12} \vphi_{23} \vphi_{34} \vphi_{41}) + \sgn (\vphi_{12} \vphi_{24} \vphi_{43} \vphi_{31} ) + \sgn(\vphi_{13} \vphi_{32} \vphi_{24} \vphi_{41}) = 1$.

We will see that we can obtain these expressions from the field theory. In the case of free hypermultiplets in Section~\ref{FREEHYPERS} we will find that we need to set $w_4 = 1/8$, while in the case of ABJM theory $w_4 =0$ as shown in Section~\ref{BULKDUAL}.

\section{Topological theories in 1d}
\label{TOPOLOGICAL}

We now move on to the boundary side of our proposed duality, and start by reviewing how the TQM models arise from 3d ${\cal N} = 4$ SCFTs.  In Section~\ref{ENHANCED} we will explain why these models have an enhanced global symmetry, and later on in Sections~\ref{VECTOR} and~\ref{MATRIX} we will provide explicit examples.

\subsection{Review of topological theories in 1d from 3d ${\cal N} = 4$ SCFTs}
\label{REVIEW}

As noticed in \cite{Beem:2013sza} and explored in more detail in \cite{Chester:2014mea,Beem:2016cbd,Dedushenko:2016jxl}, 3d ${\cal N} =4$ SCFTs contain 1d topological sectors representing the cohomology of an appropriately chosen nilpotent supercharges.  For a given ${\cal N} = 4$ SCFT, there are two inequivalent such constructions, one corresponding to the Higgs branch and one to the Coulomb branch;  they are exchanged by 3d mirror symmetry.  Let us focus on the Higgs branch case, and consider a choice for the nilpotent supercharge ${\cal Q}$ such that the 1d theory lives on a line in 3d that can be identified with the $x_1$ axis.  

The 3d local operators which also correspond to operators in the 1d theory are position-dependent linear combinations of the 3d ``Higgs branch operators'' inserted anywhere along the $x_1$ axis.  The Higgs branch operators are half BPS Lorentz scalar operators with scaling dimension $\Delta = j$, where $j$ is the spin under the $\mathfrak{su}(2)_H$ subalgebra of the R-symmetry algebra $\mathfrak{su}(2)_H \times \mathfrak{su}(2)_C$ of the ${\cal N} = 4$ SCFT\@.  In particular, if ${\cal O}_{(a_1 \ldots a_{2j})}(\vec{x})$ is a Higgs branch operator, where $a_i = 1, 2$ are $\mathfrak{su}(2)_H$ fundamental indices, then the 1d operator in the topological theory is
 \es{1dOp}{
  {\cal O}(x) = u^{a_1}(x) \ldots u^{a_{2j}}(x) {\cal O}_{(a_1 \ldots a_{2j})}(x, 0, 0) \,, \qquad
   u^a(x) = \begin{pmatrix}
    1&
    \displaystyle{\frac{x}{2 r} }
   \end{pmatrix} \,,
 }
where $r$ is a length scale that was introduced for dimensional reasons.  We refer to the 1d operators ${\cal O}(x)$ as twisted Higgs branch operators.  They have topological correlation functions because their translation in $x$, which in 3d corresponds to a translation along $x_1$ combined with an $\mathfrak{su}(2)_H$ transformation, is ${\cal Q}$-exact.  For details on this construction, we refer the reader to  \cite{Chester:2014mea,Beem:2016cbd,Dedushenko:2016jxl}. 

The simplest example of a 1d topological sector arises from the 3d SCFT of a free hypermultiplet.  The hypermultiplet has 4 real scalars that can be grouped into the complex combinations $q_a$ and $\tq_a$, with $\tq^a = (q_a)^*$, which represent the Higgs branch operators with the lowest dimensions in this theory:  they have $\Delta = j = 1/2$.  The twisted combinations of hypermultiplet scalars appearing in the 1d topological theory are
 \es{twistedHyper}{
  Q(x) = q_1(x, 0, 0) + \frac{x}{2r} q_2 (x, 0, 0) \,, \qquad
  \tQ(x) = \tq_1(x, 0, 0) + \frac{x}{2r} \tq_2 (x, 0, 0) \,.
 } 
All other twisted Higgs branch operators in this theory are products of $Q(x)$ and $\tQ(x)$.  If one starts with $N$ hypermultiplets, then the twisted Higgs branch operators will be products of $Q_I(x)$ and $\tQ^I(x)$, where $I = 1, \ldots, N$.  If one further gauges a subgroup of the $\mathfrak{u}(N)$ flavor symmetry under which the $Q_I$ and $\tQ^I$ transform as ${\bf N}$ and $\overline{\bf N}$, respectively, then the twisted Higgs branch operators will simply be gauge invariant products of $Q_I(x)$ and $\tQ^I(x)$.

Ref.~\cite{Dedushenko:2016jxl} derived a Lagrangian description of the 1d topological theory mentioned above in the case where the 3d ${\cal N} = 4$ SCFT has a Lagrangian UV description in terms of hypermultiplets and vector multiplets.  The most general setup considered in \cite{Dedushenko:2016jxl} is where the 3d gauge group is $G$, whose algebra is $\mathfrak{g}$, under which $Q$ and $\tQ$ transform in the possibly reducible representations ${\cal R}$ and $\overline{{\cal R}}$.  For the purposes of writing down the 1d theory explicitly, it was convenient to put the 3d ${\cal N} = 4$ SCFT on a round three-sphere of radius $r$, and thus consider the 1d topological theory as being defined on a great circle on this sphere as opposed to the line parameterized by $x_1$ embedded in $\R^3$, as was discussed above.  The 1d theory is of course topological, so one can simply map all correlation functions between the circle and the line by taking, for instance, $x_1 = 2r \tan \frac{\vphi}{2}$.  Upon performing the mapping to the circle, it is easy to see that $Q(\vphi)$ and $\tQ(\vphi)$ are now anti-periodic.  The partition function is 
 \es{Part1d}{
  Z = \frac{1}{\abs{\cal W}} \int_\text{Cartan of $\mathfrak{g}$} d\sigma 
   \det{}'_\text{adj}(2 \sinh(\pi \sigma)) Z_\sigma\,, 
  }
where
 \es{ZsDef}{   
    Z_\sigma \equiv \int DQ \, D\tQ\, 
    \exp\left[{4 \pi r \int d\vphi\,  \tQ (\partial_\vphi + \sigma) Q } \right]\,.
 }
As explained in \cite{Dedushenko:2016jxl}, the theory \eqref{Part1d} can be interpreted as a gauge-fixed quantum mechanics with a non-standard integration cycle in the path integral.  (In \eqref{ZsDef}, the integral over $Q$ and $\tQ$ is over a middle-dimensional cycle in the space of complex-valued fields.)  The value of $Z_\sigma$ appearing in \eqref{Part1d} obtained by integrating over $Q$ and $\tQ$ with no additional insertions is \cite{Dedushenko:2016jxl}
 \es{Zsigma}{
  Z_\sigma = \frac{1}{\det_{\cal R} (2 \cosh (\pi \sigma) )} \,.
 }
This result can also be obtained from the supersymmetric localization calculation performed by Kapustin, Willett, and Yaakov \cite{Kapustin:2009kz}.

The theory \eqref{Part1d} can be used for computing correlation functions of the 1d operators built out of $Q$ and $\tQ$ by simply inserting such operators in \eqref{Part1d}.  Normalized correlators can be written as
 \es{NormalizedCorr}{
  \langle {\cal O}_1(\vphi_1) \cdots {\cal O}_n(\vphi_n) \rangle 
   = \frac{1}{Z}  \int_\text{Cartan of $\mathfrak{g}$} d\sigma 
   \det{}'(2 \sinh(\pi \sigma)) Z_\sigma \langle {\cal O}_1(\vphi_1) \cdots {\cal O}_n(\vphi_n) \rangle_\sigma \,.
 }
This formula should be interpreted as a two-step process:  1)  one first calculates the expectation value $\langle {\cal O}_1(\vphi_1) \cdots {\cal O}_n(\vphi_n) \rangle_\sigma$ at fixed $\sigma$;  and  2) one then calculates the correlator $ \langle {\cal O}_1(\vphi_1) \cdots {\cal O}_n(\vphi_n) \rangle$ using \eqref{NormalizedCorr}.  Since at fixed $\sigma$, the theory \eqref{Part1d} is quadratic, the correlation function $\langle {\cal O}_1(\vphi_1) \cdots {\cal O}_n(\vphi_n) \rangle_\sigma$ can be calculated using the Wick theorem with a propagator derived from $Z_\sigma$:
 \es{Propag}{
  G_{\sigma} (\vphi_1- \vphi_2) \equiv \langle Q(\vphi_1) \tQ(\vphi_2) \rangle 
   = -\frac{\sgn (\vphi_1 - \vphi_2) {\bf 1} + \tanh (\pi \sigma )}{8 \pi r} e^{- \sigma (\vphi_1 - \vphi_2) } \,.
 }
Here, we should think of $Q$ as a column vector of dimension $\dim {\cal R}$ and of $\tQ$ as a row vector of the same dimension. Thus, ${\bf 1}$ is the $\dim {\cal R} \times \dim {\cal R}$ identity matrix and $\sigma$ is understood to be also represented by a $\dim {\cal R} \times \dim {\cal R}$ matrix in the representation ${\cal R}$.   The Green's function $G_\sigma(\vphi)$ in \eqref{Propag} can be easily inferred from \eqref{Part1d} as being the anti-periodic solution of the equation
 \es{eomG}{
  (\partial_\vphi + \sigma) G_\sigma(\vphi) = - 4 \pi r \delta(\vphi) \,,
 }
which is the Green's function equation following from \eqref{Part1d}.

An important generalization of the discussion above is to include real mass parameters, which is realized by replacing $Z_\sigma$ with $Z_{\sigma + mr}$ in \eqref{Part1d}, and consequently calculating the Wick contractions at fixed $\sigma$ using the propagator $G_{\sigma + mr}(\vphi)$ instead of $G_\sigma(\vphi)$.

In Sections~\ref{VECTOR} and~\ref{MATRIX}, we will use the results reviewed above in order to study specific 1d theory that arise from 3d ${\cal N} =4 $ SCFTs.  From now on we set $r = 1$.

\subsection{Enhancement of symmetries} \label{ENHANCED}

From the model \eqref{Part1d}--\eqref{ZsDef} we see that the equations of motion for $Q$ and $\tQ$ read $\partial_\vphi Q = -\sigma Q$ and $\partial_\vphi \tQ = + \sigma \tQ$, which implies that all gauge invariant operators $j$ built out of $Q$ and $\tQ$ obey $\partial_\vphi j = 0$.  This equation implies that the correlation functions of the operators $j$ are topological, and it also shows that these operators can be interpreted as conserved currents.  There is of course an infinite number of such currents, so the 1d theory is expected to have an infinite-dimensional emergent global symmetry.  Such a symmetry needs to be emergent because it is not visible in \eqref{Part1d}--\eqref{ZsDef}.

Let us explore this emergence of this infinite-dimensional global symmetry more generally.  Let us consider a topological theory of currents $j^A$. The OPE of two currents is:
\es{CurrentOPE}{
j^A(0)  j^B(\vphi)=B\,\de^{AB}+\frac12\, f^{ABC} j^C(0)\, \sgn(\vphi)+d_3^{ABC}j^C(0)\,,
}
where $B$ is a normalization constant that determines the scale of all correlation functions,\footnote{It scales like $N$ in the vector large $N$  and $N^{3/2}$  in the matrix large $N$  theories.} and by writing $\sgn(\vphi)$ we assume that $\vphi\in [-\pi,\pi)$. Note that the OPE is nonsingular. From what we wrote  $f^{ABC}$ is antisymmetric while $d_3^{ABC}$ is symmetric in the first two indices. Let us consider the three-point functions 
\es{ThreepointArg}{
G_A&\equiv \langle j^{[A}(\vphi_1) j^{B]}(\vphi_2) j^{C}(\vphi_3) \rangle \,, \\
G_S&\equiv \langle j^{(A}(\vphi_1) j^{B)}(\vphi_2) j^{C}(\vphi_3) \rangle\,,
 }
with $-\pi\leq\vphi_1<\vphi_2<\vphi_3<\pi$. Using the OPE \eqref{CurrentOPE} for all possible choices of currents, we obtain
\es{AntiSymm}{
G_A&={B\ov 2}\,f^{[AB]C}={B\ov 2}\,f^{[b|c|a]}={B\ov 2}\,f^{C[AB]} \,, \\
G_S&=B\, d_3^{(AB)C}=B\, d_3^{(B|C|A)}=B\, d_3^{C(AB)}\,,
}
where in order to take the OPE of $j^C$ and $j^A$ we used that the theory is on the circle. Combining these equations with the antisymmetry (symmetry) of $f^{ABC}$ ($d_3^{ABC}$) in the first two indices, we obtain that $f^{ABC}$ is completely antisymmetric and $d_3^{ABC}$ is completely symmetric in all three indices. This is just the consequence of the associativity of the OPE\@.  

The currents lead to a large symmetry algebra, which can be seen as follows. We claim that the transformation
\es{SymTf}{
\de_\Lam j^A=f^{ABC}\Lam^B j^C
}
is a symmetry of correlation functions, namely
\es{SymCorr}{
0&=\de_\Lam \langle j^{A_1}(\vphi_1)\dots j^{A_n}(\vphi_n) \rangle\,.
}
Without loss of generality let us assume that $-\pi\leq\vphi_1<\dots<\vphi_n<\pi$, and let us write:
\es{SymCorr2}{
\de_\Lam\langle j^{A_1}(\vphi_1)\dots j^{A_n}(\vphi_n) \rangle&=\sum_{i=1}^n \langle j^{A_1}(\vphi_1)\cdots \le[\de_\Lam j^{A_i}(\vphi_i)\ri]\cdots j^{A_n}(\vphi_n) \rangle\\
&\hspace{-1.2in}=\sum_{i=1}^n f^{A_i BC}\Lam^B\, \langle j^{A_1}(\vphi_1)\cdots j^{C}(\vphi_i)\cdots j^{A_n}(\vphi_n) \rangle\\
&\hspace{-1.2in}={\Lam^B\ov2}\,\sum_{i=1}^n  \langle j^{A_1}(\vphi_1)\cdots \le[j^{B}(\vphi_i)j^{A_i}(\vphi_i+\ep)-j^{A_i}(\vphi_i-\ep)j^{B}(\vphi_i)\ri]\cdots j^{A_n}(\vphi_n) \rangle\,,
}
where in the second line we used the definition \eqref{SymTf}, and in the third line we used \eqref{CurrentOPE}, introduced an infinitesimal angle $\ep$, and wrote the terms in the correlation function in increasing order of angles.  Writing out the terms in the sum, we see that all but two terms cancel automatically, and we are left with
\es{SymCorr3}{
\de_\Lam\langle j^{A_1}(\vphi_1)\cdots j^{A_n}(\vphi_n) \rangle&={\Lam^B\ov2} \biggl[\langle j^{B}(\vphi_1)j^{A_1}(\vphi_1+\ep)\cdots j^{A_n}(\vphi_n) \rangle\\
&-\langle j^{A_1}(\vphi_1)\cdots j^{A_n}(\vphi_n-\ep) j^{B}(\vphi_n)\rangle\biggr] \,,
}
where the right-hand side vanishes for a topological theory on a circle. This proves the existence of the infinite-dimensional symmetry algebra with transformation law given by \eqref{SymTf}. The vacuum also preserves the symmetry, which follows from the vanishing of all 1-point functions, $\langle j^{A}(\vphi)\rangle=0$.\footnote{ In 1d symmetry breaking is impossible on general grounds. } 

The emergent infinite dimensional symmetry has consequences, which we now discuss.  Let us define a family of totally symmetric tensors through the connected correlation functions:
\es{SymTensorDef}{
\langle j^{(A_1}(\vphi_1)\cdots j^{A_n)}(\vphi_n) \rangle=B\, d_n^{A_1\dots A_n}\,, 
}
where there is no dependence on the positions. These can be easily matched with bulk computations in \eqref{Contribnpt}.   Invariance under the symmetry algebra \eqref{SymTf} then implies: 
\es{fdIdentities}{
0&= f^{DE[A}\,f^{BC]E} \,, \\
0&=f^{ DE (A}\, d_n^{A_1 \dots A_{n-1}) E}\,.
} 
The first equation is the Jacobi identity, while the second equation is the defining property of invariant tensors---see also \eqref{InvTensor}.  We can thus confidently interpret the $f^{ABC}$ as the structure constants of the infinite-dimensional enhanced symmetry and the tensors $d_n^{A_1 \dots A_n}$ as invariant tensors.

We can now apply this discussion to specific theories of interest.  Let us focus on Higgs branch theories at finite $N$. They have an infinite set of currents.  In the theory of $NM$ free hypermultiplets discussed in the Introduction the currents in the $\mathfrak{u}(N)$ singlet sector are
 \es{UNSingletNM2}{
  j^{i_1\dots i_n} {}_{j_1\dots j_n} = \le(\tQ^{I_1 i_1} Q_{I_1 j_1} \ri)\cdots  \le(\tQ^{I_n i_n} Q_{I_n j_n} \ri)\,.
 } 
 In the ABJM case the currents are given by
 \es{multiTr}{
j\le[ {}^{p_{11}\dots p_{1k_1}}_{q_{11}\dots q_{1k_1}}\big \vert \cdots \big\vert {}^{p_{n1}\dots p_{nk_n}}_{q_{n1}\dots q_{nk_n}}  \ri]= \tr (X^{p_{11}} \tX^{q_{11}}  \cdots X^{p_{1k_1}} \tX^{q_{1k_1}}  )\cdots \tr (X^{p_{n1}} \tX^{q_{n1}}  \cdots X^{p_{nk_n}} \tX^{q_{nk_n}}  ) \,.
 }
(Of course, not all of the operators in \eqref{multiTr} are linearly-independent because of the existence of trace relations, but we can consider a basis of linearly independent operators.)  In both cases, the symmetry generated by these currents is infinite-dimensional.  %

In the large $N$ limit, however, the field theories undergo simplification, and we restrict attention to single-trace operators. In the vector large $N$ case there are finitely many single-trace currents: $ j^{i} {}_{j}=\tQ^{I i} Q_{I j}$, and hence the symmetry algebra of the single trace sector is finite-dimensional,  while in the matrix-like ABJM case there is an infinite set of single trace currents  $j\le[ {}^{p_{1}\dots p_{k}}_{q_{1}\dots q_{k}}\ri]= \tr (X^{p_{1}} \tX^{q_{1}}  \cdots X^{p_{k}} \tX^{q_{k}}  )$, and the corresponding symmetry algebra is infinite-dimensional.

In our proposed TQM/YM$_2$ dualities, it is only the single-trace currents that should have dual bulk gauge fields, and the exchange of multi-trace operators in $n$-point functions should be automatically resummed by the bulk dynamics.  One may worry, however, that restricting to the single-trace operators does not yield a consistent symmetry algebra, but we will see that it does.  For the sake of the following discussion let us denote indices that run over single-trace currents by $a$, $b$, $c$, etc. The structure constants of the bulk gauge group are $f^{abc}$ and the higher derivative terms come with coefficients $d_n^{a_1 \dots a_{n}}$. Bulk gauge invariance requires that \eqref{fdIdentities} is satisfied with only single-trace indices. We are of course free to set the free indices to be single-trace indices, but we have to check that we do not need to sum over multi-trace indices to satisfy \eqref{fdIdentities}. To leading order in the large $N$ expansion, it can be  shown that in the OPE \eqref{CurrentOPE} of single trace operators double-trace operators only show up with $d_3^{a b C}$ coefficients, hence  $f^{abC}=0$ unless $C$ corresponds to a single trace operator, namely $C = c$ for some $c$.  See Section~\ref{MATRIX} for an explicit example where these general statements are verified in an explicit computation. At subleading orders in $1/N$ one has to deal with operator mixing. Although we have not worked out the details, it seems possible to set  $f^{abC}=0$ whenever $j^C$ is a multi-trace current to all orders in $1/N$ by an appropriate redefinition of what we mean by single trace operators.
Thus, we can rewrite  \eqref{fdIdentities} as
\es{fdIdentities2}{
 0&=f^{d E[a}\,f^{bc]E}=f^{d e[ a}\,f^{b c]e} \,, \\
0&=f^{ dE (a}\, d_n^{a_1 \dots a_{n-1}) E}=f^{ d e (a}\, d_n^{a_1 \dots a_{n-1}) e}\,.
} 
The identities on the RHS thus only contain single-trace indices, and they will provide non-trivial checks on the computations performed in the following two sections.  The identities \eqref{fdIdentities2} also show that, on general grounds, we expect the restriction of the symmetry algebra to the single-trace sector to also be a consistent symmetry at large $N$, where the multi-trace operators are suppressed.  This restricted algebra will be the symmetry of the dual bulk YM$_2$ theory.

\section{Topological theories with a vector large $N$ limit}
\label{VECTOR}

\subsection{A topological theory dual to Abelian gauge theory on $AdS_2$}
\label{VECTORABELIAN}

As described in the Introduction, the simplest 1d topological theory with a 2d dual that we study is the topological sector of the 3d SCFT of $N$ free massless hypermultiplets.  Like its parent 3d theory, this 1d theory is also free, and its partition function is 
 \es{PartFunctionFree}{
  Z = \int D\tQ^I\, DQ_I\,  \exp \left[ 4 \pi \int \tQ^I dQ_I \right] \,,
 }
where $Q_I$ and $\tQ^I$, $I = 1, \ldots, N$, are the 1d operators corresponding to the 3d hypermultiplets.  The $Q_I$ and $\tQ^I$ transform as a fundamental / anti-fundamental of the $\mathfrak{u}(N)$ flavor symmetry of the topological theory.  This $\mathfrak{u}(N)$ flavor symmetry of the topological theory has its origin in a $\mathfrak{u}(N)$ subalgebra of the $\mathfrak{usp}(2N)$ flavor symmetry of the 3d SCFT\@. 

In general, given a 3d vector model with  flavor symmetry $G$ (where $G = O(N)$ or $U(N)$ depending on the theory) in the large $N$ limit, it is customary to look for a holographic dual of the $G$ singlet sector.  Because the vector model has higher spin conserved or almost conserved currents, all of which are $G$ singlets, the holographic dual of the $G$ singlet sector is given by a higher spin theory in $AdS_4$ \cite{Klebanov:2002ja,Giombi:2009wh}.  In particular, the higher spin theory dual to the $U(N)$ singlet sector of $N$ free hypermultiplets was explored in  \cite{Vasiliev:1992av,Vasiliev:1995dn,Vasiliev:1999ba,Engquist:2002vr,Engquist:2002gy,Chang:2012kt}.   Passing to 1d, we would like to search for the holographic dual of the $\mathfrak{u}(N)$ singlet sector of \eqref{PartFunctionFree}.  Remarkably, the $\mathfrak{u}(N)$ singlet sector of \eqref{PartFunctionFree} contains only one linearly-independent single trace operator, namely
 \es{jDefAbelian}{
   j = c\, \tQ^I Q_I \,, 
 } 
where $c$ is a normalization constant that can be chosen as desired.  We thus propose that the $\mathfrak{u}(N)$ singlet sector of \eqref{PartFunctionFree} is dual to an Abelian gauge theory on $AdS_2$.

It is easy to calculate the correlation functions of $j$ as follows.  We can add a source $a(\vphi)$ for this operator and calculate 
 \es{PartFuncFreeSource}{
  Z[a] = \int D\tQ^I\, DQ_I\,  \exp \left[ 4 \pi \int  \tQ^I dQ_I -  \int d \vphi \, a(\vphi) j(\vphi)  \right] \,.
 }
This path integral should only depend on the quantity $\int d\vphi \, a(\vphi)$, so in order to obtain $Z[a]$, it is sufficient to take $a(\vphi)$ to be a constant.  Indeed, let us set  $a(\vphi) = - \frac{4 \pi}{c} m$, calculate $Z[m]$ and then replace $m$ by $-\frac{c}{8 \pi^2} \int d\vphi\, a(\vphi)$.  The parameter $m$ amounts to introducing a real mass parameter for the $N$ free hypermultiplets, as reviewed briefly in the previous section.   The partition function $Z[m]$ is just the partition function of $N$ free hypermultiplets of mass $m$,
 \es{Zm}{
  Z[m] = \frac{1}{\left[ 2 \cosh (\pi m ) \right]^N} \,.
 }
(See~\eqref{Zsigma}.)  So the generating functional of connected correlators of $j(\vphi)$ is 
 \es{GenFunc}{
   \log Z[a] = - N \log \left[ 2 \cosh \left( -\frac{c \int d\vphi\, a(\vphi)}{8 \pi} \right) \right] \,.
 }

In Section~\ref{ABELIAN} we analyzed the theory of an Abelian gauge field in $AdS_2$ with an arbitrary bulk non-linear action \eqref{BulkNonLinear} supplemented by appropriate boundary terms.  We obtained that in the saddle point approximation, the logarithm of the partition function is given by $-S_\text{on-shell}[a]$ as in \eqref{OnShellGeneral}.  Requiring that \eqref{OnShellGeneral} and \eqref{GenFunc} should match, we conclude that the bulk Abelian gauge theory dual to the $\mathfrak{u}(N)$ singlet sector of \eqref{PartFunctionFree} is 
 \es{SbulkAbelianFree}{
  S_\text{bulk} =  -\frac{N}{2 \pi} \int d^2 x \, \sqrt{g}\, \log \left[ 2 \cosh \left( \frac{c F}{4} \right) \right] \,,
 }
with $F = \frac 12 \epsilon^{\mu\nu} F_{\mu\nu}$ as in Section~\ref{ABELIAN}.   We will see in the next section that a natural choice for $c$ is $c = 4 \pi i$, in which case we have
  \es{SbulkAbelianFree2}{
  S_\text{bulk} =  -\frac{N}{4 \pi} \int d^2 x \, \sqrt{g}\, \log \left[ 4 \cos^2 \left(\pi F \right) \right] \,.
 }
We have thus determined the full non-linear Abelian gauge theory in $AdS_2$ that in the saddle point approximation reproduces all correlation functions of the single trace operator $j$ in the boundary theory!

One may wonder why the bulk action \eqref{SbulkAbelianFree2} diverges when $F \in \frac12+\Z$. This divergence does not affect the computation of correlation functions, as they are only sensitive to the small $F$ behavior of $S_\text{bulk}$. However, the divergence does affect the response to a constant boundary source, and the singularity occurs for $a(\vphi)=im \in \frac12+\Z$. For this value of $m$ the field theory partition function \eqref{Zm} diverges. The reason for this divergence is the existence of a zero mode. Indeed, the equation of motion in the presence of a constant $m$:
\es{QEom}{
\le(\p_\vphi+m\ri) Q_I=0
}
has the solution $Q_I\propto e^{-m \vphi}$, which only satisfies the anti-periodic boundary conditions for $m \in i\le(\frac12+\Z\ri)$. We then have a zero mode, and the partition function diverges in accordance with the bulk result.

\subsection{A topological theory dual to non-Abelian gauge theory on $AdS_2$}\label{FREEHYPERS}

A more complicated example is given by the 1d theory coming from the 3d theory of $NM$ free massless hypermultiplets, which involves the 1d fields $Q_{I i}$ and $\tQ^{I i}$ with lower/upper $I = 1, \ldots, N$ and $i = 1, \ldots, M$ being $\mathfrak{u}(N)$ and $\mathfrak{u}(M)$ fundamental/anti-fundamental indices. The partition function is
  \es{PartFunctionFreeNM}{
  Z = \int D\tQ^{Ii}\, DQ_{Ii}\,  \exp \left[ 4 \pi \int \tQ^{Ii} dQ_{Ii} \right] \,.
 }
We can consider the $\mathfrak{u}(N)$ singlet sector, which now consists of the linearly-independent operators
 \es{UNSingletNM}{
  j^i{}_j = c\, \tQ^{Ii} Q_{Ij} 
 } 
transforming in the adjoint representation of $\mathfrak{u}(M)$.   Thus, the $\mathfrak{u}(N)$ singlet sector of the theory \eqref{PartFunctionFreeNM} should be dual to a non-Abelian gauge theory in $AdS_2$ with $\mathfrak{u}(M)$ gauge algebra.   In \eqref{UNSingletNM}, $c$ is a constant to be adjusted when we match $j^i{}_j$ with the conserved current dual to the bulk gauge field normalized as in the previous sections.

Let us consider $j^a = (T^a)_i{}^j  j^i{}_j$, where the $T^a$ are a collection of $M\times M$ Hermitian matrices belonging to the $\mathfrak{u}(M)$ Lie algebra.  Using the Green's function
 \es{GreenNonAb}{
  \langle Q_{Ii}(\vphi_1) \tQ^{Jj}(\vphi_2) \rangle
   =  -\frac{\sgn \vphi_{12}}{8 \pi}  \delta^J_I \delta^j_i
 }
obtained from \eqref{Propag}, we can easily compute the connected correlators 
 \es{TwoPointUM}{
  \langle j^a(\vphi_1) j^b(\vphi_2) \rangle 
   &= - c^2 \frac{N \tr (T^a T^b)}{(8\pi)^2}  \,, \\
  \langle j^a(\vphi_1) j^b(\vphi_2)  j^c(\vphi_3) \rangle 
   &= - c^3 \frac{N \tr (T^a [T^b, T^c])}{(8 \pi)^3}  \sgn (\vphi_{12} \vphi_{23} \vphi_{31}) \,, \\
  \langle j^a(\vphi_1) j^b(\vphi_2)  j^c(\vphi_3) j^d(\vphi_4) \rangle 
   &=  c^4 \frac{N}{(8 \pi)^4} \Biggl[  \tr (T^a T^b T^c T^d + T^d T^c T^b T^a)  \sgn (\vphi_{12} \vphi_{23} \vphi_{34} \vphi_{41}) \\
   &{}+ \tr (T^a T^b T^d T^c + T^c T^d T^b T^a)  \sgn (\vphi_{12} \vphi_{24} \vphi_{43} \vphi_{31})  \\
   &{}+ \tr (T^a T^c T^b T^d + T^d T^b T^c T^a)  \sgn (\vphi_{13} \vphi_{32} \vphi_{24} \vphi_{41}) \Biggr] \,.
 }

While \eqref{TwoPointUM} is true for any collection of Hermitian matrices $T^a$, let us now take $T^a$ with $a = 1, \ldots, M^2 - 1$ to be traceless and form a basis for the $\mathfrak{su}(M)$ subalgebra of $\mathfrak{u}(M)$, and take $T^{M^2}$ to be proportional to the identity matrix, thus representing the $\mathfrak{u}(1)$ generator that commutes with $\mathfrak{su}(M)$.  

For simplicity, let us study the case $M=2$.  With the notation for the generators we just introduced, the correlation functions of $j^4$ are described by the same $\mathfrak{u}(1)$ gauge theory in $AdS_2$ as in the previous section, so let us focus on the correlation functions of $j^a$, with $a = 1, \ldots, 3$, which should be described by an $\mathfrak{su}(2)$ gauge theory in $AdS_2$.  Taking the $T^a = \frac{\sigma^a}{2}$, we have
 \es{fdDef}{
  T^a T^b = \frac{\delta^{ab}}{4} {\bf 1} + \frac i2 \epsilon^{abc} T^c  \,,
 } 
which can be used to rewrite \eqref{TwoPointUM} as
 \es{TwoPointU2}{
  \langle j^a(\vphi_1) j^b(\vphi_2) \rangle 
   &= - c^2 \frac{N \delta^{ab}}{2(8\pi)^2}  \,, \\
  \langle j^a(\vphi_1) j^b(\vphi_2)  j^c(\vphi_3) \rangle 
   &= - c^3 i \frac{N \epsilon^{abc}}{2 (8 \pi)^3}  \sgn (\vphi_{12} \vphi_{23} \vphi_{31}) \,, \\
  \langle j^a(\vphi_1) j^b(\vphi_2)  j^c(\vphi_3) j^d(\vphi_4) \rangle 
   &=  c^4 \frac{N}{4 (8 \pi)^4} \Biggl[  \left(  \delta^{ab} \delta^{cd} + \delta^{ad} \delta^{bc}    -\delta^{ac} \delta^{bd}  \right)  \sgn (\vphi_{12} \vphi_{23} \vphi_{34} \vphi_{41}) \\
   &{}+ \left(  \delta^{ab} \delta^{cd} + \delta^{ac} \delta^{bd}    -\delta^{ad} \delta^{bc}  \right)   \sgn (\vphi_{12} \vphi_{24} \vphi_{43} \vphi_{31})  \\
   &{}+  \left(  \delta^{ac} \delta^{bd} + \delta^{ad} \delta^{bc}    -\delta^{ab} \delta^{cd}  \right)  \sgn (\vphi_{13} \vphi_{32} \vphi_{24} \vphi_{41}) \Biggr] \,.
 }

In Section~\ref{SU2EXAMPLE} we considered the example of a bulk $\mathfrak{su}(2)$ gauge theory in which we computed the two-, three-, and four-point functions of the dual field theory operators---See \eqref{su2Correl}.  We find that \eqref{TwoPointU2} matches \eqref{su2Correl} provided that we have the parameters $g_\text{YM}$, $w_4$, and $c$ take the following values:
 \es{gYMa4}{
   \frac{1}{g_\text{YM}^2} =  \frac{N \pi}{4} \,, \qquad  w_4 = \frac18 \,, \qquad c = 4 \pi i\,.
 }
We note that $c$ can also be determined just from field theory, by extracting the OPE from \eqref{TwoPointU2} and comparing to \eqref{CurrentOPE}.  

To summarize, we have just found that the $\mathfrak{u}(N)$ singlet sector of the theory \eqref{PartFunctionFreeNM} with $M=2$ is described, at large $N$, by the product of the Abelian gauge theory of the previous section and an $\mathfrak{su}(2)$ gauge theory on $AdS_2$ given by
 \es{su2BulkFree}{
  S_\text{bulk} = \frac{N \pi}{4} \int d^2x \, \sqrt{g} \tr \left( F^2 + \frac{\pi^2}{6} F^4 + \cdots \right) 
 }
(supplemented by appropriate boundary terms).  As we will now see, in this case we will also be able to determine the full non-linear bulk action whose expansion at small $F$ is given in \eqref{su2BulkFree}.
 
  \subsection{Correlation functions in the presence of a constant source in the $\mathfrak{su}(2)$ theory}
\label{CORRELATORS2}

To extract the full non-linear action for the bulk $\mathfrak{su}(2)$ gauge theory, let us consider adding a constant source $s$ for the operator $j^3$:
  \es{PartFunctionFreeNMSource}{
  Z[s] = \int D\tQ^{Ii}\, DQ_{Ii}\,  \exp \left[ 4 \pi \int \tQ^{Ii} dQ_{Ii} - \int d\vphi\, s j^3 \right] \,.
 }
Using $j^3 =  2 \pi i (\tQ^{I 1} Q_{I 1} - \tQ^{I 2} Q_{I 2} )$ (which follows from \eqref{UNSingletNM}, \eqref{gYMa4}, and the choice $T^3 = \sigma^3 / 2$), we can compare with \eqref{ZsDef} with $\sigma \to m$  and infer that the parameter $s$ corresponds to a mass $m_1 = -i s/2$ for $(\tQ^{I1}, Q_{I 1})$ and a mass $m_2 = i s/2$ for $(\tQ^{I2}, Q_{I2})$.    Thus one can evaluate $Z[s]$ from \eqref{Zsigma}, again with $\sigma \to m$:
 \es{Zs}{
  Z[s] = \frac{1}{\left[ 4 \cosh (\pi m_1) \cosh(\pi m_2) \right]^N}
   = \frac{1}{4^N (\cos \frac{\pi s}{2} )^{2N}} \,.
 } 

In Section~\ref{NONABELIANHIGHER} we found a general expression for the on-shell action of a non-Abelian theory in $AdS_2$ with bulk action  \eqref{SbulkNonAbNonLinear}.  We can consider the particular case where this on-shell action is evaluated for a bulk $\mathfrak{su}(2)$ gauge theory with a constant source $s$ for the boundary operator $j^3$.   Indeed, plugging $a = s T^3$ into \eqref{QFormula} we obtain
 \es{QConstantSource}{
  u_0 Q u_0^{-1} = - s T^3 \,,
 }
so then the on-shell action \eqref{SbulkNonAbNonLinear} becomes 
 \es{SOnShellNonAbConstantSource}{
  S_\text{on-shell}[s] = -\frac{2 \pi}{ g_\text{YM}^2} \tr W(-s T^3) \,.
 }
We should identify \eqref{SOnShellNonAbConstantSource} with $-\log Z[s]$ from \eqref{Zs}, from which we determine the full non-linear action of our bulk $\mathfrak{su}(2)$ gauge theory:
 \es{NonLinearsu2Bulk}{
  S_\text{bulk} =  -  \frac{N}{4 \pi} \int d^2x \, \sqrt{g} \tr  \log \left[ 4 \cos^2(\pi F) \right]  \,.
 }
One can easily check that the quadratic and quartic terms in the small $F$ expansion of this expression reproduce those in \eqref{su2BulkFree}.  Note that one can easily combine the $\mathfrak{su}(2)$ gauge theory \eqref{NonLinearsu2Bulk} with the Abelian gauge theory in \eqref{SbulkAbelianFree2} into an $\mathfrak{u}(2)$ gauge theory on $AdS_2$ with the same action as \eqref{NonLinearsu2Bulk}, with $F$ being now the field strength of a $\mathfrak{u}(2)$ gauge field.  This $\mathfrak{u}(2)$ gauge theory is the full holographic dual of the $\mathfrak{u}(N)$ singlet sector of the theory \eqref{PartFunctionFreeNM} when $M = 2$ at large $N$.

One can go further and calculate various connected correlators as a function of the deformation parameter $s$, and obtain a match between the boundary and the bulk theories.  For instance, denoting $j^\pm = j^1 \pm i j^2$, we have the following boundary field theory expressions
 \es{TwoPoint}{
  \langle j^3(\vphi) \rangle_s &= (2 \pi i) \langle \tQ^{I1} Q_{I1}(\vphi)
   - \tQ^{I2} Q_{I2}(\vphi) \rangle_s
   =  -\frac N2 \tan \frac{\pi s}{2} \,, \\
  \langle j^+(\vphi_1) j^-(\vphi_2) \rangle_s 
   &= (4 \pi i)^2 \langle \tQ^{I1} Q_{I2}(\vphi_1)\tQ^{I2} Q_{I1}(\vphi_2)  \rangle_s 
    = \frac{N e^{-i s (\vphi_{12} - \pi \sgn \vphi_{12})}}{4 \cos^2 \left( \frac{\pi s}{2} \right)} \,, \\
   \langle j^3(\vphi_1) j^3(\vphi_2) \rangle_s    
   &= \frac{(4 \pi i)^2}{2} \langle \tQ^{I1} Q_{I1}(\vphi_1)\tQ^{I1} Q_{I1}(\vphi_2)  \rangle_s 
    = \frac{N}{8\cos^2  \left( \frac{\pi s}{2} \right)}\,,
 }
which were obtained by simply using Wick contractions with the propagator 
 \es{PropagDeformation}{
 \langle Q_{Ii}(\vphi_1) \tQ^{Jj}(\vphi_2) \rangle
   =  -\frac{\sgn \vphi_{12} + \tanh (\pi m_i)}{8 \pi}  e^{-m_i \vphi_{12}} \delta^J_I \delta^j_i \,.
 }  

On the bulk side, let us denote
\es{aphiConst}{
a(\vphi)\equiv s T^3+\ta(\vphi)\,,
} 
thus isolating the constant source for $j^3$ from the source $\ta(\vphi)$ in the presence of $s$. With this split, one can write the solution \eqref{uEq} as
 \es{QFormulaConstantSource}{
  Q &= \frac{i}{2 \pi} u_0^{-1} \log \left( e^{2\pi i s T^3}\,  P \exp \left[ i \int_0^{2 \pi} d\vphi\, \tilde a_I(\vphi) \right]  \right)  u_0 \,,\\
  \tilde a_I(\vphi)&\equiv e^{-is T^3\vphi} \tilde a(\vphi)e^{is T^3\vphi}\,,
 }
 where the $I$ subscript refers to the ``interaction picture.''  In the case where $W(F)$ is given by $W(F) =   -\frac{1}{\pi^2 }   \log \left[ 4 \cos^2(\pi F) \right] $ as can be deduced from \eqref{NonLinearsu2Bulk}, we can write
  \es{GotW2}{
  \tr W(Q) &= -\frac{2\log 4}{\pi^2 }  
    -\frac{2}{\pi^2 }   \log  \le[\frac12\le(1+\cos(\pi s)\cos(I_1)+\sin(\pi s)\,{I_2 \sin(I_1)\ov I_1} \ri)\ri]\,,\\
    I_1^2&\equiv-{w^a w^a\ov 4} \,, \qquad I_2\equiv {i w^3\ov 2}\,, \qquad w\equiv \log\le(P \exp \left[ i \int_0^{2 \pi} d\vphi\, \tilde a_I(\vphi) \right] \ri) \,.
 }
This expression is amenable to expansion in small $\tilde a_I(\vphi)$. In particular, the expansion of  $w$ is identical to what was described in  \eqref{QExpansion}:
 \es{wExpansion}{
 w^a &= i \int_0^{2\pi} d\vphi_1 \,\tilde a_I^a(\vphi_1)
   - \frac{i f^{abc}}{2} \int_0^{2 \pi} d \vphi_1 \, d \vphi_2 \,\tilde a_I^b(\vphi_1) \tilde a_I^c(\vphi_2) \theta( \vphi_{12})  + \cdots \,,
  }
hence
  \es{WExpansion}{ 
      \tr W(Q) &= -{2\ov \pi^2}\, \log \left[ 4 \cos^2\le({s\pi\ov2}\ri) \right]- {2 I_2\ov \pi^2}\, \tan\le({s\pi\ov2}\ri)
   +{1\ov 2\pi^2}\,\le[2(I_1^2-I_2^2)+{-I_1^2+2I_2^2\ov \cos^2\le({s\pi\ov2}\ri)}\ri] +\cdots\,, \\
   I_1^2&=\frac14\int_0^{2\pi} d\vphi_1 \, d \vphi_2 \,\tilde a_I^a(\vphi_1)\tilde a_I^a(\vphi_2)+\cdots\,,\\
   I_2&=-{1\ov 2} \int_0^{2\pi} d\vphi_1 \,\tilde a_I^3(\vphi_1)
   + \frac{ f^{3bc}}{4} \int_0^{2 \pi} d \vphi_1 \, d \vphi_2 \,\tilde a_I^b(\vphi_1) \tilde a_I^c(\vphi_2) \theta( \vphi_{12})  + \cdots \,.
 }
Let us introduce a new basis for $ \tilde a^a_I(\vphi)$ so that the coupling of the background gauge fields to the current take the form
\es{Coupling}{
\tilde a^a_I(\vphi) j^a= \tilde a_I^{+}(\vphi) j^-+\tilde a_I^{-}(\vphi) j^++\tilde a_I^{3}(\vphi) j^3\,.
}
Using that
 \es{IntPict}{
 \tilde a_I^{\pm}(\vphi)&={ \tilde a_I^1(\vphi)\pm i \tilde a_I^2(\vphi)\ov 2}=\tilde a^{\pm}(\vphi)\, e^{\pm i s \vphi}\,, \qquad
 \tilde a_I^{3}(\vphi)= \tilde a^{3}(\vphi)\,,
 }
 we finally obtain 
  \es{su2Final}{
   \tr W(Q) &= -{2\ov \pi^2}\, \log \left[ 4 \cos^2\le({s\pi\ov2}\ri) \right] +{\tan\le({s\pi\ov2}\ri)\ov \pi^2}\, \int_0^{2\pi} d\vphi_1 \,\tilde a^{3}(\vphi_1)\\
   &+{1\ov 8\pi^2\, \cos^2\le({s\pi\ov2}\ri)}\,\int_0^{2 \pi} d \vphi_1 \, d \vphi_2 \,\le[\tilde a^{3}(\vphi_1)\tilde a^{3}(\vphi_2) +4 \,\tilde a^{-}(\vphi_1)\tilde a^{+}(\vphi_2)e^{-is(\vphi_{12}-\pi \sgn \vphi_{12})}\ri] +\cdots\,.
 }
This bulk computation then precisely reproduces \eqref{TwoPoint}.  Indeed, the negative of the on-shell action
  \es{OnShell}{
   -S_\text{on-shell}[\tilde a] 
    = \frac{2 \pi}{ g_\text{YM}^2} \tr W(Q) = \frac{N \pi^2}{2} \tr W(Q)
  }
is the generating functional of connected correlators of $-j$, and it is straightforward to check that the field theory expressions \eqref{TwoPoint} can also be written as
 \es{TwoPointAnother}{
  \langle j^3(\vphi) \rangle_s &= - \frac{\delta (-S_\text{on-shell}[\tilde a])}{\delta \tilde a^3(\vphi) } \,,  \\
   \langle j^3(\vphi_1) j^3(\vphi_2) \rangle_s &=  \frac{\delta^2 (-S_\text{on-shell}[\tilde a])}{\delta \tilde a^3(\vphi_1) \delta \tilde a^3 (\vphi_2) } \,, \\
  \langle j^+(\vphi_1) j^-(\vphi_2) \rangle_s &=  \frac{\delta^2 (-S_\text{on-shell}[\tilde a])}{\delta \tilde a^-(\vphi_1) \delta \tilde a^+ (\vphi_2) } \,,
 }
thus obtaining a match between the boundary and bulk theories in our proposal.

\section{Topological theories with a matrix large $N$ limit} \label{MATRIX}

\subsection{Setup}

More complicated examples of 1d topological theories with 2d duals arise from 3d ${\cal N} = 4$ SCFTs with a matrix large $N$ limit whose bulk duals are  11d supergravity on  backgrounds of the form $AdS_4 \times X_7$, with $X_7$ being a 7d manifold.  While many backgrounds of this kind can be constructed, we will focus only on a specific example for which the SCFT has maximal supersymmetry in 3d.  In an M-theory context, it is the effective theory on $N$ coincident M2-branes in flat space, and its Lagrangian description was provided by ABJM in \cite{Aharony:2008ug}.  An interesting feature of this theory is that its effective number of degrees of freedom grows as $N^{3/2}$ at large $N$, as was seen in  \cite{Klebanov:1996un} using supergravity or in \cite{Drukker:2010nc,Herzog:2010hf} in field theory.  The field theory computations of \cite{Drukker:2010nc,Herzog:2010hf} that revealed the $N^{3/2}$ scaling were made possible by the progress in performing supersymmetric localization computations achieved in \cite{Kapustin:2009kz}.

It can be understood using string theory dualities that the theory of $N$ coincident M2-branes in flat space has an alternative microscopic description studied in \cite{Bashkirov:2010kz}:  in ${\cal N} = 4$ language, we have a $U(N)$ vector multiplet coupled to an adjoint and a fundamental hypermultiplet.  This is the description we will use here.  

The 1d topological theory corresponding to the Higgs branch of this theory can be written down using the rules reviewed in Section~\ref{REVIEW}.  In particular, Eq.~\eqref{Part1d} becomes
 \es{ZS3N8}{
  Z = \frac{1}{N!} \int \left( \prod_{i=1}^N d\sigma_i  \right) \, \prod_{i<j} \left[ 2 \sinh(\pi (\sigma_i - \sigma_j) ) \right]^2 Z_\sigma\,,
 }
where we wrote 
 \es{sigmaDef}{
  \sigma = \diag\{\sigma_1, \ldots, \sigma_N \} 
 }
in order to restrict $\sigma$ to the Cartan of $\mathfrak{u}(N)$.  
Eq.~\eqref{ZsDef} gives the formula for $Z_\sigma$:
 \es{ZsN8}{
  Z_\sigma = \int DQ\, D\tQ\, DX\, D\tX \exp\left[{4 \pi \int d\vphi\,  \tQ (\partial_\vphi + \sigma) Q +\tr \left(  \tX \partial_\vphi X + \tX [\sigma, X] \right) } \right] \,.
 }
Here,  the 1d fields $Q$ and $\tQ$ have their 3d origin in the fundamental hypermultiplet and transform under $\mathfrak{u}(N)$  as ${\bf N}$ and $\overline{\bf N}$, respectively.  The $X$ and $\tX$ have their 3d origin in the adjoint hypermultiplet and thus both transform in the adjoint of $\mathfrak{u}(N)$.  Lastly, Eq.~\eqref{Zsigma} gives the value of $Z_\sigma$ after integrating out all the 1d fields 
 \es{ZsigmaN8}{
  Z_\sigma 
   = \frac{1} 
   {\prod_{i, j} \left[ 2 \cosh (\pi (\sigma_i - \sigma_j)) \right] \prod_{i=1}^N \left[ 2 \cosh (\pi \sigma_i)\right] } \,.
 }

The local operators of the 1d theory are the $\mathfrak{u}(N)$-invariant operators built out of $Q$, $\tQ$, $X$, and $\tX$.  However, due to the D-term relations of the 3d theory, all operators involving $Q$ and $\tQ$ can be written in terms of $X$ and $\tX$.  Thus, since we will not be using $Q$ and $\tQ$, we can integrate them out in \eqref{ZsN8} and write $Z_\sigma$ as 
 \es{ZsN8Again}{
  Z_\sigma = 
   \frac{1} 
   { \prod_{i=1}^N \left[ 2 \cosh (\pi \sigma_i)\right] }\int DX\, D\tX \exp\left[{\tr \left(  \tX \partial_\vphi X + \tX [\sigma, X] \right) } \right] \,.
 }

There is a compact way of listing all gauge-invariant operators built from $X$ and $\tX$ by noticing that the 1d theory \eqref{ZS3N8} has an $\mathfrak{su}(2)$ flavor symmetry under which $X$ and $\tX$ transform as a doublet.  Indeed, the $\mathfrak{su}(2)$ symmetry becomes obvious in \eqref{ZsN8} by denoting $(X_1,\, X_2) \equiv (\tX,\,  X)$ and rewriting the terms involving $X$ and $\tX$ as
 \es{XXtRewrite}{
  \int d\vphi\, \tr \left(  \tX \partial_\vphi X + \tX [\sigma, X] \right)
   = \frac{1}{2} \int d\vphi\, \tr \left( \veps^{ij} X_i \partial_\vphi X_j -  \veps^{ij} \sigma [X_i, X_j] \right) \,,
 }
where $\veps^{ij}$ is the anti-symmetric $\mathfrak{su}(2)$ invariant tensor with $\veps^{12} = -\veps_{12} = 1$.  In order to write out compactly all the operators in the 1d theory, we can introduce a two-component polarization vector $y^i$, $i=1, 2$, and make the further definition
 \es{XDef}{
  X(\vphi, y) \equiv y^1 \tX(\vphi) + y^2 X(\vphi) \,.
 }
Then a basis for all operators consists of all products of operators of the form $\tr X(\vphi, y)^n$ modulo trace relations that become important when we have a product of at least $N$ fields.  At large $N$, we can organize the operator spectrum into single trace, double trace, triple trace, etc.  We will focus on the single trace operators with even powers of $X$, and thus integer $\mathfrak{su}(2)$ spin $\ell$, which we denote by
 \es{On}{
  {\cal O}_\ell(\vphi, y) = \frac{1}{N^{\frac{\ell-1}{2}}} \tr X(\vphi, y)^{2\ell}  \,. 
 }
This definition includes an $N$-dependent normalization factor that we will find convenient.  

For simplicity, henceforth we will focus only on the operators ${\cal O}_{\ell}$ with integer $\ell$.  We can restrict our discussion to just these operators by considering only operators that are invariant under the $\Z_2$ symmetry of the action \eqref{ZsN8Again} that acts by $X(\vphi, y) \to - X(\vphi, y)$.  The extension of our results to the $\Z_2$-odd operators is left for future work.

The $\mathfrak{su}(2)$ symmetry fixed the 2- and 3-point functions of the ${\cal O}_\ell$ up to overall factors:
 \es{OTwoThree}{
  \langle {\cal O}_{\ell_1} (\vphi_1, y_1) {\cal O}_{\ell_2}(\vphi_2, y_2) \rangle 
   &= B_{\ell_1}\, \delta_{\ell_1 \ell_2}\, \langle y_1, y_2 \rangle^{2\ell_1}  \,, \\
  \langle {\cal O}_{\ell_1} (\vphi_1, y_1) {\cal O}_{\ell_2}(\vphi_2, y_2) {\cal O}_{\ell_3}(\vphi_3, y_3) \rangle 
   &= C_{\ell_1 \ell_2 \ell_3} \langle y_1, y_2 \rangle^{\ell_1 + \ell_2 - \ell_3}
    \langle y_1, y_3 \rangle^{\ell_1 + \ell_3 - \ell_2} 
     \langle y_2, y_3 \rangle^{\ell_2 + \ell_3 - \ell_1}  \\
      &\times (\sgn \vphi_{21})^{\ell_1 + \ell_2 - \ell_3}
        (\sgn \vphi_{31})^{\ell_1 + \ell_3 - \ell_2} 
     (\sgn \vphi_{32})^{\ell_2 + \ell_3 - \ell_1} 
 }
where $\vphi_{ij} \equiv \vphi_i - \vphi_j$ as before and $\langle y_1, y_2 \rangle = \veps_{ij}\, y_1^i y_2^j$.

\begin{figure}[!htb]
	\centering
	{
		\begin{minipage}{1\textwidth}
			\centering
			\includegraphics[scale=.8 ]{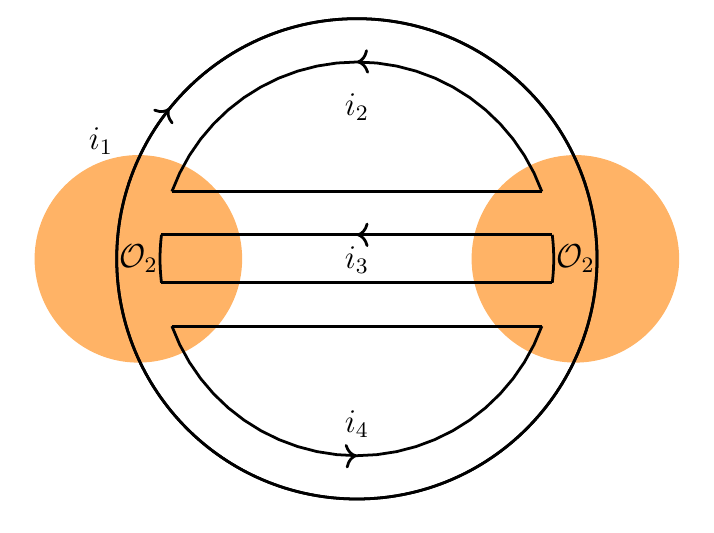}\\
		\end{minipage} 
	}
	\centering
	{
		\begin{minipage}{0.45\textwidth}
			\centering
			\includegraphics[scale=.8 ]{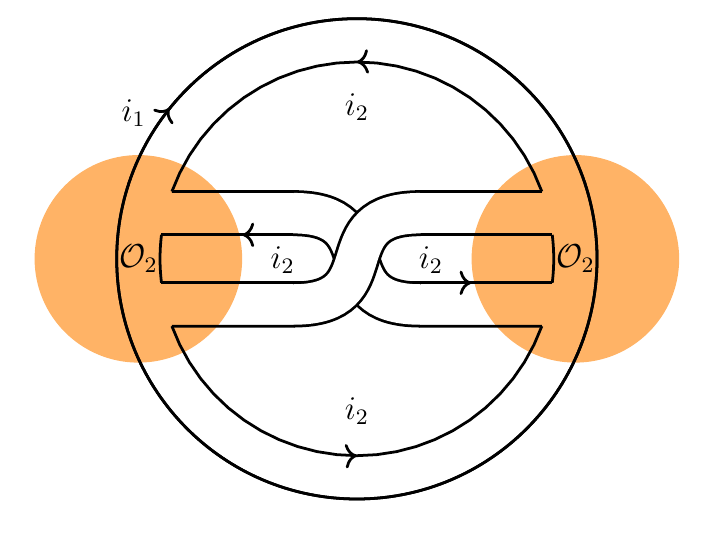}\\
		\end{minipage} 
	}
	\centering
	{
		\begin{minipage}{0.45\textwidth}
			\centering
			\includegraphics[scale=.8 ]{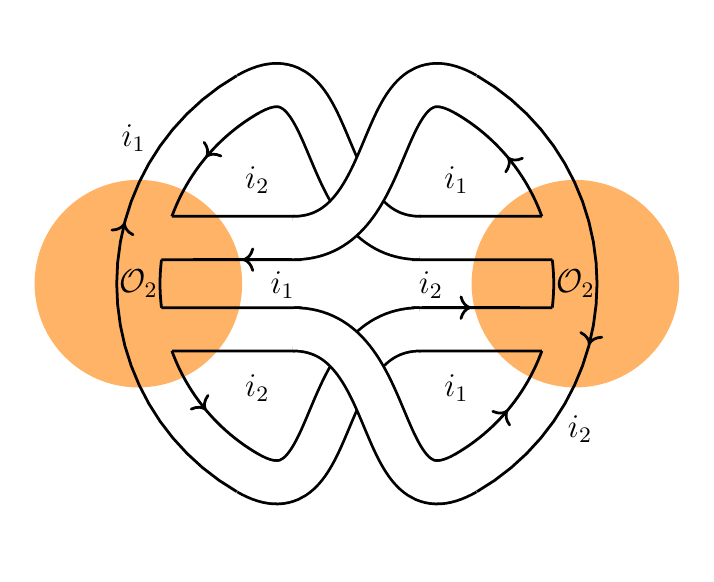}\\
		\end{minipage} 
	}
	\caption{The planar and non-planar diagrams that contribute to $B_2$.}
	\label{Fig:B2}
\end{figure}

Correlation functions such as \eqref{OTwoThree} can be computed explicitly using the propagator of the $X$ field at fixed $\sigma$ and then performing the matrix model integral with the appropriate extra insertions.  The propagator of the $X$ field is
 \es{XyTwoPoint}{
  \langle X_{ij}(\vphi_1, y_1) X_{kl} (\vphi_2, y_2) \rangle_\sigma =\langle y_1, y_2 \rangle  \delta_{il} \delta_{jk} G_{\sigma_{ji}}(\vphi_{12}) \,,
 }
where we used the compact notation $\sigma_{ji} \equiv \sigma_j - \sigma_i$ and $G_\sigma(\vphi)$ is defined in \eqref{Propag} and repeated for convenience here:
 \es{GDef}{
  G_\sigma(\vphi) \equiv -\frac{\sgn \vphi + \tanh( \pi \sigma)}{8 \pi r} e^{- \sigma \vphi} \,.
 }
For example, the two-point function of ${\cal O}_\ell$ can be calculated by first using Wick contractions at fixed $\sigma$  (see Figure~\ref{Fig:B2}), 
 \es{XX}{
  \langle {\cal O}_1 (\vphi_1, y_1) {\cal O}_1 (\vphi_2, y_2) \rangle_\sigma
    &= 2  \langle y_1, y_2 \rangle^2 \sum_{i, j} G_{\sigma_{ji}}(\vphi_{12})
     G_{\sigma_{ij}}(\vphi_{12}) \,, \\
    \langle {\cal O}_2 (\vphi_1, y_1) {\cal O}_2 (\vphi_2, y_2) \rangle_\sigma
    &= \frac{4  \langle y_1, y_2 \rangle^4}{N}  \Biggl[4 G_0^2 \sum_{i, j} G_{\sigma_{ij}} G_{\sigma_{ji}}  + \sum_{i, j} \left(  G_{\sigma_{ij}} G_{\sigma_{ji}} \right)^2  \\
     &{}+ \sum_{i, j, k, l} G_{\sigma_{ji}}(\vphi_{12})
     G_{\sigma_{kj}}(\vphi_{12})G_{\sigma_{lk}}(\vphi_{12})
     G_{\sigma_{il}}(\vphi_{12}) \Biggr] \,,  \\
   &\vdots  
 }
then inserting these expression into \eqref{ZS3N8}, computing the $\sigma_i$ integrals, and dividing the final answer by $Z$ in order to obtain a normalized correlator.  Similar expressions can be obtained for higher point functions.

It is of course rather complicated to perform the $\sigma_i$ integrals and obtain the constants $B_\ell$ and $C_{\ell_1 \ell_2 \ell_3}$ introduced in \eqref{OTwoThree} at finite $N$.  However, one can perform these computations at large $N$, where we will show that both $B_\ell$ and $C_{\ell_1 \ell_2 \ell_3}$ scale as $N^{3/2}$.  It is of course in this limit that we will discuss the bulk dual of the 1d theory \eqref{ZS3N8}.

\subsection{Large $N$ approximation}

\subsubsection{Partition function}

In preparation for performing a large $N$ computation of $B_\ell$ and $C_{\ell_1 \ell_2 \ell_3}$, let us first review the evaluation of the partition function $Z$ at large $N$, following \cite{Jafferis:2011zi}.  (See also  \cite{Herzog:2010hf} where this method was first introduced, as well \cite{Mezei:2013gqa}.)   At large $N$, the partition function can be evaluated using the saddle point approximation.   At the saddle point, the eigenvalues $\sigma_i$ approach a continuous distribution.  In particular, if we write
 \es{sigmaScaling}{
  \sigma_i = \sqrt{N} x_i \,,
 }
then the $x_i$ become dense in the large $N$ limit.  Let $\rho(x)$ be the density of the $x_i$ normalized such that $\int dx\, \rho(x) = 1$.  It can be shown \cite{Jafferis:2011zi,Mezei:2013gqa} that the integrand of \eqref{ZS3N8} (with $Z_\sigma$ given in \eqref{ZsigmaN8}) becomes
 \es{integrand}{
  \exp \left[-N^{3/2} \int dx\, \left[ \frac{\pi}{4} \rho(x)^2 + \pi \abs{x} \rho(x) \right]  \right] \,.
 }
The eigenvalue density that extremizes the exponent \eqref{integrand} under the normalization constraint that $\int dx\, \rho(x) = 1$ has a triangular shape centered at $x=0$:
 \es{rho}{
  \rho(x) = 
   \begin{cases}
    \sqrt{2} - 2\abs{x}  \,, & \text{if } \abs{x} \leq \frac{1}{\sqrt{2}} \,, \\
    0\,, & \text{otherwise} \,.
   \end{cases}
 }
Plugging this distribution back into \eqref{integrand}, one can obtain the large $N$ approximation to the partition function:
 \es{GotZN8}{
  Z \approx \exp \left[ - N^{3/2} \frac{\pi \sqrt{2}}{3} \right] \,,
 }
thus showing that the effective number of degrees of freedom scales as $N^{3/2}$ at large $N$.

\subsubsection{Two-point functions}

The coefficient $B_\ell$ appearing in the two-point function \eqref{OTwoThree} can be obtained by evaluating the expressions in \eqref{XX} on the saddle point distribution.  It is actually only one diagram that contributes to $B_\ell$ at leading order in $N$.  Indeed, in the example of $\langle {\cal O}_2 {\cal O}_2\rangle_\sigma$ presented in \eqref{XX}, one recognizes that the last term corresponds to a planar diagram, while the other two are non-planar (see Figure~\ref{Fig:B2}).  It is usually the case in matrix large $N$ theories that only planar diagrams contribute to leading order in $N$, and this is also the case here.  However, due to the unconventional $N^{3/2}$ scaling of the free energy $F = - \log Z$, these planar diagrams will also have a slightly unconventional scaling.

Let us first explain how to extract the leading large $N$ term in $B_1$, which is obtained from the $\langle {\cal O}_1 {\cal O}_1\rangle_\sigma$ correlation in \eqref{XX} that contains only one (planar) term.  Plugging in the expression for $G_\sigma(\vphi)$ and approximating the sum with an integral at large $N$, we have
 \es{X2X2LargeN}{
  B_1
   \approx \frac{2 N^2 }{(8 \pi)^2}  \int dx_1\, dx_2\, \rho(x_1) \rho(x_2)   \frac{1}{\cosh^2 (\pi \sqrt{N} (x_1 - x_2) ) } \,.
 }
At large $N$, we can approximate
 \es{coshApprox}{
  \frac{1}{\cosh^2 (\pi \sqrt{N}x ) } \approx \frac{2}{\pi \sqrt{N}}  \delta(x) \,,
 } 
which can be plugged into \eqref{X2X2LargeN} to obtain
 \es{X2X2LargeN2}{
 B_1
   \approx\frac{ N^{3/2} }{ (8 \pi)^2}  \frac{4}{\pi} \int dx_1\, \rho(x_1)^2   
    = \frac{ \sqrt{2}}{24 \pi^3}   N^{3/2} \,.
 }

The same procedure can be used to evaluate $B_\ell$.   We start with the planar term in the expression for $\langle {\cal O}_\ell {\cal O}_\ell\rangle_\sigma$, namely
 \es{OpOpPlanar}{
    \langle {\cal O}_\ell (\vphi_1, y_1) {\cal O}_\ell (\vphi_2, y_2) \rangle_{\sigma, \text{planar}}
    &= \frac{2p  \langle y_1, y_2 \rangle^{2\ell}}{N^{\ell-1}} \sum_{i_1, \ldots, i_{2\ell}} G_{\sigma_{i_2 i_1}}(\vphi_{12})
     G_{\sigma_{i_3 i_2}}(\vphi_{12}) \cdots
     G_{\sigma_{i_1 i_{2\ell}}}(\vphi_{12}) \,,
 } 
from which we can extract $B_\ell$ after plugging in the expression for the Green's function.  At large $N$, the sums can be approximated by integrals: 
 \es{XnXnLargeN}{
 B_\ell
    &\approx \frac{2\ell N^{\ell + 1} }{(8 \pi)^{2\ell} } 
     \int \prod_{a=1}^{2\ell} \frac{ dx_a\,  \rho(x_a) }{\cosh(\pi \sqrt{N} (x_a - x_{a+1})) } \,,
  }   
and we may use the approximation
 \es{prodCoshApprox}{
  \frac{ 1}{\prod_{a=1}^{2\ell} \cosh(\pi \sqrt{N} (x_a - x_{a+1})) } 
   \approx  \frac{\Gamma\left( \ell \right) }{\sqrt{\pi} \Gamma \left( \ell + \frac 12 \right) } N^{-\ell + \frac 12 } \prod_{a=1}^{2\ell-1} \delta(x_a - x_{a+1}) \,,
 }
where the constant multiplying the delta functions on the right-hand side can be determined by integrating both sides with respect to $x_a$, with $2 \leq a \leq 2\ell$.  Plugging this expression into \eqref{XnXnLargeN} and evaluating the integrals yields
  \es{XnXnLargeNAgain2}{
    B_\ell
    \approx 
     \frac{  2^{\ell + \frac 12} \Gamma\left( \ell + 1 \right) }{\sqrt{\pi}  \Gamma \left( \ell + \frac 32 \right) }  \frac{N^{3/2} }{(8 \pi)^{2\ell}} \,.
  } 
As advertised, all the $B_\ell$ scale as $N^{3/2}$ at large $N$, which is the proper normalization for boundary operators dual to bulk fields whose action is proportional to $N^{3/2}$.

\subsubsection{Three-point functions}

Next, we would like to compute the constants $C_{\ell_1 \ell_2 \ell_3}$ appearing in \eqref{OTwoThree}.  This constant will be non-zero only if $\ell_1$, $\ell_2$, and $\ell_3$ obey the triangle inequality, i.e.~$|\ell_1-\ell_2|\leq \ell_3\leq \ell_1+\ell_2$ and cyclic permutations.

\begin{figure}[!htb]
	\centering
				\includegraphics[scale=1 ]{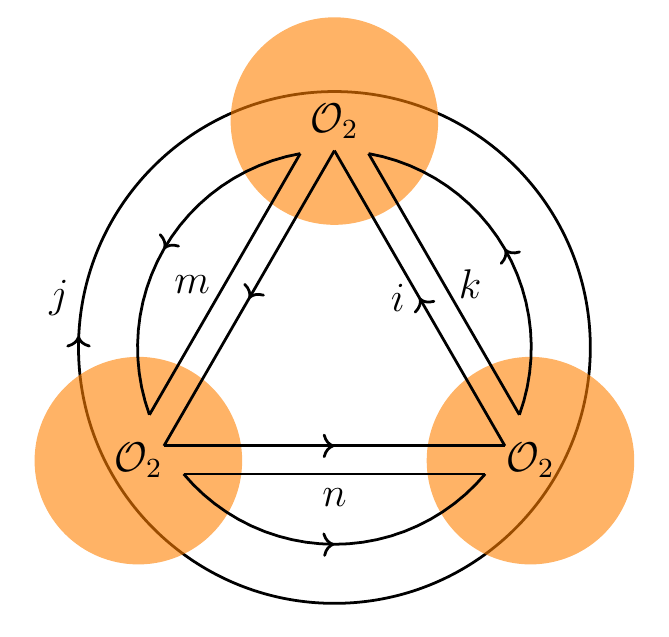}
	\caption{The planar diagram that contributes to $C_{222}$.}
	\label{Fig:C222}
\end{figure}

We again start with the planar diagram (see Figure~\ref{Fig:C222}) contributing to the three-point function in \eqref{OTwoThree}.  It gives 
\ie
& \langle {\cal O}_{\ell_1} (\vphi_1, y_1) {\cal O}_{\ell_2}(\vphi_2, y_2) {\cal O}_{\ell_3}(\vphi_3, y_3) \rangle_{\sigma, \text{planar}} 
   =   \frac{8\ell_1 \ell_2 \ell_3 \langle y_1, y_2 \rangle^{\ell_1 + \ell_2- \ell_3}
    \langle y_1, y_3 \rangle^{\ell_1 + \ell_3 - \ell_2} 
     \langle y_2, y_3 \rangle^{\ell_2 + \ell_3 - \ell_1}}{ (8\pi)^{\ell_1 + \ell_2 + \ell_3}N^  {\ell_1 + \ell_2 + \ell_3-3\over 2} }
\\
&\times \sum_{i,j }\sum_{  k_1,\dots k_{\ell_1 + \ell_2 - \ell_3-1}  } G_{i k_1}(\varphi_{12})\cdots  G_{ k_{\ell_1 + \ell_2 - \ell_3-1} j}(\varphi_{12})
\sum_{ m_1,\dots m_{\ell_2 + \ell_3 - \ell_1-1}}  G_{i m_1}(\varphi_{23})\cdots  G_{ m_{\ell_2 + \ell_3 - \ell_1-1} j}(\varphi_{23})
\\
&\times \sum_{ n_1,\dots n_{\ell_1 + \ell_3 - \ell_2-1}}  G_{j n_1}(\varphi_{13})\cdots  G_{ n_{\ell_1 + \ell_3 - \ell_2-1} i}(\varphi_{13})  \,.
\\
\fe
Plugging in the explicit form of the propagators, in the large $N$ limit we can write 
 \es{Cpqr}{
C_{\ell_1 \ell_2 \ell_3}&=\frac{8\ell_1 \ell_2 \ell_3 }{(8 \pi)^{\ell_1 + \ell_2 + \ell_3}N^\frac{\ell_1 + \ell_2 + \ell_3-3}{ 2} } N^{\ell_1 + \ell_2 + \ell_3-1}  \int dx_1\, dx_2\,  \rho (x_1) \rho(x_2)\cosh ( \pi\sqrt{N} x_{12}) \,{\cal I} \,, \\
{\cal I}  &\equiv  {\prod_{i=1}^{\ell_1 + \ell_2 - \ell_3-1} \int du_i \rho(u_i)\over \cosh(\pi \sqrt{N} (x_1-u_1)) \cdots \cosh(\pi \sqrt{N} (u_{\ell_1 + \ell_2 - \ell_3-1}-x_2)) }
\\
&\times
{\prod_{i=1}^{\ell_2 + \ell_3 - \ell_1-1} \int dv_i \rho(v_i)\over \cosh(\pi \sqrt{N} (x_1-v_1)) \cdots \cosh(\pi \sqrt{N} (v_{\ell_2 + \ell_3 - \ell_1-1}-x_2)) }
\\
&\times 
{\prod_{i=1}^{\ell_1 + \ell_3 - \ell_2-1} \int d w_i \rho(w_i)\over \cosh(\pi \sqrt{N} (x_1-w_1)) \cdots \cosh(\pi \sqrt{N} (w_{\ell_1 + \ell_3 - \ell_2-1}-x_2)) } \,.
 }

Now observe that in large $N$, we approximate
 \es{ISimple}{
I &\approx {\cal A}\,\D(x_1-x_2) \D(x_1-u_1) \D(u_1-u_2)\dots \D(u_{\ell_1 + \ell_2 - \ell_3-2}-u_{\ell_1 + \ell_2 - \ell_3-1})  
\\
&\times
\D(x_1-v_1) \D(v_1-v_2)\dots \D(v_{\ell_2 + \ell_3 - \ell_1-2}-v_{\ell_2 + \ell_3 - \ell_1-1})  
\\
&\times
\D(x_1-w_1) \D(u_1-w_2)\dots \D(w_{\ell_1 + \ell_3 - \ell_2-2}-w_{\ell_1 + \ell_3 - \ell_2-1})  
 }
with the coefficient ${\cal A}$ being fixed by integrating both sides with respect to $x_1$, the $u_i$, $v_i$, and $w_i$ to
 \es{GotA}{
{\cal A}
  =\frac{1}{ N^{\ell_1 + \ell_2 + \ell_3-2\over 2}}  \frac{\Gamma(\ell_1)\Gamma(\ell_2)\Gamma(\ell_3)}{ \sqrt{\pi} \Gamma({1+\ell_1 + \ell_2 - \ell_3\over 2})\Gamma({1+\ell_1 + \ell_3 - \ell_2 \over 2})\Gamma({1+\ell_2 + \ell_3 - \ell_1 \over 2})\Gamma({\ell_1 + \ell_2 + \ell_3\over 2})} \,.
} 

Plugging \eqref{ISimple} into \eqref{Cpqr}, we have 
 \es{GotCpqr}{
   C_{\ell_1 \ell_2 \ell_3}=\frac{8\ell_1 \ell_2 \ell_3}{(8\pi)^{\ell_1 + \ell_2 + \ell_3}N^\frac{\ell_1 + \ell_2 + \ell_3-3}{ 2} }N^{\ell_1 + \ell_2 + \ell_3-1} {\cal A}\, \int d x\, \rho(x)^{\ell_1 + \ell_2 + \ell_3-1} \,.
 }
Performing the integral in \eqref{GotCpqr} and using \eqref{GotA}, we find
 \es{CpqrFinal}{
C_{\ell_1 \ell_2 \ell_3}
=& \frac{N^\frac{3}{2}}{ (8\pi)^{\ell_1 + \ell_2 + \ell_3}}{2^{{\ell_1 + \ell_2 + \ell_3\over 2}+3}\over \ell_1 + \ell_2 + \ell_3} \frac{\Gamma(\ell_1+1)\Gamma(\ell_2+1)\Gamma(\ell_3+1)}{ \sqrt{\pi} \Gamma(\frac{1+\ell_1 + \ell_2 - \ell_3}{ 2})\Gamma({1+\ell_1 + \ell_3 - \ell_2\over 2})\Gamma({1+\ell_2 + \ell_3 - \ell_1\over 2})\Gamma({\ell_1 + \ell_2 + \ell_3\over 2})} \,.
 }
Again, it is nice to see that $C_{\ell_1 \ell_2 \ell_3}$ scales as $N^{3/2}$ at large $N$, as should have been expected from the supergravity dual.

\subsubsection{Selected four-point function}

One can use the same method as above to compute four-point functions at large $N$.  Let us simply quote the result in the two simplest cases, skipping the derivation.  We have 
 \es{O1O1O1O1}{
  \langle {\cal O}_1(\vphi_1, y_1) &{\cal O}_1 (\vphi_2, y_2)  
   {\cal O}_1 (\vphi_3, y_3)  {\cal O}_1 (\vphi_4, y_4)  \rangle
    \approx \frac{N^{\frac{3}{2} }}{288 \sqrt{2} \pi^5 } \\
    & \times \bigg( \le(2\la y_1, y_2\ra\la y_2, y_3 \ra \la  y_3, y_4 \ra \la y_4, y_1 \ra  -\la y_1, y_3\ra^2\la y_2, y_4 \ra^2\ri){\rm sgn}(\varphi_{12}\varphi_{23}\varphi_{34}\varphi_{41}) \\
  & + \le(2\la y_1, y_3 \ra\la y_3, y_4 \ra \la y_4, y_2 \ra \la y_2, y_1\ra-\la y_1, y_4\ra^2\la y_2, y_3 \ra^2\ri) {\rm sgn}(\varphi_{13}\varphi_{34}\varphi_{42}\varphi_{21})\\
  & +\le(2\la y_1, y_4 \ra\la y_4, y_2 \ra \la y_2, y_3\ra \la y_3, y_1\ra -\la y_1, y_2\ra^2\la y_3, y_4 \ra^2\ri){\rm sgn}(\varphi_{14}\varphi_{42}\varphi_{23}\varphi_{31}) 
  \bigg)  \,, \\
  \langle {\cal O}_1(\vphi_1, y_1) &{\cal O}_1 (\vphi_2, y_2) 
   {\cal O}_1 (\vphi_3, y_3)  {\cal O}_2 (\vphi_4, y_4)  \rangle
   \approx 
    \frac{N^{\frac 32}}{128 \pi^7} \\
   &\times
    \biggl( \la y_1, y_2 \ra \la y_1, y_4 \ra \la y_2, y_4 \ra \la y_3, y_4 \ra^2
      \sgn(\vphi_{12} \vphi_{24} \vphi_{41}) \\
     &{}+ \la y_1, y_3 \ra \la y_1, y_4 \ra \la y_3, y_4 \ra \la y_2, y_4 \ra^2
      \sgn(\vphi_{13} \vphi_{34} \vphi_{41}) \\
     &{}+\la y_2, y_3 \ra \la y_2, y_4 \ra \la y_3, y_4 \ra \la y_1, y_4 \ra^2
      \sgn(\vphi_{23} \vphi_{34} \vphi_{42})  \biggr)\,.
 }
 We will reproduce these results from a bulk computation in Section~\ref{BULKDUAL}.
 
\subsection{Enhanced symmetry}

So far, we have only analyzed connected correlation functions of single trace operators and found that they all scale as $N^{3/2}$.  One may also consider multi-trace operators and compute their correlation functions.  For double trace operators of the form $:{\cal O}_{\ell_1} {\cal O}_{\ell_2}:$ one finds the following scaling with $N$:
 \es{LargeNDouble}{
  \langle :{\cal O}_{\ell_1} {\cal O}_{\ell_2}: \, :{\cal O}_{\ell_1} {\cal O}_{\ell_2}: \rangle &\sim N^3 \,, \\
  \langle {\cal O}_{\ell_1} {\cal O}_{\ell_2} \, :{\cal O}_{\ell_1} {\cal O}_{\ell_2}: \rangle &\sim N^3 \,, \\
    \langle {\cal O}_{\ell_1} {\cal O}_{\ell_2} \, :{\cal O}_{\ell_3} {\cal O}_{\ell_4}: \rangle &\sim N^{3/2} \,, \qquad \text{if } (\ell_1, \ell_2) \neq (\ell_3, \ell_4) \,.
 }
This implies that two single trace operators ${\cal O}_{\ell_1}$ and ${\cal O}_{\ell_2}$ have an OPE of the schematic form
 \es{OPE}{
  {\cal O}_{\ell_1} \times {\cal O}_{\ell_2} &\sim N^{3/2}\delta_{\ell_1 \ell_2}  {\bf 1}  + N^0 \left( : {\cal O}_{\ell_1} {\cal O}_{\ell_2}: + C_{\ell_1 \ell_2}^{\ell_3} {\cal O}_{\ell_3} \right) + O(N^{-3/2}) \,,
 }
where $C_{\ell_1 \ell_2}^{\ell_3}$ has a finite limit as $N \to \infty$.\footnote{The equation \eqref{OPE} is only schematic, but we have $C_{\ell_1 \ell_2}^{\ell_3} = C_{\ell_1 \ell_2 \ell_3} / B_{\ell_3}$.}

To connect this to the discussion in Section~\ref{ENHANCED} we should be more precise.  The operators ${\cal O}_\ell(\vphi, y)$ are a shorthand notation for $2 \ell + 1$ independent operators ${\cal O}_{\ell m}(\vphi)$, $m = -\ell, \ldots, \ell$, that can be read off from the expansion of ${\cal O}_\ell (\vphi, y)$ in terms of $y$.  Let us define them by\footnote{The definition \eqref{OExpansion1} may seem overly complicated---a simpler expression one could have written down instead of \eqref{OExpansion1} is ${\cal O}_{\ell}(\vphi, y) = \sum_{m=-\ell}^\ell i^{\ell-m}\, (y^1)^{\ell +m} (y^2)^{\ell - m}\, {\cal O}_{\ell m}(y)$. We are working with \eqref{OExpansion1} instead because this choice makes the 2-point functions diagonal, as we discuss below. We note that the difference between the simpler expression just mentioned and   \eqref{OExpansion1} resembles the difference between complex and real spherical harmonics.} 
  \es{OExpansion1}{
  {\cal O}_{\ell}(\vphi, y) =& \sum_{m=1}^\ell {i^{\ell-m}\ov \sqrt{2}} \le[(y^1)^{\ell +m} (y^2)^{\ell - m}+(y^1)^{\ell +m} (y^2)^{\ell - m}\ri] \, {\cal O}_{\ell m}(\vphi)+i^{\ell}\,(y^1)^{\ell } (y^2)^{\ell} \, {\cal O}_{\ell 0}(\vphi)\\
  &+ \sum_{m=-\ell}^{-1} {i^{\ell-m-1}\ov \sqrt{2}} \le[(y^1)^{\ell +m} (y^2)^{\ell - m}-(y^1)^{\ell +m} (y^2)^{\ell - m}\ri] \, {\cal O}_{\ell m}(\vphi)\,.
 }  
Equivalently the component ${\cal O}_{\ell m}(\vphi)$ can be recovered from ${\cal O}(\vphi, y)$ by acting with a differential operator ${\cal D}_{\ell m}(y)$ on ${\cal O}(\vphi, y)$:
  \es{OExpansion}{
  {\cal O}_{\ell m}(\vphi) &= {\cal D}_{\ell m}(y)  {\cal O}_\ell (\vphi, y)\\
  {\cal D}_{\ell m}(y) &\equiv \begin{cases}
  \frac{(-i)^{\ell-m}}{\sqrt{2}(\ell + m)! (\ell - m)!} \le[\frac{\partial^{2 \ell}}{ (\partial y^1)^{\ell + m} (\partial y^2)^{\ell - m} }+\frac{\partial^{2 \ell}}{ (\partial y^1)^{\ell - m} (\partial y^2)^{\ell + m} }\ri]  \qquad &m>0 \,, \\
    \frac{(-i)^{\ell}}{(\ell !)^2} \,\frac{\partial^{2 \ell}}{ (\partial y^1)^{\ell } (\partial y^2)^{\ell } }  \qquad &m=0 \,, \\
      \frac{(-i)^{\ell-m-1}}{\sqrt{2}(\ell + m)! (\ell - m)!} \le[\frac{\partial^{2 \ell}}{ (\partial y^1)^{\ell + m} (\partial y^2)^{\ell - m} }-\frac{\partial^{2 \ell}}{ (\partial y^1)^{\ell - m} (\partial y^2)^{\ell + m} }\ri]  \qquad &m<0 \,.
  \end{cases}
 }

With this definition, we have the 2-point and 3-point functions
 \es{2pt}{
  \langle {\cal O}_{\ell_1 m_1}(\vphi_1) {\cal O}_{\ell_2 m_2}(\vphi_2) \rangle &= B_{\ell_1 m_1} \delta_{\ell_1 \ell_2} \delta_{m_1m_2} \,, \\
   \langle {\cal O}_{\ell_1 m_1}(\vphi_1) {\cal O}_{\ell_2 m_2}(\vphi_2) {\cal O}_{\ell_3 m_3}(\vphi_3) \rangle &= C_{\ell_1 m_1, \ell_2 m_2, \ell_3 m_3} \\
   &\times (\sgn \vphi_{21})^{\ell_1 + \ell_2 - \ell_3}
        (\sgn \vphi_{31})^{\ell_1 + \ell_3 - \ell_2} 
     (\sgn \vphi_{32})^{\ell_2 + \ell_3 - \ell_1} \,,  
  }
 with 
\es{BCMain}{
 B_{\ell m} &\equiv B_\ell {\cal D}_{\ell m}(y_1) {\cal D}_{\ell, -m}(y_2) \langle y_1, y_2 \rangle^{2\ell} =B_{\ell} \frac{(2 \ell)!}{(\ell + m)! (\ell -m)!}  \,, \\
 C_{\ell_1 m_1 ,\ell_2 m_2 ,\ell_3 m_3} &\equiv
  C_{\ell_1 \ell_2 \ell_3}  {\cal D}_{\ell_1 m_1}(y_1)   {\cal D}_{\ell_2 m_2}(y_2)   {\cal D}_{\ell_3 m_3}(y_3)
   \\
   &\times \langle y_1, y_2 \rangle^{\ell_1 + \ell_2 - \ell_3}
    \langle y_1, y_3 \rangle^{\ell_1 + \ell_3 - \ell_2} 
     \langle y_2, y_3 \rangle^{\ell_2 + \ell_3 - \ell_1} \,.
}
Finally, we make one additional field redefinition to have properly normalized 2-point functions that match the form of the current OPE \eqref{CurrentOPE}, and to facilitate the comparison to the bulk theory. We also unpack \eqref{2pt} a bit
\es{FieldRedef}{
 j_{\ell m}(\vphi)&\equiv {{\cal O}_{\ell m}(\vphi) \ov \sqrt{ B_{\ell m}/B}}\,,   \qquad B\equiv {N^{3/2}\ov 6\sqrt{2}\, \pi} \,,  \\
\langle j_{\ell_1 m_1}(\vphi_1) j_{\ell_2 m_2}(\vphi_2) \rangle &= B\, \delta_{\ell_1 \ell_2} \delta_{m_1m_2}  \,,\\
   \langle j_{\ell_1 m_1}(\vphi_1) j_{\ell_2 m_2}(\vphi_2) j_{\ell_3 m_3}(\vphi_3) \rangle &=\begin{cases}
 - {B\ov 2} f^{\ell_1 m_1, \ell_2 m_2, \ell_3 m_3} \, \sgn \le(\vphi_{12}\vphi_{23}\vphi_{31}\ri) \quad  &\ell_1 + \ell_2 + \ell_3=\text{odd}\,,\\
 B\,d_3^{\ell_1 m_1, \ell_2 m_2, \ell_3 m_3}  \quad &\ell_1 + \ell_2 + \ell_3=\text{even}\,,
   \end{cases}\\
   f^{\ell_1 m_1, \ell_2 m_2, \ell_3 m_3}&\equiv -{2C_{\ell_1 m_1 ,\ell_2 m_2 ,\ell_3 m_3}\ov \sqrt{ B_{\ell_1 m_1}B_{\ell_2 m_2} B_{\ell_3 m_3}/B}} \qquad \ell_1 + \ell_2 + \ell_3=\text{odd} \,, \\
    d_3^{\ell_1 m_1, \ell_2 m_2, \ell_3 m_3}&\equiv {C_{\ell_1 m_1 ,\ell_2 m_2 ,\ell_3 m_3}\ov \sqrt{ B_{\ell_1 m_1}B_{\ell_2 m_2} B_{\ell_3 m_3}/B}} \hspace{1.2cm} \ell_1 + \ell_2 + \ell_3=\text{even}\,.
}
Note that $B=O(N^{3/2})$, while the rest of the defined quantities are $O(N^0)$, and the translation between $j_{\ell m}$ and ${\cal O}_{\ell m}$ only involves a numerical factor. We have chosen the value of $B$ such that the structure constants for the $\mathfrak{su}(2)$ currents are properly normalized: $f^{1 m_1, 1 m_2, 1 m_3}=\ep^{m_1 m_2 m_3}$.  Also note that $d_3^{1 m_1, 1 m_2, 1 m_3}=0$ because $\mathfrak{su}(2)$ does not have a 3-index symmetric invariant tensor, see \eqref{d4}.

 The OPE in Eq.~\eqref{OPE} can then be written more precisely as
  \es{OPEMorePrecise1}{
 & j_{\ell_1 m_1}(\vphi_1) j_{\ell_2 m_2}(\vphi_2) \sim B\,  \delta_{\ell_1 \ell_2} \delta_{m_1 m_2} {\bf 1}  
  + N^0 \biggl( : j_{\ell_1 m_1} j_{\ell_2 m_2}:(\vphi_1) \\
   &+ 
     \sum_{\ell_3=\abs{\ell_1-\ell_2}}^{\ell_1+\ell_2}\sum_{m_3=-\ell_3}^{\ell_3} \le[
    \frac12 f^{\ell_1 m_1, \ell_2 m_2, \ell_3 m_3}\, \sgn \vphi_{21}+  d_3^{\ell_1 m_1, \ell_2 m_2, \ell_3 m_3} \ri] j_{\ell_3 m_3}(\vphi_1) \biggr) 
  + O(N^{-3/2}) \,.
 }
Let us refer to $(\ell m)\equiv a$, which runs over all single trace operators. In the above formula it is understood that $f^{a  b c}$ is only nonzero when $\ell_1 + \ell_2 + \ell_3=\text{odd}$, while $d_3^{ a  b c}$ is only nonzero for $\ell_1 + \ell_2 + \ell_3=\text{even}$. Recall that $f^{\ell_1 m_1, \ell_2 m_2, \ell_3 m_3}$ and $d_3^{\ell_1 m_1, \ell_2 m_2, \ell_3 m_3}$ vanish unless $\ell_1,\, \ell_2,\, \ell_3$ satisfy the triangle inequality. 

As explained in Section~\ref{ENHANCED}, any topological theory should have an enhanced symmetry generated by the operator algebra. The structure constants of this symmetry algebra can be read off from the terms proportional to $\sgn \vphi$, see \eqref{CurrentOPE}. 
We see that up to order $N^0$, besides the identity ${\bf 1} $, the operators that contribute to the OPE are single trace operators, as well as the double trace operator  $: j_{\ell_1 m_1} j_{\ell_2 m_2}:$. The latter does not contribute to the structure constants at large $N$, and we may consistently talk about the enhanced symmetry of the single trace sector.\footnote{As argued in Section~\ref{ENHANCED} this is a general feature of large $N$ theories, here we are seeing a particular instance of this general structure.} The double trace operator does contribute to the symmetric part of the OPE, and $d_3^{a b C}\neq 0$ in general, where the $C$ index runs over all operators of the theory.

Note that the coefficients $f^{abc}$ and $d_3^{abc}$ calculated in \eqref{FieldRedef} must obey the nontrivial relations \eqref{fdIdentities2} that are the Ward identities of the theory corresponding to the enhanced symmetry. We have explicitly checked many of these relations and found that they are indeed obeyed.  It would be interesting to understand to what extent these relations constrain the form of the structure constants.

\subsection{The algebra $\text{SDiff}(S^2)$}
\label{SDiff}

The enhanced global symmetry of the single trace sector discussed in the previous section can be identified, after complexification, with the complexified algebra  $\text{SDiff}(S^2)$ of area-preserving diffeomorphisms of $S^2$.\footnote{We work with complexified Lie algebras because we do not know which reality conditions we should impose on the generators.  The complexified Lie algebra $\text{SDiff}(S^2)$ has different real forms corresponding to the algebras of area-preserving diffeomorphisms of $S^2$ and $\HH^2$.  We suspect that when reality conditions are appropriately taken into account, we would obtain the real form corresponding to the area-preserving diffeomorphisms of $S^2$.}${}^{,}$\footnote{We thank J.~Maldacena and H.~Verlinde for suggesting that the symmetry algebra may be  $\text{SDiff}(S^2)$. }  This algebra has been studied extensively, and its structure constants are given, for example, in \cite{Pope:1990kc,Pope:1989bc}.

One way to check that the structure constants we obtained are precisely those of the $\text{SDiff}(S^2)$ algebra is to view $\text{SDiff}(S^2)$ as the limit $\lambda \to \pm i \infty$ of the 3d higher spin algebra $\mathfrak{hs}[\lambda]$ \cite{Bergshoeff:1989ns}, and to compare the normalized structure constants 
 \es{NormStruct}{
  \frac{C_{\ell_1 \ell_2 \ell_3} }{\sqrt{ B_{\ell_1} B_{\ell_2} B_{\ell_3} } }   \qquad \qquad
   \text{($\ell_1 + \ell_2 + \ell_3$ odd)}
 }
we obtained to those of the  $\lambda \to \pm i \infty$ limit of $\mathfrak{hs}[\lambda]$. We can read off  \eqref{NormStruct} for $\mathfrak{hs}[\lambda]$ from the results of \cite{Joung:2014qya}, for instance.  Indeed, if ${\cal O}_\ell(\vphi, y)$ formed a $\mathfrak{hs}[\lambda]$ algebra, then the quantities
 \es{OTwoThreeRepackage}{
   B(y_{12}) &\equiv  \sum_{\ell}^\infty B_{\ell}\, y_{12}^{2 \ell} \,, \\
  C(  y_{12},  y_{23}, y_{31}) &\equiv \sum_{\ell_1, \ell_2, \ell_3} C_{\ell_1 \ell_2 \ell_3} y_{12}^{\ell_1 + \ell_2 - \ell_3} 
   y_{23}^{\ell_2 + \ell_3 - \ell_1} y_{31}^{\ell_3 + \ell_1 - \ell_2} 
 }
 with $y_{ij}\equiv \la y_i,y_j\ra$
would be bilinear and trilinear forms on this algebra, and according to \cite{Joung:2014qya} they would have to take the form 
 \es{GotBC}{
  \mathfrak{hs}[\lambda]: \qquad \qquad\qquad
   B(y_{12}) &= {}_2F_1 \left(1 + \lambda, 1 - \lambda; \frac 32; \frac{y_{12}^2}{4} \right) \,, \\
   C(y_{12},  y_{23}, y_{31}) &=  
    {}_2F_1 \left(1 + \lambda, 1 - \lambda; \frac 32; \frac{y_{12}^2 + y_{23}^2 + y_{31}^2  + y_{12} y_{23} y_{31}}{4} \right) \,.
 }
It is straightforward to check that \eqref{GotBC} produces the same values of \eqref{NormStruct} as those obtained from \eqref{XnXnLargeNAgain2} and \eqref{CpqrFinal}, provided that we take
 \es{Gotlambda}{
  \lambda \approx \pm i \left(\frac{24}{\pi} \right)^{1/4} N^{3/4}
 }
and $N$ large.

It is tempting to give a geometric meaning to the fact that our Lie algebra corresponds to area-preserving diffeomorphisms of $S^2$.  For instance, these diffeomorphisms could represent the diffeomorphisms of a two-sphere embedded into the $S^7$ internal space of the $AdS_4 \times S^7$ supergravity background.  It would be very interesting to explore this direction further in future work.

The discussion above was limited to the $\Z_2$-invariant sector of the 1d theory \eqref{ZsN8Again}.  If we further include the $\Z_2$-odd sector of \eqref{ZsN8Again}, we expect that the higher spin algebra $\mathfrak{hs}[\lambda]$ is replaced by the supersymmetric higher spin algebra $\mathfrak{shs}[\lambda]$.  In the limit $\lambda \to \pm i \infty$, we would expect to find a geometric algebra associated with a supersymmetrized analog of a two-sphere.

\subsection{Bulk dual}  \label{BULKDUAL}

At large $N$, the bulk dual consists of a non-Abelian gauge theory in $AdS_2$ based on an infinite-dimensional gauge algebra. From \eqref{FieldRedef} it follows that the bulk Yang-Mills coupling is
\es{BulkYM}{
   \frac{1}{g_\text{YM}^2} =  \frac{N^{3/2} }{3\sqrt{2}} \,.
}
  Every  single-trace current  $j_{\ell m}$ is dual to a gauge field $A^{\ell m}$. We may read off the structure constants $f^{\ell_1 m_1, \ell_2 m_2, \ell_3 m_3}$ and the coefficients of the three-derivative couplings $d_3^{\ell_1 m_1, \ell_2 m_2, \ell_3 m_3}$ from the current 3-point functions \eqref{FieldRedef}, thus determining the bulk theory to quadratic and cubic orders.
  
It is possible to go beyond cubic order, but the computations become rather complicated, and we will limit ourselves to the determination of only one 4-point coupling of the bulk gauge theory:  $d_4^{1 m_1, 1 m_2, 1 m_3, 1 m_4}$. This object is familiar from the discussion of $\mathfrak{su}(2)$ gauge theory discussed in Section~\ref{SU2EXAMPLE}. As a consequence of  $\mathfrak{su}(2)$ gauge invariance, there is just one independent structure (see \eqref{d4}, which we repeat here for convenience)
\es{d4Copy}{   
   d_4^{1 m_1, 1 m_2, 1 m_3, 1 m_4}=w_4 \, \de^{(m_1 m_2}\de^{m_3 m_4)}
   }
proportional to a coefficient $w_4$.   The determination of $w_4$ is done by matching the 4-point function  $\langle {\cal O}_1 {\cal O}_1   {\cal O}_1  {\cal O}_{1} \rangle$ between the bulk and boundary sides.  On the boundary side, this 4-point function is given in \eqref{O1O1O1O1}. 
On the bulk side, the 4-point function can be read off from \eqref{Contrib4pt} after plugging in the 3-point data (structure constants and the coefficient $d_3$) derived above and using \eqref{OExpansion1} and \eqref{FieldRedef} to convert from $j_{\ell m}$ to ${\cal O}_{\ell m}$.  By matching the two expressions, we find 
 \es{Gota4ABJM}{
   w_4 = 0 \,.
  }  

We can also see how our bulk YM$_2$ theory reproduces the  4-point function $\langle {\cal O}_1 {\cal O}_1   {\cal O}_1  {\cal O}_{2} \rangle$ given in \eqref{O1O1O1O1}.  In the bulk, there are no 4-derivative interactions that contribute to this correlator because $\mathfrak{su}(2)$ invariance requires $d_4^{1 m_1, 1 m_2, 1 m_3, 2 m_4}=0$.  Hence, the only term in  \eqref{Contrib4pt} that contributes to it is of the form  $d_3^{1 m_1, 1 m_2, 2 m}\, f^{2 m, 1 m_3, 2 m_4}$.  Plugging in the values of the $d_3$ and $f$ constants we determined above, one can check that the bulk computation reproduces \eqref{O1O1O1O1} precisely.\footnote{ We remark that using the two explicitly worked out examples in this paper, the dual of the $\mathfrak{u}(N)$ singlet sector of $2N$ free hypermultiplets (with $\mathfrak{su}(2)$ global symmetry) in Section~\ref{FREEHYPERS} and that of ABJM theory,  we have now tested all terms in \eqref{Contrib4pt}. The  4-derivative coupling was essential to get a match with the 4-point functions in the free example, but it turned out to vanish in the  dual of ABJM theory. In the free case the $d_3^{ab e} f^{ecd}$-type terms from \eqref{Contrib4pt} all vanish in the bulk $\mathfrak{su}(2)$ Yang-Mills theory, but they represent the only contributions to the  $\langle {\cal O}_1 {\cal O}_1  
   {\cal O}_1  {\cal O}_{2} \rangle$ correlator in ABJM theory.}
   
   We conclude that we have determined the bulk dual of the topological quantum mechanical sector of ABJM theory. While in the free hypermultiplet examples discussed in Section~\ref{FREEHYPERS} in the large $N$ limit there was no enhancement of symmetry, in this interacting case the manifest  $\mathfrak{su}(2)$ symmetry was enhanced to the infinite dimensional symmetry algebra $\text{SDiff}(S^2)$. Correspondingly, the bulk gauge symmetry is also infinite dimensional. We have successfully performed computations in this theory that match the field theory result \eqref{O1O1O1O1}. The bulk theory can perhaps be reformulated in a more compact way by promoting  $A^{\ell m}(X)$ to $A(X,y)$ similar to the field theory construction \eqref{OExpansion1}, but also combining the gauge fields corresponding to different $\ell$ quantum numbers. This reformulation would necessitate the introduction of a star-product, acting on the $y$ variable. We leave such a construction for future work.   

\section{Localization from $AdS_4$}
\label{LOCALIZATION}

Since on the field theory side the 1d topological theory can be obtained from a supersymmetric localization computation in the 3d theory \cite{Dedushenko:2016jxl}, it is natural to ask whether the dual $AdS_2$ gauge theory can be obtained from localizing the $AdS_4$ or 11d supergravity in the case of the theories of Section~\ref{MATRIX} or the $AdS_4$ higher spin theories in the case of the theories of Section~\ref{VECTOR}.  Of course, an action for the higher spin theories is not available, so performing localization in these instances is not feasible. 

Since we are interested in 3d $\cN=4$ SCFTs on the boundary, we will be concerned with $\cN=4$ supergravity on $AdS_4$.
However we will not pursue the localization of the full supergravity theory here since such a computation appears to be rather involved (see nevertheless \cite{Dabholkar:2014wpa} and also \cite{Dabholkar:2010uh,Dabholkar:2012nd,Dabholkar:2011ec,Dabholkar:2014ema}).  What we will present here is a localization computation of a non-abelian vector multiplet in a fixed $AdS_4$ background metric.  This theory is certainly a consistent subsector of maximal gauged supergravity on $AdS_4$, and it would also be a consistent subsector of the higher spin theory on $AdS_4$ had an action for such a theory been written down.  See also \cite{Bonetti:2016nma} where a similar toy model approach was taken to the localization of maximally supersymmetric $AdS_5$ supergravity.

Below we will start by introducing the off-shell formulation of $\cN=4$ SYM on $AdS_4$ with supersymmetric boundary terms.  
Next we map the localization problem in the boundary SCFT to that in the bulk by identifying the relevant symmetries, and in particular the localizing supercharge $\cQ$. We then evaluate the SYM action on the BPS locus and obtain 2d YM on an $AdS_2$ slice with the desired boundary terms. In the last subsection, we comment on how the computation we presented can be applied to the holographic duality between ABJM theory and 11d supergravity on $AdS_4 \times S^7$.

\subsection{$\cN=4$ SYM on $AdS_4$ and off-shell supersymmetry}

\subsubsection{ $\cN=4$ SYM on compact four manifolds}

The maximal SYM theory in four dimensions can be obtained from dimensional reduction of 10d SYM\@.  We follow \cite{Pestun:2009nn} in using the notation from 10d SYM and split the 10d Gamma matrices as $\Gamma_M=\{\Gamma_\m,\Gamma_A\}$ with $\m=1,\dots,4$ and $A=5,\dots,9,0$ (see Appendix~\ref{App:gm} for details for the Gamma matrix conventions). The action for 4d $\cN=4$ SYM with gauge group $G$ on a general compact four manifold $\cM$ is \cite{Berkovits:1993hx,Pestun:2007rz}
\ie
S = {1\over 2 g^2}\int_{\cM} d^4 x\, \sqrt{g} \tr \Bigg(
{1\over 2}F_{MN}F^{MN}-\Psi \Gamma^M D_M\Psi  +{\cR\over 6} \Phi^A \Phi_A -K^m K_m
\Bigg)\,,
\label{SYMos}
\fe
where $\cR$ denotes the scalar curvature of $\cM$, and $K_m$ with $m=1,\dots 7$ are auxiliary fields which serve to realize the supersymmetry that we will use to localize off-shell. We discuss the symmetries of \eqref{SYMos} below.
First, the SUSY transformations are
\ie
&\D_\ve A_M= \varepsilon \Gamma_M \Psi  \,,
\\
&\D_\ve\Psi ={1\over 2}F_{MN}\Gamma^{MN}\varepsilon+{1\over 2}\Gamma_{\m A}\Phi^A \nabla^\m \varepsilon+K^m \n_m \,,
\\
&\D_\ve K^m=-\n^m\Gamma^M D_M\Psi\,,
\label{SUSYos}
\fe
where the conformal Killing spinor $\varepsilon$ is a 10d chiral spinor with 16 components and satisfies
\ie
\nabla_\m \varepsilon=\Gamma_\m \tilde{\varepsilon}\,,
\qquad  \tilde\CC^\m \nabla_\m \tilde\varepsilon= -{1\over 12}\cR {\varepsilon} \,, 
\label{CKSeqn}
\fe
for some 10d anti-chiral spinor $\tilde\ve$. Here $\Gamma_\m=e_{\hat\m \m}\Gamma^{\hat \m}$, where $e_{\hat\m}^\m$ is the vielbein and  $\Gamma^{\hat \m}$ denotes flat space 10d Gamma matrices in the chiral basis. The auxiliary 10d chiral spinors $\n^m$ with $m=1,\dots,7$ in \eqref{SUSYos} are chosen to satisfy
\ie
\varepsilon \Gamma^M \n_m=0\,,\qquad \n_m\Gamma^M\n_n=\D_{mn}\varepsilon \Gamma^M \varepsilon\,.
\label{PSpinor}
\fe

Second, the action \eqref{SYMos} are invariant under the  Weyl transformation,
\ie
g_{\m\n}\to g_{\m\n}e^{2\phi}\,,\quad A_\m \to A_\m  \,,\quad \Phi_A\to \Phi_A e^{-\phi}\,,\quad \Psi\to \Psi e^{-{3\over 2}\phi}\,,\quad K_m \to  K_m e^{-2\phi}\,.
\fe
The conformal Killing spinors also transform as
\ie
\varepsilon \to e^{{1\over 2}\phi} \varepsilon\,,\qquad \tilde\varepsilon \to  e^{-{1\over 2}\phi} \left( \tilde\varepsilon +{1\over 2} \CC^\m \pa_\m\phi \varepsilon \right)\,,
\fe
such that \eqref{CKSeqn} is invariant and the SUSY transformations \eqref{SUSYos} are also preserved.

Finally the action \eqref{SYMos} has $SO(6)$ R-symmetry which is generated by $R_{AB}$ with $A,B=5,\dots,9,0$ which act on $\Phi^C$ as $(R_{AB})^D{}_C=2\D^D_{[A}\D_{B]C} $ and on $\Psi$ as $R_{AB}={1\over 2} \CC_{AB}$. 

Anti-commutators of the supersymmetry transformations associated with two conformal Killing spinors take the form\footnote{In writing this equation we take $\D_{\varepsilon_{1,2}}$ to be the on-shell supersymmetry transformation generators (turning off the auxiliary fields $K_m$).}
\ie
\{\D_{\varepsilon_1},\D_{\varepsilon_2}\}=-2\cL_{v}-2\cR_\omega-2\cG_\zeta-2\Omega_\lambda+({\rm e.o.m})\,,
\label{SUSYac}
\fe
where $\cL_v$ is the Lie derivative along the conformal Killing vector field $v^\m=\varepsilon_{(1}\CC^\m \varepsilon_{2)}$; $\cR$ is the generator of the $SO(6)_R$ symmetry with parameter $\omega_{AB}= \varepsilon_{(1} \tilde\CC_{AB}\tilde \varepsilon_{2)}$;  $\cG$ is the gauge transformation with gauge parameter $\zeta=v^M A_M$, where  $v^M=\varepsilon_{(1}\CC^M \varepsilon_{2)}$;  and $\Omega$ denotes the Weyl transformation with  parameter $\lambda=2\varepsilon_{(1}\tilde\varepsilon_{2)}$. 

\subsubsection{ $\cN=4$ SYM on $AdS_4$ and boundary terms}
We are interested in the case where $\cM$ is the non-compact manifold $AdS_4$ with curvature radius $L$ (see Appendix~\ref{app:ads4} for the coordinate systems we use on $AdS_4$). In this case,  $\cR=-{12\over L^2}$ and the conformal Killing equation \eqref{CKSeqn} becomes
\ie
\nabla_\m \epsilon =\Gamma_\m \tilde \epsilon\,,\qquad  
\nabla_\m \tilde \epsilon ={1\over 4L^2}\Gamma_\m  \epsilon\,.
\label{CKSeqAdS}
\fe

To write down the most general solutions to \eqref{CKSeqAdS}, it is convenient to use the Poincar\'e coordinates for $AdS_4$
\ie
ds^2={L^2 \, dx_\m^2\over x_4^2}\,,
\label{PoincareAdS4}
\fe
where $x_4>0$, and the vielbein is
\ie
e_{\hat \m}^\m={L\over x_4}\D^\m_{\hat{\m}}\,.  
\fe

The conformal Killing spinors that solve \eqref{CKSeqAdS} are parametrized by two arbitrary constant 10d chiral and anti-chiral spinors $\epsilon_s$ and $\epsilon_c$,
\ie
\varepsilon(\epsilon_s,\epsilon_c) =  i\sqrt{2L\over x_4}  \left(\epsilon_s +{1\over 2L}\, x^{\hat\m} \CC_{\hat\m} \epsilon_c\right)\,.
\label{gencks}
\fe

In general, to perform localization on a non-compact space, it is essential to include appropriate boundary terms in order to preserve the off-shell supersymmetry \eqref{SUSYos}. We show in Appendix~\ref{App:bt} that when the conformal Killing spinor satisfies\footnote{This condition is compatible with \eqref{CKSeqn} and constrains the supercharges to those whose anticommutators do not involve Weyl transformations \cite{Minahan:2015jta}.}
\ie
\tilde \ve=-{1\over 2L}\Lambda \ve \,, 
\fe
with $\Lambda\equiv\CC^{789}$, the desired boundary term is given by
\ie
S_B= {1\over 2 g_4^2}\int d^3 x\,\sqrt{\ga}\, \tr \Bigg(
{1\over 2}\Psi\Lambda\Psi+2K^{a-5}\Phi_a-2D_r\Phi^a \Phi_a-4\Phi_9 F_{78}-2\Phi^a\Phi_a+\Phi^A\Phi_A
\Bigg) \,,
\label{ads4bt}
\fe
with $a=7,8,9$ and $\ga$ denotes the induced metric on the asymptotic boundary. This way we  ensure off-shell supersymmetry\footnote{This is still true if $\cM$ is only asymptotically $AdS_4$ (see Appendix~\ref{App:bt} for a detailed discussion).}
\ie
\D_\ve (S +S_B)=0\,.
\fe

\subsubsection{Variational principle and boundary conditions}\label{sec:varbc}
Now that we have the full supersymmetric action for an $\cN=4$ vector multiplet on $AdS_4$, in this subsection we study the variational problem and specify the consistent boundary conditions needed for the various fields. For this purpose we can set the auxiliary fields $K_m$ to zero. It is clear that the gauge field $A_\m$ and the fermion $\Psi$ should satisfy the standard Dirichlet boundary condition;  we will thus focus on the scalar fields $\Phi_A$. As we shall see, among the six scalars, $\Phi_i$ with $i=5,6,0$ satisfy the usual Dirichlet boundary condition, whereas $\Phi_a$ with $a=7,8,9$ satisfy a more intricate boundary condition that corresponds to the alternate quantization in the bulk \cite{Klebanov:1999tb,Freedman:2016yue}.

The variation of $S+S_B$ is
\ie
\D_{\rm on-shell}(S+S_B)
=&{1\over 2g_4^2}\int d^3 x \, \sqrt{\ga}\\
&\times \tr
\left[
2\D\Phi^i \partial_r \Phi_i
-2\partial_r \D\Phi^a \Phi_a
-4 \epsilon^{abc}\D\Phi_a [\Phi_b,\Phi_c] -2\D\Phi^a\Phi_a+2\D\Phi^i \Phi_i
\right]\,,
\label{osvar}
\fe
where we only keep terms involving $\Phi_a$ and $\Phi_i$ and have dropped terms proportional to the equation of motion.

For large $r$, the $\Phi_A$ behave as
\ie
\Phi_A\approx \Phi_A^{1} e^{-r}+\Phi_A^2 e^{-2r}.
\fe
Using the expansion above, we may then write \eqref{osvar} as
\ie
&\D_{\rm on-shell}(S+S_B)
=
{1\over 8g_4^2}\int d^3 x \sqrt{g_{S^3}} \tr
\left[ -\D \Phi_i^1  \Phi_i^2 
+ \Phi_a^1\, \D(\Phi_a^2- \epsilon^{abc} [\Phi_b^1,\Phi_c^1])
\right].
\fe
We see from this expression that a well-posed variational principle demands the scalars to satisfy the boundary conditions\footnote{See \cite{Freedman:2016yue} for an extensive discussion.}
\ie
\D \Phi_i^1=0\quad &i={5,6,0}\,,
\\
\D(\Phi_a^2- \epsilon^{abc} [\Phi_b^1,\Phi_c^1])=0\quad &a={7,8,9}\,	.
\label{SYMbc}
\fe
The former corresponds to the standard quantization of $\Phi_i$ while the latter is associated with the alternate quantization of $\Phi_a$. From the holographic dictionary, this means that $\Phi_i$ are dual to scalar operators of dimension $\Delta=2$ in the boundary SCFT whereas $\Phi_a$ are dual to scalar operators with   $\Delta=1$.

\subsection{Relation to supersymmetric localization in the boundary SCFT}
The $\cN=4$ SYM theory on $AdS_4$ is invariant under the full 4d superconformal algebra $\mathfrak{psu}(4|4)$. This super-algebra  contains an $\mathfrak{osp}(4|4)$ subalgebra, which can be interpreted either as the 4d supersymmetry algebra of ${\cal N} = 4$  gauged supergravities on $AdS_4$, or as the 3d $\cN=4$ superconformal  algebra of a boundary SCFT\@.    We would like to localize the vector multiplet on $AdS_4$ with respect to the same supercharge that was used in the boundary SCFT in order to obtain the 1d theory \cite{Dedushenko:2016jxl}, so let us first review how to identify this supercharge within $\mathfrak{osp}(4|4)$.

Recall that the 3d superconformal algebra $\mathfrak{osp}(4|4)$ has the maximal bosonic subalgebra 
\es{bosonicosp}{
	\mathfrak{osp}(4|4) \supset \mathfrak{sp}(4)_\text{conf} \times \mathfrak{su}(2)_H \times \mathfrak{su}(2)_C \,,
}
where  $\mathfrak{sp}(4)_\text{conf} \cong \mathfrak{so}(3, 2)$ is the 3d conformal algebra, while $\mathfrak{su}(2)_H$ and $\mathfrak{su}(2)_C$ are the R-symmetry algebras acting on the Higgs and Coulomb branches, respectively.  The supercharges transform in $({\bf 4}, {\bf 2}, {\bf 2})$ of \eqref{bosonicosp}.  When $\mathfrak{osp}(4|4)$ is viewed as the ${\cal N} = 4$ supersymmetry algebra in $AdS_4$, $\mathfrak{sp}(4)_\text{conf} \cong \mathfrak{so}(3, 2)$ is not a conformal symmetry, but rather it represents the isometries of $AdS_4$; $\mf{su}(2)_H\times\mf{su}(2)_C$ is a maximal subgroup of the bulk $SO(6)$ R-symmetry group generated by $\{R_{78},R_{79},R_{89}\}$ and $\{R_{56},R_{05},R_{06}\}$ respectively.

Suppose we first think of the 1d theory as living on a line in $\R^3$.  The 3d superconformal algebra $\mathfrak{osp}(4|4)$ contains an $\mathfrak{su}(2|2) \times \mathfrak{u}(1)_{\cal Z}$ subalgebra which is a supersymmetric extension of the conformal algebra on this line.  We have
\es{bosonic1d}{
	\mathfrak{su}(2|2) \times \mathfrak{u}(1)_{\cal Z} \supset \mathfrak{su}(2)_\text{conf} \times \mathfrak{su}(2)_H \times \mathfrak{u}(1)_{\cal Z} \,,
}
where $\mathfrak{su}(2)_\text{conf}$ can be identified with the global conformal algebra in 1d, while the $\mathfrak{su}(2)_H$ is just the Higgs branch R-symmetry of the 3d SCFT\@.\footnote{A similar construction is of course possible where $\mathfrak{su}(2)_H$ is replaced by $\mathfrak{su}(2)_C$ in \eqref{bosonic1d}.}     The $\mathfrak{su}(2|2)$ algebra contains four complex supercharges, among which one can find two linearly-independent nilpotent supercharges $\cQ_{1,2} $ and their conjugates, each of which defines the (same) cohomology that corresponds to the topological theory. In particular, the twisted $\mathfrak{su}(2)$ algebra
\es{hatL}{
	\hat L_0 &\equiv -D+R_1{}^1 =-{1\over 8}\{\cQ_1,\cQ_1^\dag\}=-{1\over 8}\{\cQ_2,\cQ_2^\dag\} \,,
	\\
	\hat L_- &\equiv P+ {i\over 2} R_2{}^1 \,,
	\\
	\hat L_+ &\equiv K+2 iR_1{}^2 \,,
}
satisfying $[\hat L_0,\hat L_\pm]=\pm \hat L_{\pm}$ and $[\hat L_-,\hat L_+]=2\hat L_0$,
along with the central $\mf{u}(1)_\cZ$ generated by 
\ie
\cZ\equiv iM_\perp-R^{\dot 1}{}_{\dot 1}={i\over 8}\{\cQ_1,\cQ_2\}
\fe
are exact with respect to both $\cQ_1$ and $\cQ_2$, and therefore preserve the $\cQ_{1, 2}$-cohomologies.  Here, $D$, $P$, and $K$ correspond to the dilatation, translation, and special conformal generator of $\mathfrak{su}(2)_\text{conf}$, $R_a{}^b$ are the generators of $\mathfrak{su}(2)_H$, $R^{\dot a}{}_{\dot b}$ are the generators of $\mathfrak{su}(2)_C$, and $M_\perp$ is the rotation generator that fixes the line where the 1d theory lives.\footnote{The $SU(2)_H$ generators are related to $SO(6)_R$ generators by
	\ie
	&R_{79}= -i R_1{}^1\,,\qquad  R_{89}={1\over 2}(R_2{}^1-R_1{}^2)\,,\qquad R_{78}=-{i\over 2}(R_1{}^2+R_2{}^1 )\,,
	\fe
	similarly for $SU(2)_C$ generators
	\ie
	&R_{50}=- i R^{\dot 1}{}_{\dot 1}\,,\qquad  R_{60}={1\over 2}(R^{\dot 1}{}_{\dot 2}-R^{\dot 2}{}_{\dot 1})\,,\qquad R_{56}=-{i\over 2}(R^{\dot 1}{}_{\dot 2}+R^{\dot 2}{}_{\dot 1} )\,.
	\fe 
}

More explicitly, if we denote the coordinates of $\R^3$ by $(x_1, x_2, x_3)$\footnote{These coordinates are identified with the boundary theory directions  in $AdS_4$ Poincar\'e coordinates \eqref{PoincareAdS4}.} and we fix the line on which the theory lives by $x_2 = x_3 = 0$, then in terms of vector fields,\footnote{Here we adopt the convention that the charge $Q$ associated with a vector field $v_Q$ acts on a operator $\Phi(x)$ as 
	$
	\D_{v_Q}\Phi(x)\equiv [Q,\Phi(x)]\equiv -\cL_{v_Q} \Phi(x)
	$
	so that 
	$
	[\D_v,\D_w] \Phi(x)=\D_{[v,w]} \Phi(x).
	$
	This is the reason for the overall minus signs in \eqref{DExplicit}.
}
\es{DExplicit}{
	v_D  &=- x_\m\partial_\m \,,  \\
	v_P &= -\partial_1  \,, \\
	v_K &=x_\m^2 \partial_1-2x_1 x_\m\partial_\m  \,,  \\
	v_{M_\perp} &= -x_2 \partial_3 +x_3 \partial_2 \,.
}

In \cite{Dedushenko:2016jxl}, the line on which the 1d topological sector lives was mapped to a great circle on $S^3$. One can choose the convenient coordinates,
\ie
ds^2_{S^3} = d\theta^2 +\cos^2\theta d\tau^2+\sin^2\theta d\varphi^2\,,
\fe
with $\theta\in [0,{\pi\over 2}]$ and $\tau,\phi\in [0,2\pi)$ such that the 1d theory lives on the circle at $\theta={\pi\over 2}$ and $\tau=0$. This coordinate system makes obvious the fibration of a $S^1_\tau$ over a disk $D^2$, with the circle shrinking to zero size at the boundary of the disk, $\theta={\pi\over 2}$.  We can then rewrite the twisted $\mathfrak{su}(2)$ generators \eqref{hatL} as
\es{hatLAgain}{
	\hat L_i = L_i + R_i \,, 
}
where  $L_1 = P-{1\over 4} K  $, $L_2 =   D $, $L_3=i(P+{1\over 4} K) $ and  $R_1={i\over 2}(R_2{}^1-R_1{}^2)$, $R_2 = - R_1{}^1$, $R_3 =-{1\over 2}(R_1{}^2+R_2{}^1) $  such that $[\hat L_i, \hat L_j]= i \epsilon_{ijk}\hat L_k$.  In these coordinates, the conformal Killing vector fields on $S^3$ corresponding to $L_i$ are
\es{VectorFields}{
	v_{L_1}
	&=  - \cos \theta\sin \varphi  \partial_\theta-{\cos\varphi \over \sin \theta}\partial_\varphi    \,,
	\\
	v_{L_2}
	&=  \cos \theta\cos \varphi  \partial_\theta-{\sin\varphi \over \sin \theta} \partial_\varphi    \,,
	\\
	v_{L_3} &=- i \partial_\varphi  \,, \\
	v_{M_\perp} &= \partial_\tau \,.
}

Ref.~\cite{Dedushenko:2016jxl} obtained the partition function for the 1d theory by localizing the 3d theory with respect to the linear combination of the nilpotent supercharges 
\ie
\cQ_\B=\cQ_1+\B\cQ_2\,,\qquad \B\neq 0 \,.
\fe
Without loss of generality we will take $\B=1$ from now on and define $\cQ\equiv \cQ_1+\cQ_2$  which satisfies
\es{Qsq}{
	\cQ^2={8i\over L}(P_\tau+R_C)\,,
}
where we defined $R_C\equiv R^{\dot 1}{}_{\dot 1}$.  Note that up to a discrete set of equivalent choices, the properties described above uniquely identify $\cQ$ within $\mathfrak{osp}(4|4)$, provided that we fix the embedding of $\mathfrak{su}(2|2) \times \mathfrak{u}(1)_{\cal Z}$ within it.

To proceed, our immediate task is to identify the Killing spinor in $AdS_4$ that corresponds to $\cQ$.  To begin, we can write the Euclidean $AdS_4$ metric with $S^3$ asymptotic boundary as 
\ie
ds^2=d r^2+  L^2\sinh^2{r\over L} ds^2_{S^3} \,.
\fe 
So far, we have interpreted $\mathfrak{osp}(4|4)$ as the superconformal algebra of the 3d boundary theory, and consequently the vector fields \eqref{VectorFields} were conformal vector fields on $S^3$.  If we now interpret $\mathfrak{osp}(4|4)$ as the supersymmetry algebra of the $AdS_4$ theory, we should extend the vector fields \eqref{VectorFields} into $AdS_4$ in such a way that they correspond to isometries of $AdS_4$.  
Such an extension is unique, and it is given by
\es{VectFieldsExtended}{
	v^{AdS_4}_{L_1}&=
	-\sin \theta \sin \varphi  \partial_r-{1\over L}\cos \theta\sin \varphi \coth {r\over L} \partial_\theta-{1\over L}{\cos\varphi \over \sin \theta}\coth {r\over L} \partial_\varphi  \,,
	\\
	v^{AdS_4}_{L_2} &=
	\sin \theta \cos \varphi  \partial_r+{1\over L}\cos \theta\cos \varphi \coth {r\over L} \partial_\theta-{1\over L}{\sin\varphi \over \sin \theta} \coth {r\over L}\partial_\varphi  \,,
	\\
	v^{AdS_4}_{L_3} &=-{i\over L}\partial_\varphi \,, \\
	v^{AdS_4}_{M_\perp} &= \frac 1L \partial_\tau \,.
}

The $AdS_2$ slice, which we would like to localize onto, sits at the fixed locus of $M_\perp$ at $\theta= \frac{\pi}{2}$. The vector fields $	v^{AdS_4}_{L_i}$, when restricted to $\theta = \pi/2$, generate the isometries of this $AdS_2$.  

Once we have identified $L_i$ with Killing vectors on $AdS_4$, we should identify ${\cal Q}$ with a Killing spinor in a way consistent with the equation \eqref{Qsq}. Comparing  \eqref{SUSYac} with \eqref{Qsq}, we are looking for $\varepsilon_{1,2}=\varepsilon$ with 
\es{Parameters}{
v={8\over L}\partial_\tau\,,\qquad \omega_{AB}=-{4 \over L} (\D_{A 5}\D_{B 0}-\D_{A 0}\D_{B 5}) \,,\qquad \lambda=0\,.
}
The desired solution for $\varepsilon$  is\footnote{
	Since Euclidean $AdS_4$ is conformally equivalent to half of an $S^4$, from the Weyl invariance of the conformal Killing spinor equations \eqref{CKSeqn}, we can map conformal Killing spinors on $AdS_4$ to those on $S^4$. It is then not surprising that the conformal Killing spinor $\varepsilon$ we use to localize onto $AdS_2$ in $\cN=4$ SYM on $AdS_4$ is directly related to what Pestun used in \cite{Pestun:2009nn} to localize onto $S^2$ from the same theory on $S^4$, by a simple conformal map 
	\ie
	e^{2\phi}={L^2(1 + {x^2 \over 4L^2})^2\over   x_4^2}\,,
	\label{S4toAdS4}
	\fe
	where we take $S^4$ to be in its stereographic coordinates,
	\ie
	ds^2_{S^4}={ dx^2\over (1+{x^2\over 4L^2})^2}\,,
	\fe
	and the fixed $S^2$ is located at $x_2=x_3=0$. (Note that the stereographic coordinates $x_\m$ we use here are different from the ones in \cite{Pestun:2009nn} where the $S^2$ is located at $x_1=0\,,\,x_2^2+x_3^2+x_4^2=4L^2$.)
	
	There the corresponding supersymmetry transformation generator also squares to a linear combination of the Lie derivative along the Killing vector $\partial_\tau$ and $SO(6)_R $ rotation by $R_{05}$. It is important that $\partial_\tau \phi=0$ so that under the conformal mapping \eqref{S4toAdS4}, $\partial_\tau$ remains a Killing vector. Many of the formulas we write here are directly related to those in \cite{Pestun:2009nn} via \eqref{S4toAdS4}.
}
in the form of \eqref{gencks} with 
\setcounter{MaxMatrixCols}{20}
\ie
\label{localizeQ}
\begin{alignedrows}{16}
	\arow{\epsilon_s=}{{-1}  & 0 & 0 &  0 &  0 &  0 &  i&  0 & 0 &  {-1} &  0 &  0 & 0 & 0 &0&  -i} \,, \\
	\arow{\epsilon_c =}{0 & 0 & 0 & 1 &  0 &  -i &  0&  0 & 0 &  0 &  1 &  0 & -i & 0 &  0& 0}.
\end{alignedrows}
\fe
The twisted $\mf{su}(2)$ generators $\hat L_a$ in \eqref{hatLAgain} are identified with the 10d Killing spinors $\epsilon_a$ such that acting on bulk fields,
\ie
\hat L_a= {1\over 2}\{\D_{\varepsilon},\D_{\varepsilon_a}\}\,,\qquad a=1,2,3\,.
\fe
The explicit expressions for $\ve_a$ are given in \eqref{twistedcks}.

\subsection{The BPS equations and 2d Yang-Mills} \label{Sec:to2d}
In this section, we will study the BPS locus of the supercharge $\cQ$ in the $\cN=4$ SYM,
\ie
\Psi=0\,,\qquad \D_\ve \Psi=0\,,
\label{BPSeqs}
\fe
and show that the 4d $\cN=4$ SYM action \eqref{SYMos} on $AdS_4$ reduces to $\rm YM_2$ on $AdS_2$ when restricted to BPS configurations. In particular, we will derive the boundary terms \eqref{YMBdry} necessary for the alternate quantization of 2d gauge fields. We emphasize here that our localization procedure is in the same spirit as the recent work \cite{Bonetti:2016nma}, where we are not specific about the reality conditions of the 4d fields. Consequently we do not specify the integration contour for the emergent 2d gauge fields in the $\rm YM_2$ path integral. In principle one can add an  appropriate localizing term $t \cQ V(A,\Phi,\Psi)$ to the 4d action and determine the integration contour that guarantees the convergence of the path integral as $t\to \infty$.\footnote{The choice of integration contour is subtle here due to the non-compactness of the spacetime manifold $\cM$. See \cite{Assel:2016pgi} for discussions in 3d.} This will be important for example to compute the full partition function. Here we restrict ourselves to the classical theory and leave the incorporation of quantum corrections to future investigation.

Our strategy for finding the BPS loci \eqref{BPSeqs} will be to impose the vanishing of the square of the off-shell supersymmetry variations of all fields. Let us first recall that in $\cN=4$ SYM they square to a combination of bosonic symmetries and a field dependent gauge transformation  as given in \eqref{SUSYac}. In \eqref{Parameters} we have already determined the vector field $v$ parametrizing translations (by Lie derivative) and $\om$ parametrizing the $R$-symmetry rotation. The gauge parameter and the parameter of the $SO(7)$ rotation of the auxiliary fields is given by 
\ie
&v^A \Phi_A= {8i\over L} \sinh{r\over L} \cos \theta (\sin \tau \Phi_0 -\cos \tau \Phi_5)\,,
\\
& M^{mn}\equiv \n^{[m}\CC^\m\nabla_\m \n^{n]}\,,
\fe
so that the square of the supersymmetry variations of all the fields is given by
 \cite{Pestun:2009nn}
\ie
&\D_\ve^2 A_\m=-v^\n F_{\n\m}+[D_\m,v^A\Phi_A] \,,
\\
&\D_\ve^2 \Phi_A=-v^\n D_\n \Phi_A+[\Phi_A,v^B\Phi_B]- 2\om_{AB}\Phi^B \,,
\\
&\D_\ve^2 \Psi= -v^N D_N \Psi-{1\over 4} \nabla_\m \n_\n \CC^{\m\n} \Psi-\frac12 \om_{AB} \CC^{AB}\Psi \,,
\\
&\D_\ve^2 K^m=-v^M D_M K^m-M^{mn}K_n\,. 
\label{dsq}
\fe

Setting all equations in \eqref{dsq} to zero, we obtain the following constraints on the bosonic fields:
\ie
& {8\over L} F_{\tau \m}=  [D_\m,v^A \Phi_A] \,,
\\
& {8\over L}[D_\tau,\Phi_A]=[\Phi_A, v^B \Phi_B]-2\omega_{AB}\Phi^B  \,,
\\
& {8\over L}[D_\tau,K^m]= [K^m, v^B \Phi_B]-M^{mn}K_n\,,
\label{covconst}
\fe
These constraints imply that all fields are covariantly constant along $\tau$ up to field dependent gauge transformations and bosonic symmetry transformations. To make the $\tau$ invariance more manifest, we can define the twisted combinations
\ie
\hat\Phi_5\equiv\cos\tau \Phi_0+\sin\tau \Phi_5\,,\qquad \hat\Phi_0\equiv \sin \tau \Phi_0 -\cos\tau \Phi_5\,, 
\label{tphi05}
\fe 
and a modified gauge connection, giving a new covariant derivative $\cD_\m$:
\ie
\cD_\tau\equiv D_\tau+i \sinh {r\over L} \cos\theta \hat \Phi_0\,,\qquad \cD_{r,\theta,\varphi}\equiv D_{r,\theta,\varphi}\,.
\fe 
This allows us to rewrite \eqref{covconst} as
\ie
\cF_{\tau\m}\equiv [\cD_\tau,\cD_\m]=0\,,\qquad [\cD_\tau,\Phi_{6,7,8,9}]=[\cD_\tau,\hat\Phi_{0,5}]=0\,,\qquad [\cD_\tau, K^m]= - {L\over 8}M^{mn}K_n\,.
\fe

We would like to compute the bosonic SYM action $S_{tot}\equiv   S+S_B$ on the BPS locus 
\ie
S_{tot}
=& {1\over 2 g_4^2}\int_{AdS_4} d^4 x\, \sqrt{g} \, \tr \Bigg(
{1\over 2}F_{MN}F^{MN} -{2\over L^2} \Phi^A \Phi_A -K^m K_m
\Bigg)
\\
&
+{1\over 2 g_4^2}\int_{r=r_0} d^3 x\,\sqrt{\ga}\, \tr \Bigg(
2K^{a-5}\Phi_a-2D_r\Phi^a \Phi_a-4\Phi_9 F_{78}-2\Phi^a\Phi_a+\Phi^A\Phi_A
\Bigg)\,,
\label{SYMtbos}
\fe
where $a=7,8,9$ are $SU(2)_H$ adjoint indices.

\begin{figure}
	\centering
	\includegraphics[scale=1]{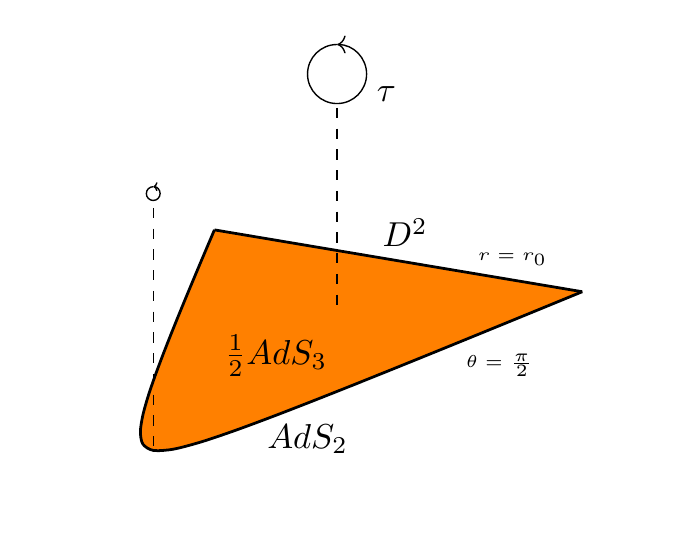}
	\caption{$AdS_4$ as a $S^1_\tau$ fibration over ${1\over 2}AdS_3$.}
	\label{Fig:halfh3}
\end{figure}

As a consequence of the above argument demonstrating $\tau$-invariance, we may simply evaluate the field configurations at $\tau=0$. To see geometrically what is going on, it is convenient to write the $AdS_4$ metric as
\ie
ds^2=ds^2_{{1\over 2}AdS_3}+L^2 \sinh^2 {r\over L}\cos^2\theta  d\tau^2 \,,
\fe
with
\ie
ds^2_{{1\over 2}AdS_3}=dr^2+L^2\sinh^2 {r\over L}  (d\theta^2 +\sin^2\theta d\varphi^2)\,,
\fe
which makes manifest the $S^1_\tau$ fibration over the half hyperbolic space which we denote by ${1\over 2}AdS_3$ (see Figure~\ref{Fig:halfh3}).

The $\tau$-invariance of the fields implies that it suffices to evaluate the integrands of \eqref{SYMtbos} on the ${1\over 2}AdS_3$ with two boundary components at $\theta={\pi\over 2}$ and $r=r_0$ respectively. 
Furthermore, $\hat\Phi_0$, if nonvanishing, has a first order pole approaching $\theta={\pi\over 2}$, which corresponds to a surface defect wrapping $AdS_2$ \cite{Pestun:2009nn}.\footnote{Strictly speaking the analysis in \cite{Pestun:2009nn} is about potential surface defect wrapping $S^2$ in $S^4$. We assume the same is also true here with an appropriate choice of the integration contour.} We will set $\hat{\Phi}_0=0$ from now on and leave the analysis of defects in $AdS_4$ to future work.

As we show in the Appendix~\ref{App:bps}, on the BPS locus, the first term in \eqref{SYMtbos} becomes a total derivative on ${1\over 2}AdS_3$, so it gives an integral over the boundary $AdS_2$ at  $\theta={\pi\over 2}$ and an integral over the boundary $D^2$ at $r=r_0$. The latter combined with the second term in \eqref{SYMtbos} is again a total derivative on $D^2$, which then reduces to an integral over its boundary $S^1$. This $S^1$ is also the asymptotic boundary of the $AdS_2$, as shown in Figure~\ref{Fig:halfh3}. After some algebra (see Appendix~\ref{App:bps}), we end up with the action 
\ie
S_{tot} 
=& -{ \pi \over   g_4^2}\int_{AdS_2} dr d\varphi\,\sqrt{g_{AdS_2} } \,
\tr \Bigg( 
2\tilde \Phi {*\hat F} +\tilde\Phi^2-\hat \Phi_5^2 
\Bigg)
+{ \pi \over   g_4^2}\int_{r=r_0} d\varphi\,{ \hat A_\varphi(\hat A_\varphi-2\hat F_{r\varphi})\over \sinh r}\,,
\label{PhiF}
\fe
where for simplicity we have set $L=1$. We have also defined twisted scalar fields
\ie
&\tilde\Phi \equiv i\sinh r (\Phi_8  \sin  \varphi +  \Phi_7 \cos \varphi -i \coth r \Phi_9 ) +i \Phi_6 \,,
\\
&\hat \Phi_r \equiv i(\cos \varphi \Phi_8-\sin \varphi \Phi_7) \,,
\\
&\hat \Phi_\varphi \equiv -i\sinh r \cosh  r(  \sin \varphi \Phi_8 +  \cos \varphi \Phi_7-i\tanh r \Phi_9)\,,
\label{tphi6789}
\fe
and twisted connection\footnote{It is easy to check that the emergent 2d gauge field $\hat A$ is annihilated by the SUSY transformation \eqref{SUSYos} generated by $\cQ$.}
\ie
&\hat A_\varphi \equiv A_\varphi+\hat \Phi_\varphi \,,
\\
&\hat A_r \equiv A_r+\hat \Phi_r\,,
\fe
with curvature $\hat F$
\ie
\hat F_{r\varphi}
=&F_{r\varphi}+[\hat\Phi_r,\hat\Phi_\varphi]+D_r \hat \Phi_\varphi-D_\varphi \hat \Phi_r\,,
\fe
and used that the 2d Hodge dual is $*\hat F={\hat F_{r\varphi}\over \sinh r}$. The emergent 2d twisted connection $\hat A$ will be identified with the 2d YM field in the rest of the paper.\footnote{Note that we have not specified the integration contour for the emergent 2d gauge fields $\hat A$. Nonetheless this does not affect the classical analysis in the previous sections.} As we observe, the 2d YM gauge potential has its origin as a linear combination of the gauge fields and scalar fields of the 4d parent theory.

We observe that $\hat\Phi_5$  decouples in \eqref{PhiF}, and we can integrate out $\tilde\Phi$  to obtain the bosonic Yang-Mills action on $AdS_2$,
\ie
S= {1\over  g^2_{\rm YM} }\int_{AdS_2} dr d\vphi\, \sqrt{g_{AdS_2} }    \, \tr (*\hat F)^2+
{1\over   g_{\rm YM}^2}\int_{r=r_0}  d\varphi \,  { \hat A_\varphi(\hat A_\varphi-2\hat F_{r\varphi})\over \sinh r}\,,
\fe
where the 2d Yang-Mills coupling is related to the 4d coupling by
\es{gYMFinal}{
	g_{\rm YM}^2={ g_4^2 \over \pi L^2 }\,.
}

\subsection{Embedding in the ABJM holographic duality}

We focus on the Higgs branch topological theory of $\cN=4$ SCFTs. The $\cQ$ cohomology is generated by the $SU(2)_H$ highest weight states of the Higgs branch chiral primaries $\cO_{(a_1\dots a_{2j})}$ which we take to be those with $a_1=\dots =a_{2j}=1$. In the holographic dual, they correspond to $\cN=4$ chiral superfields which contain scalar fields that furnish the same $SU(2)_H$ spin $j$ representation. In particular, the dual of the $\Delta=1$ chiral primaries $\cO_{(11)}$ are $\cN=4$ vector multiplets that constitute a subsector of the full supergravity theory that we localized in the previous subsections. As we have explained in Section~\ref{sec:varbc},  with our bulk action, the scalar fields $\Phi_a$ are dual to the operators $\cO_{(a_1 a_2)}$ with $\Delta=1$, so they obey  alternate boundary conditions in the $AdS_4$ bulk. The other three scalars fields $\Phi_i$ in the vector multiplet are dual to $\Delta=2$ superconformal descendant scalar operators in the same superconformal multiplet of the boundary SCFT, and thus they obey standard boundary conditions in $AdS_4$.

In the ABJM case, the $SO(8)$ R-symmetry of the $\cN=8$ SCFT decomposes into
\ie
\mf{so}(8)\supset \mf{su}(2)_H\times \mf{su}(2)_C\times \mf{su}(2)_1\times \mf{su}(2)_2
\label{so8dec}
\fe
in such a way that the eight supercharges of the ${\cal N} = 8$ theory decompose as ${\bf 8}_s = ({\bf 2}, {\bf 2}, {\bf 1}, {\bf 1}) \oplus ({\bf 1}, {\bf 1}, {\bf 2}, {\bf 2})$.   In $\cN=4$ language, $\mf{su}(2)_1\times \mf{su}(2)_2$ is viewed as a flavor symmetry in the SCFT but, as we will see, only $\mf{su}(2)_1$ becomes a global symmetry of the Higgs branch topological theory.

Under the decomposition \eqref{so8dec}, the $\Delta=1$ operators, which transform as ${\bf 35}_v$ under the $\mf{so}(8)$ R-symmetry, decompose as
\ie
{\bf 35}_v=\bf (3,1,3,1)\oplus(1,3,1,3)\oplus(2,2,2,2)\oplus(1,1,1,1) \,.
\label{35vdec}
\fe
Similarly, the $\Delta=2$ operators transform as ${\bf 35}_c$ of $\mathfrak{so}(8)$ and become
\ie
{\bf 35}_c=\bf (1,3,3,1)\oplus(3,1,1,3)\oplus(2,2,2,2)\oplus(1,1,1,1)\,.
\label{35cdec}
\fe
On the supergravity side, the $\cN=8$ supergravity multiplet contains 70 scalars which are dual to the above operators. Under the $\cN=4$ subalgebra, the $\cN=8$ supergravity multiplet decomposes into 1 supergravity multiplet, 4 gravitino multiplets and $6$ vector multiplets. The vector multiplets are associated with the $6$ generators of the $\mf{su}(2)_1 \times  \mf{su}(2)_2$ gauge algebra. 
Now the scalar fields $\Phi_a$ in the $\mf{su}(2)_2$ vector multiplets are dual to operators of quantum numbers $\bf(3,1,1,3)$ in \eqref{35cdec} with $\Delta=2$, incompatible with the boundary condition \eqref{SYMbc}. Hence our localization computation only applies to the  $\mf{su}(2)_1$ vector multiplets.  Regarding the $\mf{su}(2)_2$ vector multiplets, we will not study them here, but we expect that with appropriate boundary terms one can perform supersymmetric localization for them too and obtain that their classical action vanishes on the localizing locus.

Now that we have clarified how our localization computation embeds in the full ABJM holographic duality, we can use the known dictionary to determine the YM coupling $g_{\rm YM}$ of the $\mathfrak{su}(2)_1$ gauge theory in 2d in terms of the parameters of the boundary theory.  In the dual $AdS_4$ gauged supergravity the $\mathfrak{so}(8)$ (and hence the  $\mathfrak{su}(2)_1$) gauge coupling is related to the AdS radius and Newton's constant by ${1\ov g_4^2}={L^2\ov 4\pi G_4}$ \cite{deWit:1982bul}, and from the duality we also know that ${L^2\ov G_4}={2\sqrt{2}\ov 3}\, N^{3/2}$ \cite{Aharony:2008ug}. Combining these two, we get
\es{gYMFinal2}{
	\frac{1}{g_{\rm YM}^2L^2}=\frac{N^{3/2} }{3\sqrt{2}}\,, 
}
which is exactly what we got in the bottom-up construction of the bulk dual of ABJM theory in \eqref{BulkYM}. Here we gave a first-principle derivation of this relation.

\section{Conclusions and open questions} 
\label{CONCLUSIONS}

To summarize, we have proposed new instances of the holographic duality in 2d bulk / 1d boundary dimensions.  In our examples, the bulk theories are 2d gauge theories on $AdS_2$ with potentially infinite gauge algebras, and they are dual to topological theories on the boundary that have vector or matrix large $N$ limits.  Conceptually, topological theories in 1d can be thought of as theories of conserved currents, so they have a large global symmetry generated by these currents.  At large $N$, this global symmetry algebra has a consistent subalgebra $\mathfrak{g}$ generated by the ``single trace'' conserved currents.  In AdS/CFT a global symmetry on the boundary is gauged in the bulk, hence our 2d bulk theories are gauge theories.  Our proposal is that the gauge symmetry in the bulk is given by $\mathfrak{g}$.

In Sections~\ref{ABELIAN} and \ref{NONABELIAN}, we discuss in detail such 2d gauge theories in $AdS_2$, with the aim of explaining how to use the AdS/CFT dictionary to read off the CFT correlation functions.  An important subtlety is which boundary terms should supplement the bulk action in order for the boundary operators dual to our bulk gauge fields to be conserved currents.  We argue that in order to have conserved currents on the boundary, the bulk gauge fields must obey an analog of the ``alternate quantization'' boundary condition of \cite{Klebanov:1999tb}, and therefore our choice of boundary terms is guided by this principle. 

In Sections~\ref{TOPOLOGICAL}--\ref{MATRIX}, we discuss the 1d topological theories, at first generally in Section~\ref{TOPOLOGICAL}, and then in specific examples in Sections~\ref{VECTOR} and~\ref{MATRIX}.  In all cases, these 1d theories are exactly solvable, and they can be embedded as subsectors of 3d ${\cal N} = 4$ SCFTs with holographic duals.  As reviewed in Section~\ref{TOPOLOGICAL}, their partition functions can be obtained using supersymmetric localization of the parent 3d theories, when these theories are placed on a round $S^3$.   In the case of 1d theories with vector large $N$ limits discussed in Section~\ref{VECTOR}, the dual 2d gauge theories have finite-dimensional gauge algebras, and in some cases we determine their full non-linear actions.  In Section~\ref{MATRIX}, we study a 1d theory with matrix large $N$ limit.  This theory comes from the 3d ABJM theory, and like ABJM theory, has a number of degrees of freedom that scales as $N^{3/2}$ at large $N$.  In this case we use matrix model techniques to determine the 2-point, 3-point, and a couple of 4-point correlation functions of $\Z_2$-even single trace operators, from which we infer the first few terms in the derivative expansion of the 2d bulk action, and obtain that the 2d gauge algebra is the algebra of area-preserving diffeomorphisms of a two-sphere, $\text{SDiff}(S^2)$.  We expect that the inclusion of $\Z_2$-odd operators extends the $\text{SDiff}(S^2)$ algebra to a super Lie algebra whose precise form is left for future explorations.  While $\text{SDiff}(S^2)$ can be thought of as the $\lambda \to \pm i \infty$ limit of the 3d higher spin algebra $\mathfrak{hs}[\lambda]$, we expect that the super Lie algebra obtained after the inclusion of $\Z_2$-odd operators is the $\lambda \to \pm i \infty$ limit of the supersymmetric higher spin algebra $\mathfrak{shs}[\lambda]$.

Lastly, in Section~\ref{LOCALIZATION}, we argue in a simple example that the 2d gauge theories, whose dual 1d theories are embedded as sectors of 3d ${\cal N} = 4$ SCFTs with $AdS_4$ duals, could be obtained from the corresponding $AdS_4$ theories via supersymmetric localization.  While we do not perform the supersymmetric localization of the full $AdS_4$ theories, we show that an ${\cal N} = 4$ Yang-Mills theory on $AdS_4$ with gauge algebra $\mathfrak{g}_\text{vec}$ localizes to a 2d Yang-Mills theory on $AdS_2$ with the same gauge algebra $\mathfrak{g}_\text{vec}$.  The 4d SYM theory is a consistent subsector of the $AdS_4$ theory and, correspondingly, the 2d YM theory we obtain is a subsector of the full 2d YM theory with gauge algebra $\mathfrak{g}$.  Quite nicely, the supersymmetric localization computation reproduces the boundary terms that supplement the $AdS_2$ Yang-Mills action as expected from our general analysis in Sections~\ref{ABELIAN} and~\ref{NONABELIAN}.

We conclude with some topics worthy of further exploration:

{\bf Beyond leading order in $1/N$.}  Our computations involved classical gauge theory in $AdS_2$ that matched computations at leading order in $N$ in the boundary.  It would be interesting to understand how to define the bulk theory at the quantum level, and hence how to obtain a duality that works beyond the leading order in the $1/N$ expansion.  Perhaps the precise integration cycle both in the boundary and in the bulk path integrals, which so far we mostly ignored, may now become important.  While perturbative $1/N$ corrections can plausibly be encoded in an effective action in $AdS_2$ whose couplings receive $1/N$ corrections, the complete finite $N$ theory is probably much more complicated.  A simple test of the correct finite $N$ theory is that in the case of the bulk theories dual to the $U(N)$ singlet sector of our large $N$ vector models, the correlation functions of single trace operators should not receive any perturbative or non-perturbative corrections in $1/N$.  Hopefully this fact will be reproduced from the appropriate bulk theory.

Along similar lines, it would be interesting to understand whether there are any connections between the finite $N$ theory and the work of Cattaneo and Felder \cite{Cattaneo:1999fm}, which describes 1d topological theories of the kind studied here by a bulk 2d Poisson sigma model.  The theory of \cite{Cattaneo:1999fm} is, however, purely topological.  It is plausible that the large $N$ limit of this topological theory agrees with the zero gauge coupling limit of our bulk $AdS_2$ gauge theory, which also becomes topological at vanishing gauge coupling.

{\bf Supersymmetric localization.}  We showed in Section~\ref{LOCALIZATION} how to obtain an $AdS_2$ Yang-Mills theory from localizing an ${\cal N} = 4$ non-abelian vector multiplet in $AdS_4$.  It would be interesting to generalize this computation further in two directions.  The first direction would be to construct an appropriate localizing term and compute the one-loop determinant associated with it.  The second direction would be to consider other multiplets on $AdS_4$ obtained by reducing the $AdS_4 \times S^7$ supergravity on $S^7$.  It would be very interesting if a localization computation of these other multiplets could be performed, and if, as we expect, it would result in a 2d gauge theory with an infinite-dimensional gauge symmetry that has $\text{SDiff}(S^2)$ as a bosonic subalgebra.  From this computation one could perhaps also infer which higher-derivative couplings in 11d supergravity are encoded in our $AdS_2$ gauge theory.  

Along similar lines, one can attempt to perform supersymmetric localization of 11d supergravity on $AdS_4 \times S^7$.  Such a computation could provide a geometric interpretation of the $\text{SDiff}(S^2)$ global symmetry of the $\Z_2$-even sector of the corresponding 1d theory.  In particular, it would be interesting to understand whether the $AdS_2$ gauge theory we obtained is secretly a theory on $AdS_2 \times S^2$.

{\bf Bosons vs.~fermions.}  While the boundary theories we have studied involve bosonic fields which are anti-periodic on $S^1$ and come with kinetic terms of the form $\int \tQ d Q$, one may wonder what would change had the fields $Q$ and $\tQ$ been replaced with Grassmann-valued fields $\widetilde{\psi}$ and $\psi$ respectively, that are still anti-periodic on $S^1$.    In the free case, the two-point function of two bosonic fields $Q(\vphi_1)$ and $\tQ(\vphi_2)$ is proportional to $\sgn(\vphi_1 - \vphi_2)$, and so would the two-point function of two fermionic fields $\psi(\vphi_1)$ and $\widetilde{\psi}(\vphi_2)$. However, simply replacing bosons with fermions in any of the theories in Sections~\ref{VECTOR} and~\ref{MATRIX} certainly results in a theory that is different from the original bosonic one.  Indeed, bosonic theories have operators constructed as products of large numbers of bosons, while the fermionic theories obtained from them do not.  (Equivalently, the fermionic theory has a finite-dimensional Hilbert space when continued to Lorentzian signature,  but the bosonic theory does not, and it is probably not true that it can be continued to a unitary Lorentzian theory.)  However, it is possible that in the large $N$ limit, the replacement of bosons with fermions in a given boundary theory does not change the type of 2d bulk dual it has, and that the 2d/1d correspondence analyzed in this paper holds for theories with fermions too.  

{\bf Deforming to a non-topological theory.} Turning on sources in the boundary theory makes correlation functions non-topological. We demonstrated this effect in Section~\ref{CORRELATORS2} in a simple case where in the boundary we turned on source for a single trace operator interpreted as a mass term. Besides sources for single-trace operators, one may turn on sources for multi-trace operators. It would be very interesting to understand how complicated the dynamics of these deformed theories can become. 

{\bf Comparison to CS/WZW duality.}   As can be seen from \cite{Bonetti:2016nma}, the 3d/2d analog of the duality we proposed here is a particular case of the duality between Chern-Simons (CS) theory and the chiral half of a Wess-Zumino-Witten (WZW) model.  Indeed, Ref.~\cite{Bonetti:2016nma} started with 4d ${\cal N} = 4$ supersymmetric Yang-Mills theory with gauge group $SU(N)$, which at large $N$ is dual to string theory on $AdS_5 \times S^5$.  They argued that upon performing supersymmetric localization with the supercharge used in \cite{Beem:2013sza}, the bulk theory reduces to a CS theory on $AdS_3$ with a higher-spin algebra as its gauge algebra, while the single-trace sector of the boundary theory reduces to the WZW model dual to this CS theory.  It would be interesting to understand whether the higher spin algebra obtained in \cite{Bonetti:2016nma} also has a geometric interpretation in terms of the diffeomorphism algebra of some (super)manifold.  It is worth stressing that while the CS/WZW duality appeared in \cite{Bonetti:2016nma} in a similar way to our 2d/1d duality, the CS/WZW duality is more general and does not rely on large $N$.

In the case of the duality between Chern-Simons theory on $D^2\times \R$ and the chiral half of the Wess-Zumino-Witten model on $S^1\times \R$, it can be shown in \cite{Elitzur:1989nr} that the boundary values of the gauge field $A_\vphi$ become boundary current operators that satisfy a Ka\v{c}-Moody algebra.  As discussed in  Appendix~\ref{WITTENDIAG}, in our case the leading piece of $A_\vphi$ can also be identified with the boundary current operators following the philosophy of the holographic extrapolate dictionary  \cite{Banks:1998dd,Harlow:2011ke}.   In the CS/WZW case, the theory of the boundary theory currents can be written in terms of gauge-invariant observables as a WZW theory.  It would be interesting to see whether our boundary TQM theories, which we have written as gauge theories, can also be formulated in terms of gauge-invariant degrees of freedom.

{\bf Bulk reconstruction.} The complete set of gauge invariant bulk operators are Wilson loops, and Wilson lines ending on the boundary. The latter are gauge invariant, because we treat large gauge transformations as global symmetries. One can use the bulk-to-boundary propagator (discussed in Appendix~\ref{WITTENDIAG}) as a smearing function to express the bulk gauge field (in some gauge) in terms of an integral of the currents in the boundary theory. Once we have the bulk gauge field, we can build a Wilson loop out of it perturbatively. This is a rather convoluted procedure, and so it is natural to ask whether there is a simpler construction.  In addition, in the ABJM example it would be interesting to understand the bulk dual of the boundary Wilson line operator built from the gauge field ${\cal A}$ in~\eqref{Top}.

{\bf Lorentzian duality.}  While so far we have worked only in Euclidean signature, it would be interesting to understand the Lorentzian version of the 2d/1d dualities presented here.  On the boundary side, the possibility of continuing the Euclidean path integral to Lorentzian signature to provide a definition for a unitary Lorentzian theory requires a better understanding of the integration cycle in the Euclidean path integral---such a continuation may be possible for some integration cycles and impossible for others.  Correspondingly, on the bulk side, the continuation of the YM theory on hyperbolic space to a theory on $AdS_2$ also requires a better understanding of the integration cycle.  It is conceivable that the Euclidean theories cannot be continued to unitary Lorentzian theory in the case of interest in this paper where the boundary 1d theory contains scalar fields anti-periodic on $S^1$, but it is possible that such a continuation would be possible in the case of boundary fermions.  While we do not study this issue in detail here, let us make some comments about a possible Lorentzian continuation.

In Lorentzian signature, we would have to face the fact that $AdS_2$ has two disconnected boundaries. In global coordinates, the metric takes the form
\es{GlobalAdS2}{
ds^2= \frac{-dt^2+d\sig^2}{\sin^2 \sig}\,,
}  
where the two boundaries are located at $\sig=0, \pi$.\footnote{Starting from the metric \eqref{Back}, we perform the coordinate transformation \cite{Sen:2011cn}
\es{CoordTf}{
\sig+i \tau&\equiv 2\arctan\le(\tanh\le({\log\coth\le(r\ov 2\ri)+i\vphi\ov 2}\ri)\ri) \,,\\
ds^2&={d\tau^2+d\sig^2\ov \sin^2 \sig}\,,
}
which maps the boundary circle at $r=\infty$ onto two lines at $\sig=0, \pi$. We can Wick rotate \eqref{CoordTf} to obtain \eqref{GlobalAdS2}.
}
In usual two-sided setups in AdS/CFT such as the thermofield double \cite{Maldacena:2001kr}, the two boundary CFTs are independent, and the Hilbert space is a tensor product. In this instance of the $AdS_2$/CFT$_1$ duality, the two boundaries cannot host two independent copies of the topological quantum mechanics theories. One reason is that the two boundaries are causally connected. Another reason is that the Hilbert space of the bulk theory is not a tensor product, and instead it is given by
\es{YMHilbert}{
{\cal H}=\bigoplus_{R} V_R \otimes V_{\bar R}\,,
}
where $R$ is an irreducible representation of the gauge group, $\bar R$ is its conjugate, and $V_R$ is the $\dim(R)$-dimensional vector space on which the representation $R$ acts. For a nice discussion, see \cite{Donnelly:2014gva}.

 We note that these results are analogous to the CS/WZW duality.  Indeed, if the Chern-Simons theory is defined on ${\cal A}\times \R$, where ${\cal A}$ is an annulus, the Hilbert space is not a tensor product either: it is
\es{CSHilbert}{
{\cal H}=\bigoplus_{R} {\cal H}_R \otimes {\cal H}_{\bar R}\,,
}
where $R$ is restricted to those representations that label the Hilbert space on $T^2\times \R$, and  ${\cal H}_R$ is the Hilbert space on a disk with a source in the representation $R$  \cite{Elitzur:1989nr}.

We hope to come back to some of these questions in the future.

\subsection*{Acknowledgments}

We thank S.~Giombi, J.~Gomis, D.~Harlow, I.~Klebanov, B.~Le Floch, J.~Maldacena, E.~Perlmutter, A.~Sen, S.-H.~Shao, D.~Stanford, H.~Verlinde, R.~Yacoby, and E.~Witten for very useful discussions. The research of MM was supported in part by the U.S. Department of Energy under grant No.~DE-SC0016244.  SSP was supported in part by the US NSF under Grant No.~PHY-1418069 and by the Simons Foundation Grant No.~488653.  The work of YW was supported in part by the US NSF under Grant No.~PHY-1620059 and by the Simons Foundation Grant No.~488653.

\appendix

\section{Perturbative computations in the bulk}\label{WITTENDIAG}

\subsection{Bulk gauge profile at leading order}

In this Appendix we will work in Poincar\'e coordinates
\es{Poincare}{
ds^2={dt^2+dz^2\ov z^2}\,.
}
One can translate between the global coordinates \eqref{Back} and \eqref{Poincare} using the coordinate transformation:
\es{CoordTrans}{
t&={\sinh r \sin\vphi\ov \cosh r +\sinh r \cos \vphi} \,, \\
z&={1\ov \cosh r +\sinh r \cos \vphi}\,.
}
In these coordinates the asymptotic behavior of the gauge field is given by
\es{AsympExpansionApp}{
  A_t &= {{\cal Q}(t)\ov z} + b(t) + O(z)  \,, \qquad A_z =c(t) \,,\\
 0&= \p_t {\cal Q}-i\big[b-i\le[c,{\cal Q}\ri],{\cal Q}\big] \,, 
 }
where the differential equation follows from the Yang-Mills equations, and the remaining gauge freedom is restricted to gauge transformations that decay at the boundary, and do not change these functions.\footnote{In the approach of the main text, one can obtain the most general (regular) solution of  the field equations by starting from a simple solution
 \es{SimpleSolP}{
  F_{zt} &= -{Q\ov z^2}\,, \qquad
  A_{t} = {Q \ov z} \,, \qquad A_z = 0 \,,
 }
 and then gauge transforming this solution to 
 \es{GeneralSolP}{
  F_{z t} &= -U {Q\ov z^2} U^{-1} \,, \qquad A_t = U {Q\ov z} U^{-1} + i U \partial_t U^{-1} \,, \qquad
  A_z =  i U \partial_z U^{-1} \,.
 }
 Expanding $U$ near the boundary as $U(t,z)=u(t)\le[1+i \lam(t)\, z\ri]+O(z^2)$, the functions $E,\, b,\, c$ are given by
\es{Ebc}{
E&=uQu^{-1} \,, \qquad
b=i u \p_t u^{-1}+ i u\le[\lam,Q\ri]u^{-1} \,, \qquad
c=u \lam u^{-1}\,.
} 
One can now check that the differential equation in \eqref{AsympExpansionApp} is satisfied for any choice of $Q,\, u,\, \lam$.
} 
It is very important to note that $b$ is not equal to $a$, the boundary source kept fixed by our boundary conditions
\es{abRelation}{
a=\lim_{z\to 0} \le(A_t+z F_{zt}\ri)=b-i\le[c,{\cal Q}\ri]\,.
}
This relation will have to be taken into account when interpreting the results coming from bulk perturbation theory: the Witten diagrams give terms in the effective action as a function of $b$, whereas the boundary source is $a$. We will see that \eqref{abRelation} is easily inverted perturbatively.

From the knowledge of the boundary conditions on the asymptotic boundary we can reconstruct the bulk field configurations perturbatively using Witten diagrams. To do so, we need to know the bulk-to-boundary propagator $G_\mu(X; t')$ that
 is defined by the requirement that a linearized solution of the bulk gauge field for which $b(t)$ is prescribed is given by 
 \es{Bulk}{
  A_\mu(X) = \int dt' \, G_{\mu}(X; t') b(t') +O(b^2)\,,
 }
where $X$ represents the coordinate of a bulk point.  Because $G_{\mu}(X; t')$ is proportional to the identity matrix, we will just treat it as a number.  In the $( z,t)$ coordinates introduced in \eqref{Poincare} we would write $G_\mu(z, t; t')$.

The canonical momentum \eqref{CanMom} in these coordinates is $\pi^t(z, t)=-\frac{1}{e^2}z\le(z F_{zt}+A_t\ri)$ and approaches $-\frac{1}{e^2} z b(t)$ at the boundary, from which we infer we see that the function $G_\mu(z, t; t')$ must obey
 \es{eqForGmu}{
   2z \partial_{[z} G_{t]}(z, t; t') - G_t (z, t; t')  \to \delta(t - t')  \,, \qquad
   \text{as $z \to 0$}
 } 
and must solve the field equations \eqref{eoms} with vanishing sources. 

In addition, we require that $G_\mu(z, t; t')$ is normalizable, which means $G_\mu(z, t; t') = o(z^{-1/2})$ as $ z \to  \infty$. A good solution obeying all these requirements is the dimensional continuation of the bulk-to-boundary propagator given in  \cite{Freedman:1998tz}:
 \es{BulkToBound}{
  G_{\mu}(z, t; t') = \frac{1}{2\pi\, z}  \left[ z^2+ (t - t')^2  \right] \partial_\mu \frac{ t -t'}{z^2 + (t - t')^2}  \,.
 }
This solution is not unique, because one can perform gauge transformations $G_\mu (z, t; t') \to G_\mu (z, t; t') + \partial_\mu \Lambda(z, t-t')$, with $\Lambda(z, t-t')$ decaying as $ z \to 0,\infty$. 
(As $z \to 0$ any rate of decay is acceptable, but as $z \to \infty$ the decay should be $o(z^{-1/2})$.) We can pick out this solution from the many gauge equivalent ones, by choosing the gauge
\es{GaugeChoice}{
0=\nabla_\mu A^\mu={1\ov \sqrt{g}} \, \p_\mu A_\mu\,.
}
Note that \eqref{BulkToBound} implies that
 \es{FtrReconstruct}{
  F_{zt}(r, t) = \frac{1}{ 2\pi z^2} \int dt'\, b(t')+O(b^2) \,,
 }
and we can thus show that the Maxwell equation $\partial_\mu (z^2 F_{zr}) = 0$ is satisfied.

The statement of the extrapolate dictionary \cite{Banks:1998dd,Harlow:2011ke} is that if we take the  bulk point of the  bulk-to-boundary propagator to the boundary, we get the boundary two-point function.  Because the fluctuating component of the gauge field is the leading piece $A_t\approx {{\cal Q}\ov z}$, we have to compensate for this with a factor of $z$. Note that in higher dimensions the translation between bulk-to-boundary propagators and boundary two-point functions is a factor $z^{2-d}$, so our result is also the dimensional continuation of the analogous prescription in higher dimensions. In formulas, we have
\es{Extrapolate}{
{1\ov g_\text{YM}^2}\lim_{z\to 0}z\, A_t(t_1,z)=  \langle j(t_1) \rangle_{a}\,,
}
where the $a$ subscript indicates that the one-point function is evaluated in the presence of sources. According to \eqref{Correl} we can compute correlation functions using a hybrid method:
\es{Correl2}{
  \langle j^{a_1}(t_1) \cdots j^{a_n}(t_n)  \rangle
   &= (-1)^{n-1} \frac{\delta \langle j(t_1) \rangle_{a}}{\delta a^{a_2}(t_2) \cdots \delta a^{a_n}(t_n)  }\Bigg\vert_{a=0}  \\
   &={(-1)^{n-1}\ov g_\text{YM}^2}\, \frac{\delta \le[\lim_{z\to 0}z\, A_t(t_1,z)\ri]}{\delta a^{a_2}(t_2) \cdots \delta a^{a_n}(t_n)  }\Bigg\vert_{a=0} \,.
 }
Using that $b=a+O(a^2)$ \eqref{Bulk} enables us to compute the two-point function 
\es{BulkPointLimit}{
\langle j^a(t_1) j^b(t_2) \rangle
&=-{\de^{ab}\ov g_\text{YM}^2} \lim_{z\to 0}z\,   G_{t}(z,t;t')\\
&={\de^{ab}\ov 2\pi g_\text{YM}^2}\,,
}
which agrees with the result \eqref{TwoPointYM} computed both from the bulk using the GKPW dictionary \cite{Gubser:1998bc,Witten:1998qj} and field theory.

\subsection{Bulk gauge profile at second order}

\begin{figure}[!htb]
	\centering
	\includegraphics[scale=1.7]{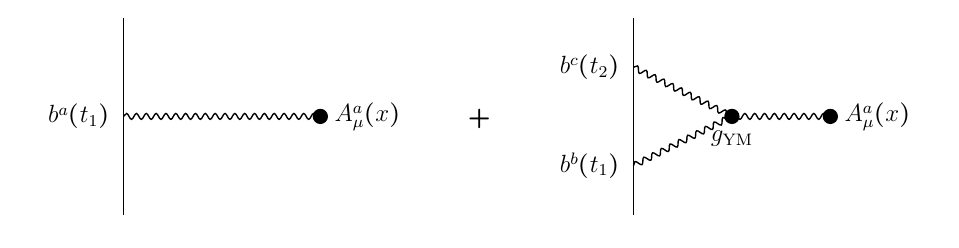}
	\caption{Reconstruction of classical solution for bulk gauge fields from Witten diagrams.}
	\label{Fig:bdybulk}
\end{figure}

To go to next order, we have to evaluate a Witten diagram with a three-point vertex, see Figure~\ref{Fig:bdybulk}. In this appendix we will restrict our attention to the two-derivative Yang-Mills action, but extensions to higher-derivative Lagrangians is straightforward.

We will need the bulk-to-bulk propagator to compute Witten diagrams. We follow the procedure presented in \cite{DHoker:1999bve}. First, we write
\es{FandS}{
{\bf G}_{\mu\nu'}(X;X')&=-\le(\p_\mu\p_{\nu'}u\ri) F(u)+ \p_\mu\p_{\nu'} S(u) \,, \\
u&\equiv {(X-X')^2\ov 2 z z'}\,,
}
where $F(u)$ is the physical and $S(u)$ is the pure gauge component. The equation satisfied by the propagator is
\es{eqGBB}{
\nabla^\mu\p_{[\mu}\, {\bf G}_{\nu]\nu'}(X;X')=g_{\nu\nu'}\de(X,X')+\p_{\nu'}\le((\p_\nu u)\, \Lam(u)\ri)\,.
}
Plugging in \eqref{FandS} into \eqref{eqGBB} we get a set of ODEs for $F$ and $\Lam$, whose solution is:
\es{FLamSol}{
F(u)&=A_1+A_2\log\le(u(2+u)\ri)+A_3 \log u \,, \\
\Lam(u)&={2A_2\ov u(2+u)}+{A_3\ov u}\,,
}
where the normalization of the delta function in \eqref{eqGBB} fixes $A_2={1\ov 4\pi}$. We observe that $A_1$ can be moved to $S(u)$, and henceforth we will eliminate it from $F(u)$. 

Because the gauge field obeys the  boundary conditions $\lim_{z\to 0}A_t+z F_{zt}=0$ and $\lim_{z\to 0}z A_z=0$, we impose
\es{BC}{
\lim_{z'\to 0} \le( {\bf G}_{\mu t}+2z' \p_{[z'}{\bf G}_{|\mu | t]} \ri)(X;X')&=0 \,, \\
\lim_{z'\to 0}z'\,  {\bf G}_{\mu z}(X;X')&=0\,.
}
The solution obeying these boundary conditions is
\es{FinalGBB}{
F(u)&={\log\le(u(2+u)\ri)\ov 4\pi} \,, \\
S(u)&={(u+1)\log u -2 u\ov 2\pi}\,.
}
Here we have chosen $S(u)$ such that we also satisfy
\es{BC2}{
\lim_{z'\to \infty}  {\bf G}_{\mu \mu'}(X;X')&=0\,.
}
This condition still does not fix $S(u)$ uniquely, as there is still some gauge freedom left, but in \eqref{FinalGBB} we made a choice based on simplicity. 
Note that this form of $F(u)$ is the dimensional continuation of the formula from \cite{DHoker:1999bve}, up to a divergent constant $A_1$ that is a gauge artifact.\footnote{There is also a sign difference between our result and the dimensional continuation of the formula of \cite{DHoker:1999bve} stemming from the difference in sign choice in \eqref{eqGBB}. }
Alternatively, instead of the expression ${\bf G}_{\mu\nu'}(X;X')$ obtained from \eqref{FandS} and \eqref{FinalGBB}, one may use the perhaps a simpler form
\es{FinalGBB2}{
{\bf G}_{\mu\nu'}(X;X')&=\le(\p_\mu \p_{\nu'} u\ri)\, {u\log\le(u\ov 2+u\ri)-2(u-1)\ov 4\pi u}+\le( \p_\mu u\,  \p_{\nu'} u\ri) \, {u-1\ov 2\pi u^2}\,,
}
but we will not do so here.  Note that regardless of which expression for \eqref{FinalGBB2} we use, 
as we take the one of the bulk points to the boundary, we get the bulk-to-boundary propagator $G_{\mu t}(X;t')$:
\es{BulkPointLimit2}{
\lim_{z'\to 0}z'\,  {\bf G}_{\mu t}(X;X')&=G_{\mu t}(X;t')\,.
}
(See the explanation around \eqref{BulkPointLimit}.)

An important ingredient in evaluating Witten diagrams is the bulk-boundary-boundary three-point function. Using the bulk three-point vertex of Yang-Mills theory, we get:
\es{BBdyBdy}{
F^{abc}_\mu\le(t_1,t_2,Y\ri)=-2{f^{abc}}\int dX \ \sqrt{g}\, &\le[\p_{[\al} G_{\beta]}(X;t_1) G^\al(X;t_2)  {\bf G}^\beta_{\,\,\mu }(X;Y)  \ri.\\
& \le.- G^{\al}(X;t_1) \p_{[\al} G_{\beta]}(X;t_2)  {\bf G}^\beta_{\,\,\mu }(X;Y) \ri.\\
& \le.+G^{\al}(X;t_1) G^\beta(X;t_2)\p_{[\al}{\bf G}_{\beta]\mu }(X;Y)\ri]
}
where we dropped the superfluous index $t$ from the bulk-to-boundary propagator.\footnote{This equation can also be thought of as the solution of Yang-Mills equations to second order in a  perturbative expansion. The bulk-to-bulk propagator is the Green's function of the Maxwell equation \eqref{eqGBB} and has to be convolved with the current of the non-Abelian gauge fields determined by the first order solution \eqref{Bulk}. }
At first sight it seems that there is  another three-point vertex hiding in   the boundary term $-{2\ov g_\text{YM}^2} \int_\partial dt\, z^2\, \tr \left( A^i F_{\mu i} n^\mu \right)$. This is not the case however, because  $\tr\le(A_\mu\le[A_\nu, A_\rho\ri]\ri)=0$, as  in two dimensions $\mu,\nu,\rho$ cannot all be different. It is worth noting that because the gauge field couples to a conserved current,  \eqref{BBdyBdy} is insensitive to the choice of $S(u)$ in \eqref{FandS},  the pure gauge part of the bulk-to-bulk propagator.

It is now possible to show that the bulk gauge field constructed using \eqref{BBdyBdy} satisfies the gauge condition \eqref{GaugeChoice} implementing the proof given in \cite{DHoker:1999mqo} to our setup. We first make use of the fact that $\nabla_Y^\mu{\bf G}_{\alpha\mu }(X;Y)$ is a rank-1 bitensor in a maximally symmetric space, hence it has to be expressible as
\es{Expressible}{
\nabla_Y^\mu{\bf G}_{\alpha\mu }(X;Y)=\p_\alpha u\, g(u)=\p_\alpha \le(\int^u du' g(u)\ri)\,.
}
Making use of the fact that the gauge field couples to a conserved current it now follows that
\es{FCond}{
\nabla_Y^\mu F^{abc}_\mu\le(t_1,t_2,Y\ri)=0\,,
}
which we wanted to show.

Now we follow the technology developed in \cite{Freedman:1998tz,DHoker:1999bve}. We will use the flat metric to contract indices from now on, use only lower indices, and write out the explicit powers of $z$. We introduce the notation
\es{JDef}{
J_{\mu\nu}(X)&=\de_{\mu\nu}-2{X_\mu X_\nu\ov X^2}\,,\\
J_{\mu\nu}(X) J_{\nu\lam}(X)&=\de_{\mu\lam}\,.
}
In this notation the bulk-to-boundary propagator takes the form
\es{GExpJ}{
G_{\mu }(X;t')={1\ov 2\pi z}\, J_{\mu t}(X-t')\,. 
}

We make the following change of coordinates:
\es{CoordChange}{
X'_{\mu}={X_\mu\ov X^2}\,, \qquad t'={1\ov t}\,,
}
which is a simple change of coordinates in the bulk, but a conformal transformation in the boundary.
Under this transformation  the objects of interest transform as
\es{TfProp}{
G_{\al }(X;t)&={\p X'_\mu\ov \p X_\al} \, G_{\mu }(X';t') \,, \\
{\bf G}_{\al\beta }(X;Y) &={\p X'_\mu\ov \p X_\al}{\p Y'_\nu\ov \p Y_\beta}\, {\bf G}_{\mu\nu }(X;Y) \,, \\
dX &={dX' \ov X'^4}\,,
}
where
\es{CoordWithJ}{
{\p X'_\mu\ov \p X_\al}=X'^2 \, J_{\mu \al}(X')\,.
}
The transformation of derivatives of these objects also follows analogous rules.
We use translation invariance to set $t_1=0$. Plugging all this into~\eqref{BBdyBdy} we get
\es{BBdyBdy2}{
F^{abc}_\mu\le(0,t_2,Y\ri)=-2{f^{abc}}\int dX' \ z'^2\, &\le[\p_{[\al} G_{\beta]}(X';\infty) G_\al(X';t_2')  {\bf G}_{\beta\nu }(X';Y')  \ri.\\
& \le.- G_{\al}(X';\infty) \p_{[\al} G_{\beta]}(X';t_2')  {\bf G}_{\beta\nu }(X';Y')  \ri.\\
& \le.+G_{\al}(X';\infty) G_\beta(X';t_2')\p_{[\al}{\bf G}_{\beta]\nu }(X';Y')\ri]\, Y'^2 J_{\nu \mu}(Y')\,.
}
Now we can use the extreme simplicity of 
\es{Simp}{
G_{\al}(X;\infty)&=-{\de_{\al t}\ov 2\pi z} \,,\\
\p_{[\al} G_{\beta]}(X;\infty)&=\p_{[\al} G_{\beta]}(X;t)=-{ \ep_{\al\beta}\ov 4\pi z^2} \,, \qquad \ep_{tz}\equiv 1
}
to simplify~\eqref{BBdyBdy2}:
\es{BBdyBdy3}{
F^{abc}_\mu\le(0,t_2,Y\ri)&={f^{abc}\ov \pi}\int dX' \  \le[  G_{[t}(X';t_2')  {\bf G}_{z]\nu }(X';Y') + {1\ov 4\pi z'}  {\bf G}_{z\nu }(X';Y')  \ri.\\
& \le.+{z'} \,G_z(X';t_2')\p_{[t}{\bf G}_{z]\nu }(X';Y')\ri]\, Y'^2 J_{\nu \mu}(Y')\,.
}
This can be brought to the final form
\es{BBdyBdy4}{
F^{abc}_\mu\le(0,t_2,Y\ri)={f^{abc}\ov \pi }\, I_\nu(Y'-t_2')\, {1\ov Y^2} J_{\nu \mu}(Y)\,,
}
where $I_\nu$ is the integral in~\eqref{BBdyBdy3}.
We can evaluate this integral explicitly. The strategy is simple, but tedious to implement. The integrand has (integrable) singularities at $X'=Y'$ and $X'=t_2'$, and decays fast enough for $X'\to\infty$.  We subtract a carefully designed function from the integrand that has the same singularity structure, decays fast enough at infinity, and is easier to integrate than the original integral. After this subtraction, we first integrate with respect to $X'_t$, and then with respect to $X'_z$. Then we integrate the function that we subtracted by cutting out small disks around its singularities, and then take the radii of the disks to zero. The final result is:
\es{BBdyBdy5}{
I_\nu(Z)&={Z_\mu\ov 2\pi\, Z^2}\, \le[ 1+{t\ov z}\, \arctan\le({t\ov z}\ri)\,\ri] \,.
}

We can perhaps also arrive at this result in a slicker way. 
It is argued in  \cite{DHoker:1999mqo} about a similar integral that it has to take the form
\es{Integral}{
I_\mu(Z)={Z_\mu\ov Z^2}\, f(\tau)+{ \de_{\mu z}\ov z}\, h(\tau)\,, \qquad \tau\equiv{z^2\ov Z^2}\,.
}
Note that~\eqref{BBdyBdy5} is of this form. We can actually determine that $h(\tau)=0$ based on the Bose symmetry of $F^{abc}_\mu$. We are then still left with the task of determining $f(\tau)$, which may be simpler than the route that we took in obtaining \eqref{BBdyBdy5}. 

With this choice, we have from \eqref{BBdyBdy4}
 \es{GotF}{
  F^{abc}_t(t_1; t_2; t, z) &= -{f^{abc}\ov2 \pi^2}\,\frac{\left[ (t_1 - t) (t_2 - t) - z^2 \right]\le[(t_1-t_2)z+\le((t_1 - t) (t_2 - t) + z^2\ri)  \arctan \frac{(t_1 - t) (t_2 - t) + z^2}{(t_1 - t_2) z}\ri] }{ z \left[ (t_1 - t)^2 + z^2 \right] \left[  (t_2 - t)^2 + z^2 \right]  } \,, \\
   F_z^{abc}(t_1; t_2; t, z) &= {f^{abc}\ov2 \pi^2}\,\frac{[t_1 + t_2 - 2 t] \left[(t_1-t_2)z+ \le((t_1 - t) (t_2 - t) + z^2\ri)  \arctan \frac{(t_1 - t) (t_2 - t) + z^2}{(t_1 - t_2) z} \right]}{
   \left[ (t_1 - t)^2 + z^2 \right] \left[  (t_2 - t)^2 + z^2 \right]  } \,. 
 }
 
 Finally, we have to remind ourselves what these expressions really mean. As discussed above, the Witten diagram (see Figure~\ref{Fig:bdybulk}) computation determines the bulk gauge field 
 \es{WittenSummary}{
  A^a_\mu(X) = \int dt_1 \, G_{\mu}(X; t_1) b^a(t_1) +\frac12\int dt_1dt_2 \, F^{abc}_{\mu}(X; t_1;t_2) b^b(t_1)b^c(t_2)+O(b^3)\,.
 }
$A^a_\mu(X)$ also obeys the gauge condition \eqref{GaugeChoice}, which serve as a useful check of the computation. It is easy to check that $A^a_t(X)={\#\ov z}+b(t)+O(z)$, as we wanted. This is however not the boundary condition that the variational principle of the action imposes. We instead impose
\es{abRelation2}{
a^a(t)=\lim_{z\to 0} \le(A^a_t+z F^a_{zt}\ri)=b^a(t)+{f^{abc}\ov 4\pi^2}\int dt_1 dt_2\ {t_1-t_2\ov (t-t_1)(t-t_2)}\, b^b(t_1)b^c(t_2)+O(b^3)\,.
}
This relation between $a$ and $b$ can also be deduced from \eqref{abRelation}.\footnote{Explicitly
 \es{abRelation3}{
a(t)&=b-i\le[c,{\cal Q}\ri]\\
&=b-i\le[c,Q\ri]+O(b^3)\,.
}
Using the asymptotic expansion of the bulk-to-boundary propagator we have:
\es{AzAsymp}{
A_t(z, t) &= -{1\ov z}\,{ 1\ov 2\pi} \int dt_1\, b(t_1) + b(t) +O(z,b^2) \,,\\
 A_z(z, t) &=-{1\ov \pi}\int dt_1 \  {b(t_1)\ov t-t_1}+O(z,b^2)\,,
}
giving
 \es{abRelation4}{
a^a(t)
&=b^a(t)+{f^{abc}\ov 4\pi^2}\int dt_1 dt_2\ {t_1-t_2\ov (t-t_1)(t-t_2)}\, b^b(t_1)b^c(t_2)+O(b^3)\,.
}
This is identical to to \eqref{abRelation2}.
}
Inverting \eqref{abRelation2}, we can finally express the bulk gauge field in terms of $a$:
 \es{WittenSummary2}{
  A^a_\mu(X)& = \int dt_1 \, G_{\mu}(X; t_1) a^a(t_1) +\frac12\int dt_1dt_2 \, \widetilde F^{abc}_{\mu}(X; t_1;t_2) a^b(t_1)a^c(t_2)+O(a^3) \,,\\
  \widetilde F^{abc}_t(t_1; t_2; t, z) &= -{f^{abc}\ov2 \pi^2}\,\frac{\left[ (t_1 - t) (t_2 - t) - z^2 \right]\le((t_1 - t) (t_2 - t) + z^2\ri)  \arctan \frac{(t_1 - t) (t_2 - t) + z^2}{(t_1 - t_2) z}}{ z \left[ (t_1 - t)^2 + z^2 \right] \left[  (t_2 - t)^2 + z^2 \right]  } \,, \\
  \widetilde F_z^{abc}(t_1; t_2; t, z) &= {f^{abc}\ov2 \pi^2}\,\frac{[t_1 + t_2 - 2 t] \le((t_1 - t) (t_2 - t) + z^2\ri)  \arctan \frac{(t_1 - t) (t_2 - t) + z^2}{(t_1 - t_2) z}}{
   \left[ (t_1 - t)^2 + z^2 \right] \left[  (t_2 - t)^2 + z^2 \right]  } \,.
 }
 
We can now use \eqref{Correl2} to obtain the three-point function from \eqref{WittenSummary2}
  \es{BdyLimit}{
  \langle j^a(t_1) j^b(t_2) j^c(t) \rangle&={1\ov g_\text{YM}^2}
  \lim_{z\to 0} z \widetilde F^{abc}_t(t_1; t_2; t, z) \\
  &= -{f^{abc}\ov4 \pi}\, \sgn(t_1 - t_2) \sgn(t_2 - t) \sgn(t - t_1)\,,
 }
 which agrees with \eqref{ThreePointYM}. We note that having a formula for the bulk gauge field \eqref{WittenSummary} takes us halfway towards computing the 4-point function using Witten diagrams. We do not finish this computation here.
 
 Finally, we comment on the resolution of the puzzle raised in Section~\ref{NONABELIANHIGHER}. There it was pointed out that the higher derivative bulk theory has interaction vertices  with strength $d_n^{a_1\dots a_n}$, and Witten diagrams containing multiple such vertices will give a polynomial result in $d_n^{a_1\dots a_n}$.
 However, boundary theory correlation functions are strictly linear in these coefficients. The resolution is that we have to convert the source for the Witten diagrams, $b^a(t)$ to the boundary theory source, $a^a(t)$ as in \eqref{abRelation4}. The higher derivative terms in the action contribute to this relation, and as a result the boundary theory correlation functions become linear in $d_n^{a_1\dots a_n}$.

\subsection{Comparison with the result from solution generating large gauge transformation}

In the main text instead of working with propagators we used large gauge transformations to obtain solutions for the Yang-Mills equations with prescribed boundary conditions. Let us see, how that approach reproduces what we have obtained here. Let us start with the gauge field configuration
\es{GaugeStart}{
A_t &= {Q\ov z}\,, \qquad A_z=0\,.
}
Under a gauge transformation
\es{PertGauge}{
U(z,t)=e^{i \Lam(z,t)}=1+i \Lam-\frac12\, \Lam^2+\dots
}
the gauge field changes into
\es{Gauge2}{
A_\mu &= {Q\ov z}\,\de_{\mu t}+\p_\mu \Lam+i\le[\Lam,{Q\ov z}\,\de_{\mu t}\ri]+{i\ov 2}\,\le[\Lam,\p_\mu\Lam\ri]+\dots\,.
}
where we assumed that $Q=O(\Lam)$. More systematically,
\es{GaugeSecond}{
A_\mu &= A_\mu^{(1)}+A_\mu^{(2)}\,,\\
A_\mu^{(1)}&= {Q^{(1)}\ov z}\,\de_{\mu t}+\p_\mu \Lam^{(1)}\,,\\
A_\mu^{(2)}&={Q^{(2)}\ov z}\,\de_{\mu t}+\p_\mu \Lam^{(2)}+i\le[\Lam^{(1)},{Q^{(1)}\ov z}\,\de_{\mu t}\ri]+{i\ov 2}\,\le[\Lam^{(1)},\p_\mu\Lam^{(1)}\ri]\,.
}
We want to satisfy the gauge condition \eqref{GaugeChoice}, which imposes that $\Lam^{(1)}$ is a harmonic function: 
\es{LamCond}{
0&=\p_\mu \p_\mu \Lam^{(1)}\,.
}
We also want  the resulting gauge field to satisfy the conditions:
\es{LamFurth}{
\lim_{z\to 0}\p_t \Lam^{(1)}(z,t)=a(t)\,, \qquad \lim_{z\to 0}z \p_z \Lam^{(1)}(z,t)=0\,.
}
 We have to solve the Laplace equation on half space, which can be done with the Green's function method
\es{Laplace}{
\p_\mu \p_\mu F(z,t)&=s(z,r)\,, \qquad F(0,t)=f(t)\,,\\
F(z,t)&=\int dz' dt' \ G(z,t; z',t') s(z',t')+\int dt' \ g(z,t; t') \, f(t')\,,\\
G(z-z', t-t') &\equiv{1\ov 4\pi} \, \log \le[{(t-t')^2+(z-z')^2\ov (t-t')^2+(z+z')^2}\ri]\,,\\
g(z,t; t')&=\lim_{z'\to 0}\p_{z' }G(z,t; z',t') =-{1\ov \pi}\, {z\ov (t-t')^2+z^2}\,.
}

Applying the general technology \eqref{Laplace} to the equation \eqref{LamCond} with the boundary condition \eqref{LamFurth} we obtain:
\es{LamSolGreen}{
\Lam^{(1)}(z,t)&=\int dt_1 \ \le[g(z,t; t_1) \, \int^{t_1} dt_2\ a(t_2) \ri]\\
&=-\int dt_1 \ \le[\int^{t_1} dt_2 \ g(z,t; t_2)  \ri]\, a(t_1)\\
&={1\ov \pi}\int dt_1 \ \arctan\le(t-t_1\ov z\ri)\, a(t_1)\,.
}
To determine the value of $ Q^{(1)}$, we go to global coordinates using the coordinate transformation \eqref{CoordTrans}. Let us set $\phi=\pi+\de \phi$. We find that
\es{Problem}{
A_r\approx-e^r\,\le(Q^{(1)}+{1\ov 2 \pi}\int dt_1 \  a(t_1)\ri)\,  \de \phi+O(\de \phi^3)\,,
} 
which violates the boundary condition that $A_r$ has to decay at the boundary, unless
 we choose
\es{GaugeTf}{
 Q^{(1)}=-{1\ov 2 \pi}\int dt_1 \  a(t_1)\,.
}
With these choices \eqref{Gauge2} reproduces \eqref{Bulk} with the propagator \eqref{BulkToBound}. We conclude that to first order we have successfully shown the equivalence of the method of large gauge transformations to the more conventional propagator approach. We will solve the problem at second order below.

We want to obtain the second order piece of $A_\mu$ of \eqref{GaugeSecond} to compare to the Witten diagram computation above. The gauge condition and the appropriate boundary behavior  of the gauge field impose the following on $\Lam^{(2)}$:
\es{A2Bdy}{
\p_\mu\p_\mu \Lam^{(2)}+i\le[\p_t\Lam^{(1)},{Q^{(1)}\ov z}\ri]&=0\,,\\
\lim_{z\to 0 }\p_t \Lam^{(2)}+{i\ov 2}\,\le[\Lam^{(1)},\p_t\Lam^{(1)}\ri]&=0\,, \qquad \lim_{z\to 0 }z\le(\p_z \Lam^{(2)}+{i\ov 2}\,\le[\Lam^{(1)},\p_z\Lam^{(1)}\ri]\ri)=0\,.
}
Again, using \eqref{Laplace}, we can determine $\Lam^{(2)}$:
\es{Lam2Laplace}{
\p_\mu \p_\mu F(z,t)&=s(z,r)\,, \qquad F(0,t)=f(t) \,, \\
\Lam^{(2)}=&-i\int dz_1 dt_1 \ G(z,t; z_1,t_1) \le[\p_{t_1}\Lam^{(1)}(z_1,t_1),{Q^{(1)}\ov z_1}\ri] \\
&-{i\ov 2}\int dt_1 \ \le[g(z,t; t_1) \, \int^{t_1} dt_2\ \le[\Lam^{(1)}(0,t_2),\p_{t_2}\Lam^{(1)}(0,t_2)\ri]\ri] \,.
}
Let us first manipulate the second line:
\es{SecondLineApp}{
&-{i\ov 2}\int dt_1 \ \le[g(z,t; t_1) \, \int^{t_1} dt_2\ \le[\Lam^{(1)}(0,t_2),\p_{t_2}\Lam^{(1)}(0,t_2)\ri]\ri]\\
&={i\ov 2} \int dt_1 \ \le[\int^{t_1} dt_2 \ g(z,t; t_2)  \ri]\, \le[\Lam^{(1)}(0,t_1),\p_{t_!}\Lam^{(1)}(0,t_1)\ri]\\
&=-{i\ov 4\pi} \int dt_1 dt_2 \ \arctan\le(t-t_1\ov z\ri)\, \sgn(t_1 -t_2) \, \le[ a(t_2),a(t_1)\ri]\\
&=-{f^{abc}\,T^a\ov 4\pi}  \int dt_1 dt_2 \ \arctan\le(t-t_1\ov z\ri)\, \sgn(t_1 -t_2) \,  a^b(t_1)a^c(t_2)\,.
}
The first line is given by
\es{FirstLine}{
&-i\int dz' dt' \ G(z,t; z',t') \le[\p_{t_1}\Lam^{(1)}(z',t'),{Q^{(1)}\ov z'}\ri]\\
&=-{f^{abc}\,T^a\ov 2\pi^2} \int dz' dt' dt_1 dt_2 \ G(z,t; z',t') \,{1\ov (t'-t_1)^2+z'^2} \, a^b(t_1)a^c(t_2)\,.
}
Because we only need $\p_\mu \Lam^{(2)}$, and it is easier to compute than $\Lam^{(2)}$ we differentiate the above expressions, and perform the integrals to obtain:\footnote{In this case we do not get extra contributions from the different methods of integrations unlike in the case explained around \eqref{BBdyBdy5}.}
\es{pLam2}{
\p_t \Lam^{(2)}&=-{f^{abc}\,T^a\ov 2\pi^2}  \int dt_1 dt_2\ {z\, \le[\arctan\le(t-t_1\ov z\ri)+{\pi\ov 2}\sgn(t_1 -t_2)\ri]\ov (t-t_1)^2+z^2} \,a^b(t_1)a^c(t_2)\,,\\
\p_z \Lam^{(2)}&={f^{abc}\,T^a\ov 2\pi^2}  \int dt_1 dt_2\ {(t-t_1)\, \le[\arctan\le(t-t_1\ov z\ri)+{\pi\ov 2}\sgn(t_1 -t_2)\ri]\ov (t-t_1)^2+z^2} \,a^b(t_1)a^c(t_2)\,.
}
We write down all terms in \eqref{GaugeSecond} with the result
\es{A2}{
A^{(2)}_t=&{Q^{(2)}\ov z}-{f^{abc}\,T^a\ov 2\pi^2} \int dt_1 dt_2\ a^b(t_1)a^c(t_2)\\
&\times \le[{z\, \le[\arctan\le(t-t_1\ov z\ri)+{\pi\ov 2}\sgn(t_1 -t_2)\ri]\ov (t-t_1)^2+z^2} - {\arctan\le(t-t_1\ov z\ri)\,(t-t_2)^2\ov z\le[(t-t_2)^2+z^2\ri]}\ri] \,, \\
A^{(2)}_z=&{f^{abc}\,T^a\ov 2\pi^2}  \int dt_1 dt_2\ a^b(t_1)a^c(t_2)\\
&\times \le[{(t-t_1)\, \le[\arctan\le(t-t_1\ov z\ri)+{\pi\ov 2}\sgn(t_1 -t_2)\ri]\ov (t-t_1)^2+z^2}+{\arctan\le(t-t_1\ov z\ri) \, (t-t_2)\ov (t-t_2)^2+z^2} \ri]\,.
}
A similar argument to what was presented around \eqref{Problem} fixes
\es{Q2Fix}{
Q^{(2)}={f^{abc}\,T^a\ov 8\pi} \int dt_1 dt_2\ \sgn(t_1 -t_2)\,a^b(t_1)a^c(t_2)\,.
}

To facilitate comparison with \eqref{GotF}, we antisymmetrize the integrand in $t_1,\, t_2$. A few lines of algebra convinces us that
\es{FA2}{
A^{(2)}_\mu(t,z)=\frac12 \int dt_1 dt_2\ \widetilde F^{abc}_\mu (t_1;t_2; t,z) \,T^a \, a^b(t_1)a^c(t_2)\,,
}
where $\widetilde F^{abc}_\mu$ is defined in \eqref{WittenSummary2}. This concludes the demonstration that the solution generating large gauge transformation method used in the main text is just a convenient way of performing Witten diagram computations in 2d Yang-Mills theory.

\section{Localization on $AdS_4$}\label{LOCAPP}

\subsection{Coordinates on $AdS_4$}\label{app:ads4}

In terms of the embedding coordinates in $\mR^{1,4}$, $AdS_4$ is given by 
\ie
X_1^2+X_2^2+X_3^2 +X_4^2-X_5^2=-L^2\,.
\fe
We can parametrize 
\ie
&X_1=L\sinh {r\over L} \sin  \theta\sin  \phi \,, 
\\
&X_2=-L\sinh {r\over L}\cos \theta \cos \tau  \,,
\\
&X_3=L\sinh {r\over L} \cos \theta \sin \tau  \,,
\\
&X_4=-L\sinh {r\over L}\sin  \theta \cos  \phi \,,
\\
&X_5=L\cosh {r\over L}\,.
\fe
This gives rise to the familiar metric
\ie
ds^2=d r^2+ L^2 \sinh^2{r\over L} ds^2_{S^3}\,,
\fe
where the boundary $S^3$ has metric
\ie
ds^2_{S^3}=d\theta^2 +\cos^2\theta d\tau^2+\sin^2\theta d\varphi^2  \,,
\fe
with $\theta\in [0,{\pi \over 2}]$ and $\tau,\varphi \in [-\pi,\pi]$.

We rewrite the bulk metric as follows 
\ie
ds^2=ds^2_{{1\over 2}AdS_3}+L^2\cos^2\theta d\tau^2\,,
\fe
and
\ie
ds^2_{{1\over 2}AdS_3}=dr^2+\sinh^2 {r\over L} ds^2_{D^2}\,,
\fe
which is half of  hyperbolic space and we will denote it by ${1\over 2}AdS_3$. Note that the boundary of ${1\over 2}AdS_3$ consists of $AdS_2$ at $\theta={\pi\over 2}$ with metric
\ie
ds^2_{AdS_2}=dr^2+L^2\sinh^2 {r\over L} d\varphi^2\,,
\fe
and $D^2$ at $r=r_0\to \infty$. See Figure~\ref{Fig:halfh3} for a cartoon of the geometry.

Alternatively we have in Poincar\'e coordinates with $x_4>0$ that
\ie
&X_{1,2,3}={L\, x_{1,2,3}\over x_4} \,, 
\\
&X_4={x_i^2+x_4^2-4L^2\over 4x_4}  \,, 
\\
&X_5={x_i^2+x_4^2+4L^2\over 4x_4}\,,
\fe
and the metric becomes
\ie
ds^2={L^2\,  dx^2\over x_4^2} \,.
\label{AdS4Poincare}
\fe 
 
We collect some useful formulas about the differential geometry on $AdS_4$ below.  In Poincar\'e coordinates, the metric is manifestly conformally flat, and we can rewrite \eqref{AdS4Poincare} as
\ie
g_{\m\n}=\D_{\m\n} e^{2\Omega}\,,\qquad e^{ \Omega}={L\over x_4}\,.
\fe
In terms of the vielbein we have
\ie
g_{\m\n}=\D_{\hat\m\hat \n} e^{\hat\m}_\m e^{\hat \n}_\n\,,\qquad e^{\hat \m}_\m=\D^{\hat \m}_\m e^\Omega\,.
\fe
The spin connection is then
\ie
\omega^{\hat \m}_{\hat\n \lambda}=(e^{\hat \m}_\lambda e^\n_{\hat \n}-e_{\hat \n \lambda} e^{\hat \m \n}) \pa_\n \Omega\,.
\fe
It satisfies
\ie
d e^{\hat \m}+\omega^{\hat\m}{}_{\hat\n} \wedge e^{\hat \n}=0\,.
\fe

\subsection{Supersymmetric boundary term for $\cN=4$ SYM on $AdS_4$}
\label{App:bt}

The SYM action \eqref{SYMos} is only invariant under SUSY transformation \eqref{SUSYos} up to boundary terms. In this appendix, we derive the boundary term $S_B$ in \eqref{ads4bt} for the theory on asymptotically Euclidean $AdS_4$, which preserves an off-shell supercharge $\D_\ve$ with any conformal Killing spinor $\ve$ satisfying the condition
\ie
\tilde \ve=-{1\over 2L}\Lambda \ve\,.
\label{16susy}
\fe
We will set $L=1$ below.

We first perform the explicit SUSY variation of the SYM action $S$  \eqref{SYMos}, and obtain after some algebra,
\ie
\D_\ve S
=&{1\over 2  g_4^2}\int_{AdS_4}d^4 x\, \sqrt{g} \,  \pa_\m \tr
\Bigg(
\Psi \CC^\m \n_m K^m+2 \tilde\ve \CC_A{}^\m\Psi \Phi^A
+{1\over 2}  F_{MN} \ve\CC^{MN\m} \Psi
+F^{\m N}\Psi \CC_N \ve
\Bigg)
\\
=&{1\over 2  g_4^2}\int_{\partial AdS_4} d^3 x\, \sqrt{\ga}  \, n_\m \tr
\Bigg(
\Psi \CC^\m \n_m K^m+2 \tilde\ve \CC_A{}^\m\Psi \Phi^A
+{1\over 2}  F_{MN} \ve\CC^{MN\m} \Psi
+F^{\m N}\Psi \CC_N \ve
\Bigg)\,,
\label{bdyvar}
\fe
where $n^\mu$ is the outward unit normal at the asymptotic boundary of $AdS_4$ which satisfies $n_\m \CC^\m=-\CC^{\hat 4}$, and $\ga$ is the induced metric on the boundary.

The most general asymptotic behavior of the bulk fields consistent with the equations of motion and supersymmetry is
\ie
&A_\m \approx A_\m^1 +A_\m^2 e^{-r}  \,,
\\
&\Phi_A \approx \Phi_A^1e^{-r}  +\Phi_A^2 e^{-2r} \,,
\\
&\Psi \approx\Psi^1e^{-{3\over 2}r} +\Psi^2 e^{-{5\over 2}r}  \,,
\\
&K_m\approx K_m^1 e^{-2r}+K_m^2 e^{-3r}\,.
\label{fieldsab}
\fe 
In addition the conformal Killing spinor $\ve$ has asymptotic form
\ie
\ve\approx \ve^1 e^{{1\over 2}r}+\ve^2 e^{-{1\over 2}r}\,,\qquad \tilde \ve\approx \tilde\ve^1 e^{{1\over 2}r}+\tilde\ve^2 e^{-{1\over 2}r}\,.
\fe
An immediate consequence of \eqref{16susy} and the conformal Killing spinor equations \eqref{CKSeqn} is the relation
\ie
\CC^{\hat 4}\ve^1=\Lambda \ve^1=-2\tilde \ve^1\,,
\fe
which we will use repeatedly to simplify expressions below.

For general boundary conditions, the boundary variation \eqref{bdyvar} does not vanish, and we want to the find boundary terms $S_B$ on the boundary $S^3$ such that
\ie
\D_\ve( S+S_B)=0\,.
\fe
Below we check explicitly that \eqref{ads4bt}, which we record below, fulfills this purpose,
\ie
S_B={1\over 2 g_4^2}\int d^3 x \,\sqrt{\ga}\,\tr \Bigg(
{1\over 2}\Psi\Lambda\Psi+2K^{a-5}\Phi_a-2D_r\Phi^a \Phi_a-4\Phi_9 F_{78}-2\Phi^a\Phi_a+\Phi^A\Phi_A
\Bigg)\,,
\fe
where $a=7,8,9$. Over the course of the derivation, we will drop terms that vanish in the $r=r_0 \to \infty$ limit.

Next, we want to determine the large $r$ behavior of the auxiliary spinors $\n_m$. There are 7 independent solutions of  \eqref{PSpinor} rotated into each other by an $SO(7)$ transformation that acts on the $m$ indices. One convenient set of  $\n_m$ that satisfies \eqref{PSpinor} is given by
\es{AuxSpinorSol}{
\n^m=\{
\CC^{\hat 46}\ve, \CC^{\hat  4a}\ve, \CC^{\hat  6a}\ve
\}\,,\qquad a=7,8,9\,.
}
Choosing the above solution fixes the $SO(7)$ freedom mentioned above. 
\eqref{AuxSpinorSol} implies
\ie
K^m \n_m \CC^{\hat 4} \Psi -K^m \n_m \Lambda \Psi-2  K^{a-5} \ve\CC_{a}\Psi=0\,.
\fe
Using the above identity and Gamma matrix identities in Appendix~\ref{App:gm}, we have 
\ie
{1\over 2}\D_\ve(\Psi \Lambda\Psi)
&= 
-{1\over 2}F_{MN}\ve \tilde\CC^{MN}\Lambda\Psi-2 \Phi^A\tilde\ve \tilde \CC_A \Lambda \Psi+ K^m \n_m \Lambda \Psi
\\
&={1\over 2} F_{MN} \ve \CC^{MN2} \Psi -  F_{r N}\ve \CC^N \Psi+12F^{[78} \ve \CC^{9]} \Psi-  \Phi^A \ve \CC_A  \Psi+2\Phi^a \ve \CC_a  \Psi
\\
&-2F_{Ma}\ve \CC^{Ma2} \Psi +2 F_{r a}\ve \CC^a \Psi
+ (K^m \n_m \CC^r \Psi -2  K^{a-5} \ve\CC_{a}\Psi)\,,
\label{plp}
\fe
where $
D_r=n^\m D_\m=-x_4 D_4
$.

Next, we have
\ie
2\D_\ve ( K^{a-5} \Phi_a)&=
2(K^{a-5} \ve \CC_a \Psi +\Phi_a  \ve \CC^{\hat 4a} \CC^N D_N \Psi)
\\
&=2(K^{a-5} \ve \CC_a \Psi +\Phi_a  \ve \CC^{\hat 4aN} D_N \Psi+\Phi_a  \ve \CC^{\hat 4} D_a \Psi
+\Phi_a  \ve \CC^{a} D_r\Psi)
\\
&=2K^{a-5} \ve \CC_a \Psi +2F_{Na} \ve \CC^{\hat 4Na} \Psi+3\Phi_a\ve \CC^a\Psi
+2\Phi_a  \ve \CC^{a} D_r \Psi\,,
\fe
where in the last equality we used that the boundary total derivative
\ie
&- \Phi_{a}\ve \CC^{\hat 4a N} D_N\Psi +F_{Na} \ve \CC^{\hat 4Na} \Psi+{3\over 2}\Phi_a\ve \CC^a\Psi=-D_\m(\Phi_a\ve \CC^{\hat 4 a \m}\Psi)
\fe
integrates to $0$ on $S^3$. 

Moving on to the third term in \eqref{ads4bt}, we obtain
\ie
&-2\D_\ve (D_r \Phi^a \Phi_a)
=-2D_r \Phi^a  \ve \CC_a\Psi-2\ve \CC_a D_r \Psi  \Phi_a -\ve  \CC_a  \Psi  \Phi_a\,.
\fe
We also have
\ie
-2\D_\ve(\Phi_9[\Phi_7,\Phi_8])=-6 \D_\ve \Phi_{[9} F_{78]}=-6F_{[78} \ve \CC_{9]}\Psi \,.
\fe
Therefore, the total SUSY variation of the integrand of $S_B$ is 
\ie
&\D_\ve\le({1\over 2}\Psi\Lambda\Psi+2K^{a-5}\Phi_a-2D_r\Phi^a \Phi_a-4\Phi_9 F_{78}-2\Phi^a\Phi_a+\Phi^A\Phi_A\ri)
\\
&=K^m \n_m \CC^{\hat 4} \Psi+{1\over 2}F_{MN} \ve \CC^{MN{\hat 4}} \Psi -  F_{r N}\ve \CC^N \Psi+  \Phi^A \ve  \CC_A  \Psi \,,
\fe
which is exactly $-\D S$ given in \eqref{bdyvar}. This is what we wished to show.

\subsection{Twisted $\mf{su}(2)$ and Killing spinors in $AdS_4$}

The twisted $\mf{su}(2)$ generators $\hat L_a$ \eqref{hatLAgain} are identified with the 10d Killing spinors $\ve_a$ such that acting on bulk fields,
\ie
\hat L_a= {1\over 2}\{\D_\ve,\D_{\ve_a}\} \,, \quad a=1,2,3 \,.
\fe
Here $\ve_a$ are given in the form \eqref{gencks} with 
\ie
\begin{alignedrows}{16}
	\arow{\epsilon_s^1={1\over 8}}{{-1}  & 0 & 0 &  0 &  0 &  0 &  -i&  0 & 0 &  1 &  0 &  0 & 0 & 0 & 0  &-i}\,, \\
	\arow{\epsilon_c^1 ={1\over 8}}{0 & 0 & 0 & -1 &  0 &  -i&  0&  0 & 0 &  0 &  1 &  0 &  i & 0 & 0 &0} \,,\\
	\arow{\epsilon_s^2={1\over 8}}{0 & 0 & 0 &  1 &  0 & i &  0&  0& 0 & 0 &  1 &  0 &  i & 0 &   0 &0} \,, \\
	\arow{\epsilon_c^2 ={1\over 8}}{-1 & 0 & 0 & 0 &  0 &  0 &  -i &  0 & 0 &  -1 &  0 &  0 & 0 & 0 & 0 &i} \,, \\
	\arow{\epsilon_s^3={1\over 8}}{{-i}  & 0 & 0 &  0 &  0 &  0 &  1&  0 & 0 &  i &  0 &  0 & 0 & 0 & 0 &1} \,,\\
	\arow{\epsilon_c^3 ={1\over 8}}{0 & 0 & 0 & i &  0 &  -1 &  0&  0 & 0 &  0 &  -i &  0 & 1 & 0 & 0 &0} \,.
\end{alignedrows}
\label{twistedcks}
\fe 

\subsection{Action on the BPS locus}\label{App:bps}
In this appendix, we give some details for the evaluation of the full $AdS_4$ action on the BPS locus. As explained in Section~\ref{Sec:to2d}, it suffices to evaluate the fields at $\tau=0$ and we may also drop all covariant derivatives along $\tau$ (due to the absence of surface defects on $AdS_2$). Once again we will set $L=1$ and restore the units in the end.

We will start by computing the bosonic part of the bulk action.
\ie
S_{b}
=& {1\over 2 g_4^2}\int_{AdS_4}  d^4 x\, \sqrt{g}\, \tr \Bigg(
{1\over 2}F_{MN}F^{MN} -2 \Phi^A \Phi_A -K^m K_m
\Bigg)\,.
\label{ba1}
\fe 
Let us define for convenience
\ie
\CC_\tau\equiv {v^\m \CC_\m\over v_\n v^\n}\,,
\fe
which  at $\tau=0$ satisfies
\ie
\ve \CC_\tau \ve=1\,,
\qquad
\CC_\tau  = {1\over 8 \cos \theta \sinh r }\CC_{\hat 3}   \,.
\fe
Using Gamma matrix identities from Appendix~\ref{App:gm} and the BPS equation for the fermion \eqref{SUSYos}, we can simplify
\ie
K^m K_m
=&-\le({1\over 2}F_{PQ}\varepsilon\tilde \Gamma^{PQ}+ \Phi^B \varepsilon \Lambda \tilde \Gamma_{  B} \ri)  \CC_\tau \le({1\over 2}F_{MN}\Gamma^{MN}\varepsilon+ \Phi^A \tilde \Gamma_{  A} \Lambda \varepsilon\ri)
\\
=&-\Bigg(
\Phi^A \Phi^B \varepsilon \Lambda\tilde \CC_B\CC_\tau\tilde \CC_A \Lambda \varepsilon
+{1\over 4} F_{PQ} F_{MN}\ve \tilde \CC^{PQ} \CC_\tau \CC^{MN} \ve
+  \Phi^B F_{MN} \ve \Lambda\tilde \CC_B \CC_\tau \CC^{MN} \ve
\Bigg)
\\
=&-\Bigg(
-  \Phi^A \Phi_A
+{1\over 8  \cos\theta\sinh r}\Phi_B F_{MN} \ve\Lambda \CC^{BMN \hat 3}\ve +{1\over 4  \cos\theta\sinh r}\Phi^B F_{BN} \ve\Lambda \CC^{N\hat 3}\ve
\\
&
-{1\over 2 } F_{MN} F^{MN}+{1\over 32 \cos\theta \sinh r}F_{PQ}F_{MN} \ve   \CC^{MNPQ\hat 3} \ve
\Bigg)\,.
\fe
From
\ie
&{1\over 4} \tr( F_{MN}F_{PQ} )\ve  \CC^{MNPQ\hat 3} \ve
\\
&=  \ve  \CC^{ \A A\B B\hat 3}\ve D_\A \tr (\Phi_A F_{\B B})
+{1\over 3} \ve \CC^{\A ABC\hat 3}\ve D_\A \tr (\Phi_A F_{BC})
+ \ve  \CC^{\A A \B\ga\hat 3}\ve D_\A \tr (\Phi_A F_{\B\ga})\,,
\fe
where $\A,\B,\ga=1,2,4$ labels the directions along ${1\over 2}AdS_3$,
and using \eqref{CKSeqn} restricted to ${1\over 2}AdS_3$, we have
\ie
& D_\A \tr \Bigg(
\ve \CC^{ \A A\B B\hat 3}\ve  \Phi_A F_{\B B} 
+{1\over 3} \ve \CC^{\A ABC\hat 3}\ve   \Phi_A F_{BC} 
+ \ve \CC^{\A A \B\ga\hat 3}\ve  \Phi_A F_{\B\ga} 
\Bigg)
\\
&={1\over 4} \tr( F_{MN}F_{PQ} )\ve  \CC^{MNPQ\hat 3} \ve+ 
\ve \Lambda \CC^{  A MN \hat 3}\ve  \Phi_A F_{MN} \,.
\label{totald1}
\fe
We also have
\ie
D_\A \tr (\Phi^B \Phi_B \ve \Lambda  \CC^{\A\hat 3} \ve)
=2\tr (\Phi^B F_{\A B})\ve \Lambda   \CC^{\A\hat 3}\ve +3\tr (\Phi^B \Phi_B) \ve\CC^{\hat 3}\ve\,.
\label{totald2}
\fe
Using \eqref{totald1} and \eqref{totald2}, we can write \eqref{ba1} as a total derivative on ${1\over 2}AdS_3$,
\ie
S_b
=& {1\over 8}{2\pi   \over 2 g_4^2}\int_{{1\over 2}AdS_3}   d\theta dr d\phi\, \sqrt{g_{{{1\over 2}AdS_3}} }
\\
&
\times D_\A \tr \Bigg( 
\ve   \CC^{ \A A\B B\hat 3}\ve  \Phi_A F_{\B B} 
+{1\over 3} \ve \CC^{\A ABC\hat 3}\ve   \Phi_A F_{BC} 
+ \ve \CC^{\A A \B\ga\hat 3}\ve  \Phi_A F_{\B\ga} 
-  \Phi^B \Phi_B \ve \Lambda \CC^{\A\hat 3} \ve 
\Bigg)
\\
\equiv& S_I +S_{II}\,.
\label{bulktd}
\fe
Recall that the boundary of the ${1\over 2} AdS_3$ has two components: the $AdS_2$ slice at $\theta={\pi\over 2}$ and a two-disk $D^2$ at $r=r_0$, which intersect at the asymptotic boundary $S^1$ of $AdS_2$. Evaluation of the total derivative gives rise to two terms on $AdS_2$ and $D^2$ respectively, which we denote by $S_{I}$ and $S_{II}$ in \eqref{bulktd}. 

Using the explicit form of the Killing spinor $\ve$ \eqref{localizeQ}, we obtain
\ie
S_{I} 
=& -{ \pi \over   g_4^2}\int_{AdS_2} dr d\varphi\, \sqrt{g_{AdS_2} }    \,
\tr \Bigg( 
2\tilde \Phi {*\hat F} +\tilde\Phi^2-\hat \Phi_5^2 
\Bigg)
-{ \pi \over   g_4^2}\int_{r=r_0} d\varphi\,{ \hat \Phi_\varphi(\hat \Phi_\varphi-  i\sinh r \Phi_6 )\over \sinh r}\,,
\fe
where the fields on $AdS_2$ are twisted combination of the 4d gauge fields and scalar fields \eqref{tphi05} and \eqref{tphi6789}.

On the other hand
\ie
S_{II}
 =& -{1\over 8}{2\pi \over 2 g_4^2}\int_{D_2}d\theta d\phi\,  \sqrt{g_{D_2} }  \, \tr \le( 
\ve \CC^{ \hat 3\hat 4 A\m B}\ve  \Phi_A F_{\m B} 
+16\cos\theta\sinh r\Phi_9 F_{78}\ri.\\
&\le.
+ \ve \CC^{\hat 3\hat 4 a ij}\ve   \Phi_a F_{ij} 
+\ve \CC^{ \hat 3\hat 4 a\m\n}\ve  \Phi_a F_{\m\n} 
+8\cos\theta \sinh r \Phi^B \Phi_B 
\ri) \,,
\fe
where $i,j=5,6,0$. As we will see below, this will combine with $S_B$
\ie
S_{B }
=
{1\over 2 g_4^2}\int_{S_3}d^3 x\,  \sqrt{\ga}\, \tr \Bigg({1\over 2}\Psi \Lambda\Psi+
2K^{a-5}\Phi_a-2D_r\Phi^a \Phi_a-4\Phi_9 F_{78}-2\Phi^a\Phi_a+\Phi^A\Phi_A
\Bigg)
\fe
to become yet another total derivative on $D^2$. 
Let us denote $S_B$ evaluated on the BPS locus by $S_{III}$. 

From the $\D_\ve\Psi=0$ on the boundary $D^2$, we have
\ie
K^{a-5}&
=\n^{a-5}\CC_\tau \n_m K^m
\\
&=
-{1\over 2}F_{MN}\ve \CC^{a\hat 4}\CC_\tau\CC^{MN}\ve+2\Phi_A\ve \CC^{a\hat 4}\CC_\tau\tilde \CC^{A}\tilde \ve
\\
&=
-{1\over 2}F_{MN}\ve \CC^{a\hat 4}\CC_\tau\CC^{MN}\ve+\Phi_a\,.
\fe
Hence
\ie
2\Phi_aK^{a-5} 
=&
- F_{MN}\Phi_a\ve \CC^{a\hat 4}\CC_\tau\CC^{MN}\ve +2 \Phi^a\Phi_a \,,
\fe
where the first term can be further simplified by Gamma matrix identities,
\ie
&F_{MN}\Phi_a\ve \CC^{a\hat 4}\CC_\tau\CC^{MN}\ve
\\
&=F_{ij}\Phi_a\ve \CC^{a\hat 4 ij}\CC_\tau\ve+F_{\m\n}\Phi_a\ve \CC^{a\hat 4 \m\n}\CC_\tau\ve
+2F_{\m i}\Phi_a\ve \CC^{a\hat 4 \m i}\CC_\tau\ve- 2F_{  ra}\Phi_a -6F_{78}\Phi_9\,.
\fe
Therefore
\ie
S_{III}
 =&{1\over 8}{2\pi\over 2 g_4^2}\int d\theta d\varphi\, \sqrt{g_{D_2}} 
 \tr \le(
F_{ij}\Phi_a\ve \CC^{\hat 3\hat 4a ij} \ve+F_{\m\n}\Phi_a\ve \CC^{\hat 3\hat 4 a \m\n} \ve\ri.\\
&\le.+2F_{\m i}\Phi_a\ve \CC^{\hat 3\hat 4 a \m i}\ve+16 \cos\theta \sinh r\Phi_9 F_{78}+8\cos\theta \sinh r \Phi^A\Phi_A
\ri) \,.
\fe
Putting together the two terms on $D^2$, we obtain
\ie
S_{II}+S_{III}
=& {1\over 8}{2\pi\over 2 g_4^2}\int d\theta d\varphi \,\sqrt{g_{D^2}}
\tr \left(
F_{\m i}\Phi_a\ve \CC^{\hat3\hat 4 a  \m i}\ve 
+\ve\tilde\CC^{\hat 3\hat 4  a\m i}\ve  \Phi_i F_{\m a} 
\right)
\\
=& {1\over 8}{2\pi\over 2 g_4^2}\int d\theta d\varphi \,\sqrt{g_{D^2}}
D_\m \tr(
\Phi_i\Phi_a\ve \CC^{\hat 3\hat 4 a\m i}\ve 
)
\\
=& {2\pi\over 2 g_4^2}\int d\varphi\,\sqrt{g_{S^1}}  \, i\hat \Phi_\varphi \Phi_6 \,,
\fe
where we dropped terms that vanish in the large $r$ limit in the second and third equalities.

Now the total 2d action is given by
\ie
S_\text{2d}=&S_{I}+S_{II}+S_{III}
\\
=& -{ \pi \over   g_4^2}\int_{AdS_2} dr d\varphi\, \sqrt{g_{AdS_2} }    \,
\tr \Bigg( 
2\tilde \Phi {*\hat F} +\tilde\Phi^2-\hat \Phi_5^2 
\Bigg)
-{ \pi \over   g_4^2}\int_{r=r_0} d\varphi\,{ \hat \Phi_\varphi(\hat \Phi_\varphi-2 i\sinh r \Phi_6 )\over \sinh r}\,.
\label{2daction}
\fe
Recall that the bulk equation of motion in $AdS_2$ says that
\ie
\tilde\Phi=\Phi_9 \cosh r+i (\Phi_8  \sin \phi +i \Phi_7 \cos \phi)\sinh r +i \Phi_6=-{\hat F_{r\varphi}\over \sinh r}\,.
\label{tPhieom}
\fe
Hence we should choose the boundary condition for $\Phi_6$ that is consistent with \eqref{tPhieom}.
So asymptotically we have\footnote{Here we used that $\tilde\Phi\approx {\hat\Phi_{\varphi}\over \sinh r}+i \Phi_6$ up to $\cO(e^{-2r})$.}
\ie
&i\Phi_6\approx {\hat \Phi_{\varphi}- \hat F_{r\varphi} \over \sinh r }\,.
\fe
Therefore the boundary terms in \eqref{2daction} becomes 
\ie
S_\text{2d}^B
=&-{ \pi\over   g_4^2}\int_{r=r_0}  d\varphi \, {\hat \Phi_\varphi  (\hat \Phi_{\varphi}- 2\hat F_{r\varphi})\over  \sinh r}\,,
\fe
which can also be rewritten as 
\ie
S_\text{2d}^B
=&-{ \pi\over   g_4^2}\int_{r=r_0}  d\varphi \, {\hat A_\varphi  (\hat A_{\varphi}- 2\hat F_{r\varphi})\over  \sinh r}
\fe
using $\hat A_\varphi=A_\varphi+\hat{\Phi}_{\varphi}$ and the asymptotic behavior of the fields \eqref{fieldsab}.

\subsection{Gamma matrix conventions}\label{App:gm}
Here we record our Gamma matrix conventions. We largely follow \cite{Pestun:2009nn}. In this subsection all indices are taken to be flat. The 10d gamma matrices in the chiral basis are chosen to be symmetric $16\times 16$ matrices $\CC_M$ and $\tilde\CC_N$ related by complex conjugation
\ie
\tilde \CC_N=\CC_N^*\,,
\fe
which satisfy
\ie
\CC_M \tilde \CC_N+\CC_N \tilde \CC_M=2\D_{MN} {\bf 1}_{16}\,.
\fe
More explicitly
\ie
\CC_{M=2,\dots,9}=&\begin{pmatrix}
	0 && E_M^T \\ E_M && 0
\end{pmatrix}
\,,\qquad
\CC_1=&\begin{pmatrix}
	1_{8} && 0  \\ 0  && -1_8
\end{pmatrix}
\,,\qquad
\CC_0=&\begin{pmatrix}
	i 1_{8} && 0  \\ 0  && - i 1_8
\end{pmatrix}\,.
\fe
Here, $E_M$ with $M=2,3,\dots,9$ are $8\times 8$ matrices representing the left multiplication of octonions $e_M$ in the octonion algebra $\mathbb O$ with $e_9$ chosen to be the identity. The explicit form of $E_M$ is given below:\footnote{For careful readers, our convention for $E_6$ and $E_8$ are different from that in \cite{Pestun:2009nn}.}
 \es{GotEs}{
E_2=&
\footnotesize\begin{pmatrix}
	0 & -1 & 0 & 0 & 0 & 0 & 0 & 0 \\
	1 & 0 & 0 & 0 & 0 & 0 & 0 & 0 \\
	0 & 0 & 0 & -1 & 0 & 0 & 0 & 0 \\
	0 & 0 & 1 & 0 & 0 & 0 & 0 & 0 \\
	0 & 0 & 0 & 0 & 0 & -1 & 0 & 0 \\
	0 & 0 & 0 & 0 & 1 & 0 & 0 & 0 \\
	0 & 0 & 0 & 0 & 0 & 0 & 0 & 1 \\
	0 & 0 & 0 & 0 & 0 & 0 & -1 & 0 \\
\end{pmatrix}\,,
\qquad E_3=
\footnotesize\begin{pmatrix}
	0 & 0 & -1 & 0 & 0 & 0 & 0 & 0 \\
	0 & 0 & 0 & 1 & 0 & 0 & 0 & 0 \\
	1 & 0 & 0 & 0 & 0 & 0 & 0 & 0 \\
	0 & -1 & 0 & 0 & 0 & 0 & 0 & 0 \\
	0 & 0 & 0 & 0 & 0 & 0 & -1 & 0 \\
	0 & 0 & 0 & 0 & 0 & 0 & 0 & -1 \\
	0 & 0 & 0 & 0 & 1 & 0 & 0 & 0 \\
	0 & 0 & 0 & 0 & 0 & 1 & 0 & 0 \\
\end{pmatrix}\,,
}
\es{GotE4}{
E_4=&
\footnotesize\begin{pmatrix}
	0 & 0 & 0 & -1 & 0 & 0 & 0 & 0 \\
	0 & 0 & -1 & 0 & 0 & 0 & 0 & 0 \\
	0 & 1 & 0 & 0 & 0 & 0 & 0 & 0 \\
	1 & 0 & 0 & 0 & 0 & 0 & 0 & 0 \\
	0 & 0 & 0 & 0 & 0 & 0 & 0 & -1 \\
	0 & 0 & 0 & 0 & 0 & 0 & 1 & 0 \\
	0 & 0 & 0 & 0 & 0 & -1 & 0 & 0 \\
	0 & 0 & 0 & 0 & 1 & 0 & 0 & 0 \\
\end{pmatrix}
\,,\qquad
E_5=
\footnotesize\begin{pmatrix}
	0 & 0 & 0 & 0 & -1 & 0 & 0 & 0 \\
	0 & 0 & 0 & 0 & 0 & 1 & 0 & 0 \\
	0 & 0 & 0 & 0 & 0 & 0 & 1 & 0 \\
	0 & 0 & 0 & 0 & 0 & 0 & 0 & 1 \\
	1 & 0 & 0 & 0 & 0 & 0 & 0 & 0 \\
	0 & -1 & 0 & 0 & 0 & 0 & 0 & 0 \\
	0 & 0 & -1 & 0 & 0 & 0 & 0 & 0 \\
	0 & 0 & 0 & -1 & 0 & 0 & 0 & 0 \\
\end{pmatrix}\,,
}
 \es{GotE6}{
E_6=&
\footnotesize\begin{pmatrix}
	0 & 0 & 0 & 0 & 0 & 0 & 0 & -1 \\
	0 & 0 & 0 & 0 & 0 & 0 & 1 & 0 \\
	0 & 0 & 0 & 0 & 0 & -1 & 0 & 0 \\
	0 & 0 & 0 & 0 & -1 & 0 & 0 & 0 \\
	0 & 0 & 0 & 1 & 0 & 0 & 0 & 0 \\
	0 & 0 & 1 & 0 & 0 & 0 & 0 & 0 \\
	0 & -1 & 0 & 0 & 0 & 0 & 0 & 0 \\
	1 & 0 & 0 & 0 & 0 & 0 & 0 & 0 \\
\end{pmatrix}\,, 
\qquad
E_7=
\footnotesize\begin{pmatrix}
	0 & 0 & 0 & 0 & 0 & 0 & -1 & 0 \\
	0 & 0 & 0 & 0 & 0 & 0 & 0 & -1 \\
	0 & 0 & 0 & 0 & -1 & 0 & 0 & 0 \\
	0 & 0 & 0 & 0 & 0 & 1 & 0 & 0 \\
	0 & 0 & 1 & 0 & 0 & 0 & 0 & 0 \\
	0 & 0 & 0 & -1 & 0 & 0 & 0 & 0 \\
	1 & 0 & 0 & 0 & 0 & 0 & 0 & 0 \\
	0 & 1 & 0 & 0 & 0 & 0 & 0 & 0 \\
\end{pmatrix}\,,
}
\es{GotE8}{
E_8=&
\footnotesize\begin{pmatrix}
	0 & 0 & 0 & 0 & 0 & 1 & 0 & 0 \\
	0 & 0 & 0 & 0 & 1 & 0 & 0 & 0 \\
	0 & 0 & 0 & 0 & 0 & 0 & 0 & -1 \\
	0 & 0 & 0 & 0 & 0 & 0 & 1 & 0 \\
	0 & -1 & 0 & 0 & 0 & 0 & 0 & 0 \\
	-1 & 0 & 0 & 0 & 0 & 0 & 0 & 0 \\
	0 & 0 & 0 & -1 & 0 & 0 & 0 & 0 \\
	0 & 0 & 1 & 0 & 0 & 0 & 0 & 0 \\
\end{pmatrix}\,,
\qquad
E_9=
\footnotesize\begin{pmatrix}
	1 & 0 & 0 & 0 & 0 & 0 & 0 & 0 \\
	0 & 1 & 0 & 0 & 0 & 0 & 0 & 0 \\
	0 & 0 & 1 & 0 & 0 & 0 & 0 & 0 \\
	0 & 0 & 0 & 1 & 0 & 0 & 0 & 0 \\
	0 & 0 & 0 & 0 & 1 & 0 & 0 & 0 \\
	0 & 0 & 0 & 0 & 0 & 1 & 0 & 0 \\
	0 & 0 & 0 & 0 & 0 & 0 & 1 & 0 \\
	0 & 0 & 0 & 0 & 0 & 0 & 0 & 1 \\
\end{pmatrix}\,.
}
The higher rank Gamma matrices are defined by
\ie
&\CC^{MN}\equiv \tilde \CC^{[M}\CC^{N]}\,,\qquad \tilde\CC^{MN}\equiv \CC^{[M} \tilde\CC^{N]}\,,\qquad 
\CC^{MNP}\equiv  \CC^{[M}\tilde\CC^{N}\CC^{P]}\,,\qquad \tilde\CC^{MNP}\equiv \tilde \CC^{[M}\CC^{N}\tilde\CC^{P]}\,,
\\
&\CC^{MNPQ}\equiv \tilde \CC^{[M}\CC^{N}\tilde\CC^{P}\CC^{Q]}\,,\qquad \tilde\CC^{MNPQ}\equiv \CC^{[M}\tilde \CC^{N}\CC^{P}\tilde \CC^{Q]}\,, 
\\
&
\CC^{MNPQR}\equiv  \CC^{[M}\tilde\CC^{N}\CC^{P}\tilde\CC^{Q}\CC^{R]}\,,\qquad \tilde\CC^{MNPQR}\equiv \tilde \CC^{[M}\CC^{N}\tilde\CC^{P}\CC^{Q}\tilde\CC^{R]}\,.
\fe
In particular $\CC^{MNP},\,\tilde\CC^{MNP} $ are anti-symmetric and $\CC^{MNPQR},\,\tilde\CC^{MNPQR}$ are symmetric. We also have
\ie
(\CC^{MN})^T=-\tilde \CC^{MN}\,,\qquad (\CC^{MNPQ})^T=\tilde \CC^{MNPQ}\,.
\fe

Below we list some useful Gamma matrix identities,
\ie
&\CC_{PQ}\CC^{MN}=-2\D^M_{[P}\D^N_{Q]}-4\D^{[M}_{[P}\CC_{Q]}{}^{N]}+\CC_{PQ}{}^{MN}\,,
\\
&\CC_M\CC_{NP}=2\D_{M[N}\CC_{P]}+\CC_{MNP}\,,
\\
&\CC_{ABC}\CC_{MN}=\CC_{ABCMN}-3(\D_{N[A}\CC_{BC]M}-\D_{M[A}\CC_{BC]N})+6\D_{M[A}\CC_B \D_{C]N}\,.
\fe

\bibliographystyle{ssg}
\bibliography{LargeN}

\begingroup\raggedright\begin{thebibliography}{10}

\bibitem{Maldacena:1997re}
J.~M. Maldacena, ``{The Large N limit of superconformal field theories and
  supergravity},'' {\em Int. J. Theor. Phys.} {\bf 38} (1999) 1113--1133,
  \href{http://xxx.lanl.gov/abs/hep-th/9711200}{{\tt hep-th/9711200}}. [Adv.
  Theor. Math. Phys.2,231(1998)].

\bibitem{Witten:1998qj}
E.~Witten, ``{Anti-de Sitter space and holography},'' {\em Adv. Theor. Math.
  Phys.} {\bf 2} (1998) 253--291,
  \href{http://xxx.lanl.gov/abs/hep-th/9802150}{{\tt hep-th/9802150}}.

\bibitem{Gubser:1998bc}
S.~S. Gubser, I.~R. Klebanov, and A.~M. Polyakov, ``{Gauge theory correlators
  from noncritical string theory},'' {\em Phys. Lett.} {\bf B428} (1998)
  105--114, \href{http://xxx.lanl.gov/abs/hep-th/9802109}{{\tt
  hep-th/9802109}}.

\bibitem{Aharony:2008ug}
O.~Aharony, O.~Bergman, D.~L. Jafferis, and J.~Maldacena, ``{N=6 superconformal
  Chern-Simons-matter theories, M2-branes and their gravity duals},'' {\em
  JHEP} {\bf 10} (2008) 091, \href{http://xxx.lanl.gov/abs/0806.1218}{{\tt
  0806.1218}}.

\bibitem{Klebanov:2002ja}
I.~R. Klebanov and A.~M. Polyakov, ``{AdS dual of the critical O(N) vector
  model},'' {\em Phys. Lett.} {\bf B550} (2002) 213--219,
  \href{http://xxx.lanl.gov/abs/hep-th/0210114}{{\tt hep-th/0210114}}.

\bibitem{Giombi:2009wh}
S.~Giombi and X.~Yin, ``{Higher Spin Gauge Theory and Holography: The
  Three-Point Functions},'' {\em JHEP} {\bf 09} (2010) 115,
  \href{http://xxx.lanl.gov/abs/0912.3462}{{\tt 0912.3462}}.

\bibitem{Vasiliev:1990en}
M.~A. Vasiliev, ``{Consistent equation for interacting gauge fields of all
  spins in (3+1)-dimensions},'' {\em Phys. Lett.} {\bf B243} (1990) 378--382.

\bibitem{Vasiliev:2003ev}
M.~A. Vasiliev, ``{Nonlinear equations for symmetric massless higher spin
  fields in (A)dS(d)},'' {\em Phys. Lett.} {\bf B567} (2003) 139--151,
  \href{http://xxx.lanl.gov/abs/hep-th/0304049}{{\tt hep-th/0304049}}.

\bibitem{Giombi:2016ejx}
S.~Giombi, ``{TASI Lectures on the Higher Spin - CFT duality},'' in {\em
  {Proceedings, Theoretical Advanced Study Institute in Elementary Particle
  Physics: New Frontiers in Fields and Strings (TASI 2015): Boulder, CO, USA,
  June 1-26, 2015}}, pp.~137--214, 2017.
\newblock \href{http://xxx.lanl.gov/abs/1607.02967}{{\tt 1607.02967}}.

\bibitem{Bonetti:2016nma}
F.~Bonetti and L.~Rastelli, ``{Supersymmetric Localization in AdS$_5$ and the
  Protected Chiral Algebra},'' \href{http://xxx.lanl.gov/abs/1612.06514}{{\tt
  1612.06514}}.

\bibitem{Maldacena:1998uz}
J.~M. Maldacena, J.~Michelson, and A.~Strominger, ``{Anti-de Sitter
  fragmentation},'' {\em JHEP} {\bf 02} (1999) 011,
  \href{http://xxx.lanl.gov/abs/hep-th/9812073}{{\tt hep-th/9812073}}.

\bibitem{Jensen:2011su}
K.~Jensen, S.~Kachru, A.~Karch, J.~Polchinski, and E.~Silverstein, ``{Towards a
  holographic marginal Fermi liquid},'' {\em Phys. Rev.} {\bf D84} (2011)
  126002, \href{http://xxx.lanl.gov/abs/1105.1772}{{\tt 1105.1772}}.

\bibitem{Almheiri:2014cka}
A.~Almheiri and J.~Polchinski, ``{Models of AdS$_{2}$ backreaction and
  holography},'' {\em JHEP} {\bf 11} (2015) 014,
  \href{http://xxx.lanl.gov/abs/1402.6334}{{\tt 1402.6334}}.

\bibitem{Iqbal:2011ae}
N.~Iqbal, H.~Liu, and M.~Mezei, ``{Lectures on holographic non-Fermi liquids
  and quantum phase transitions},'' in {\em {Proceedings, Theoretical Advanced
  Study Institute in Elementary Particle Physics (TASI 2010). String Theory and
  Its Applications: From meV to the Planck Scale: Boulder, Colorado, USA, June
  1-25, 2010}}, pp.~707--816, 2011.
\newblock \href{http://xxx.lanl.gov/abs/1110.3814}{{\tt 1110.3814}}.

\bibitem{Sachdev:1992fk}
S.~Sachdev and J.~Ye, ``{Gapless spin fluid ground state in a random, quantum
  Heisenberg magnet},'' {\em Phys. Rev. Lett.} {\bf 70} (1993) 3339,
  \href{http://xxx.lanl.gov/abs/cond-mat/9212030}{{\tt cond-mat/9212030}}.

\bibitem{Kitaev}
A.~Kitaev, ``{A simple model of quantum holography}.''
  http://online.kitp.ucsb.edu/online/entangled15/kitaev, April, 2015.

\bibitem{Polchinski:2016xgd}
J.~Polchinski and V.~Rosenhaus, ``{The Spectrum in the Sachdev-Ye-Kitaev
  Model},'' {\em JHEP} {\bf 04} (2016) 001,
  \href{http://xxx.lanl.gov/abs/1601.06768}{{\tt 1601.06768}}.

\bibitem{Jevicki:2016bwu}
A.~Jevicki, K.~Suzuki, and J.~Yoon, ``{Bi-Local Holography in the SYK Model},''
  {\em JHEP} {\bf 07} (2016) 007,
  \href{http://xxx.lanl.gov/abs/1603.06246}{{\tt 1603.06246}}.

\bibitem{Maldacena:2016hyu}
J.~Maldacena and D.~Stanford, ``{Remarks on the Sachdev-Ye-Kitaev model},''
  {\em Phys. Rev.} {\bf D94} (2016), no.~10 106002,
  \href{http://xxx.lanl.gov/abs/1604.07818}{{\tt 1604.07818}}.

\bibitem{Gurau:2009tw}
R.~Gurau, ``{Colored Group Field Theory},'' {\em Commun. Math. Phys.} {\bf 304}
  (2011) 69--93, \href{http://xxx.lanl.gov/abs/0907.2582}{{\tt 0907.2582}}.

\bibitem{Witten:2016iux}
E.~Witten, ``{An SYK-Like Model Without Disorder},''
  \href{http://xxx.lanl.gov/abs/1610.09758}{{\tt 1610.09758}}.

\bibitem{Klebanov:2016xxf}
I.~R. Klebanov and G.~Tarnopolsky, ``{Uncolored random tensors, melon diagrams,
  and the Sachdev-Ye-Kitaev models},'' {\em Phys. Rev.} {\bf D95} (2017), no.~4
  046004, \href{http://xxx.lanl.gov/abs/1611.08915}{{\tt 1611.08915}}.

\bibitem{Jensen:2016pah}
K.~Jensen, ``{Chaos in AdS$_2$ Holography},'' {\em Phys. Rev. Lett.} {\bf 117}
  (2016), no.~11 111601, \href{http://xxx.lanl.gov/abs/1605.06098}{{\tt
  1605.06098}}.

\bibitem{Maldacena:2016upp}
J.~Maldacena, D.~Stanford, and Z.~Yang, ``{Conformal symmetry and its breaking
  in two dimensional Nearly Anti-de-Sitter space},'' {\em PTEP} {\bf 2016}
  (2016), no.~12 12C104, \href{http://xxx.lanl.gov/abs/1606.01857}{{\tt
  1606.01857}}.

\bibitem{Engelsoy:2016xyb}
J.~Engels{\"o}y, T.~G. Mertens, and H.~Verlinde, ``{An investigation of
  AdS$_{2}$ backreaction and holography},'' {\em JHEP} {\bf 07} (2016) 139,
  \href{http://xxx.lanl.gov/abs/1606.03438}{{\tt 1606.03438}}.

\bibitem{Castro:2008ms}
A.~Castro, D.~Grumiller, F.~Larsen, and R.~McNees, ``{Holographic Description
  of AdS(2) Black Holes},'' {\em JHEP} {\bf 11} (2008) 052,
  \href{http://xxx.lanl.gov/abs/0809.4264}{{\tt 0809.4264}}.

\bibitem{Grumiller:2014oha}
D.~Grumiller, R.~McNees, and J.~Salzer, ``{Cosmological constant as confining
  U(1) charge in two-dimensional dilaton gravity},'' {\em Phys. Rev.} {\bf D90}
  (2014), no.~4 044032, \href{http://xxx.lanl.gov/abs/1406.7007}{{\tt
  1406.7007}}.

\bibitem{Grumiller:2015vaa}
D.~Grumiller, J.~Salzer, and D.~Vassilevich, ``{AdS$_{2}$ holography is
  (non-)trivial for (non-)constant dilaton},'' {\em JHEP} {\bf 12} (2015) 015,
  \href{http://xxx.lanl.gov/abs/1509.08486}{{\tt 1509.08486}}.

\bibitem{Cvetic:2016eiv}
M.~Cveti{\v c} and I.~Papadimitriou, ``{AdS$_{2}$ holographic dictionary},''
  {\em JHEP} {\bf 12} (2016) 008,
  \href{http://xxx.lanl.gov/abs/1608.07018}{{\tt 1608.07018}}.

\bibitem{Klebanov:1999tb}
I.~R. Klebanov and E.~Witten, ``{AdS/CFT correspondence and symmetry
  breaking},'' {\em Nucl. Phys.} {\bf B556} (1999) 89--114,
  \href{http://xxx.lanl.gov/abs/hep-th/9905104}{{\tt hep-th/9905104}}.

\bibitem{Chester:2014mea}
S.~M. Chester, J.~Lee, S.~S. Pufu, and R.~Yacoby, ``{Exact Correlators of BPS
  Operators from the 3d Superconformal Bootstrap},'' {\em JHEP} {\bf 03} (2015)
  130, \href{http://xxx.lanl.gov/abs/1412.0334}{{\tt 1412.0334}}.

\bibitem{Beem:2016cbd}
C.~Beem, W.~Peelaers, and L.~Rastelli, ``{Deformation quantization and
  superconformal symmetry in three dimensions},''
  \href{http://xxx.lanl.gov/abs/1601.05378}{{\tt 1601.05378}}.

\bibitem{Dedushenko:2016jxl}
M.~Dedushenko, S.~S. Pufu, and R.~Yacoby, ``{A one-dimensional theory for Higgs
  branch operators},'' \href{http://xxx.lanl.gov/abs/1610.00740}{{\tt
  1610.00740}}.

\bibitem{Pestun:2009nn}
V.~Pestun, ``{Localization of the four-dimensional N=4 SYM to a two-sphere and
  1/8 BPS Wilson loops},'' {\em JHEP} {\bf 12} (2012) 067,
  \href{http://xxx.lanl.gov/abs/0906.0638}{{\tt 0906.0638}}.

\bibitem{Giombi:2009ds}
S.~Giombi and V.~Pestun, ``{Correlators of local operators and 1/8 BPS Wilson
  loops on S**2 from 2d YM and matrix models},'' {\em JHEP} {\bf 10} (2010)
  033, \href{http://xxx.lanl.gov/abs/0906.1572}{{\tt 0906.1572}}.

\bibitem{Giombi:2009ek}
S.~Giombi and V.~Pestun, ``{The 1/2 BPS 't Hooft loops in N=4 SYM as instantons
  in 2d Yang-Mills},'' {\em J. Phys.} {\bf A46} (2013) 095402,
  \href{http://xxx.lanl.gov/abs/0909.4272}{{\tt 0909.4272}}.

\bibitem{Vasiliev:1992av}
M.~A. Vasiliev, ``{More on equations of motion for interacting massless fields
  of all spins in (3+1)-dimensions},'' {\em Phys. Lett.} {\bf B285} (1992)
  225--234.

\bibitem{Vasiliev:1995dn}
M.~A. Vasiliev, ``{Higher spin gauge theories in four-dimensions,
  three-dimensions, and two-dimensions},'' {\em Int. J. Mod. Phys.} {\bf D5}
  (1996) 763--797, \href{http://xxx.lanl.gov/abs/hep-th/9611024}{{\tt
  hep-th/9611024}}.

\bibitem{Vasiliev:1999ba}
M.~A. Vasiliev, ``{Higher spin gauge theories: Star product and AdS space},''
  \href{http://xxx.lanl.gov/abs/hep-th/9910096}{{\tt hep-th/9910096}}.

\bibitem{Engquist:2002vr}
J.~Engquist, E.~Sezgin, and P.~Sundell, ``{On N=1, N=2, N=4 higher spin gauge
  theories in four-dimensions},'' {\em Class. Quant. Grav.} {\bf 19} (2002)
  6175--6196, \href{http://xxx.lanl.gov/abs/hep-th/0207101}{{\tt
  hep-th/0207101}}.

\bibitem{Engquist:2002gy}
J.~Engquist, E.~Sezgin, and P.~Sundell, ``{Superspace formulation of 4-D higher
  spin gauge theory},'' {\em Nucl. Phys.} {\bf B664} (2003) 439--456,
  \href{http://xxx.lanl.gov/abs/hep-th/0211113}{{\tt hep-th/0211113}}.

\bibitem{Chang:2012kt}
C.-M. Chang, S.~Minwalla, T.~Sharma, and X.~Yin, ``{ABJ Triality: from Higher
  Spin Fields to Strings},'' {\em J. Phys.} {\bf A46} (2013) 214009,
  \href{http://xxx.lanl.gov/abs/1207.4485}{{\tt 1207.4485}}.

\bibitem{Bashkirov:2010kz}
D.~Bashkirov and A.~Kapustin, ``{Supersymmetry enhancement by monopole
  operators},'' {\em JHEP} {\bf 05} (2011) 015,
  \href{http://xxx.lanl.gov/abs/1007.4861}{{\tt 1007.4861}}.

\bibitem{Jafferis:2011zi}
D.~L. Jafferis, I.~R. Klebanov, S.~S. Pufu, and B.~R. Safdi, ``{Towards the
  F-Theorem: N=2 Field Theories on the Three-Sphere},'' {\em JHEP} {\bf 06}
  (2011) 102, \href{http://xxx.lanl.gov/abs/1103.1181}{{\tt 1103.1181}}.

\bibitem{Klebanov:2011gs}
I.~R. Klebanov, S.~S. Pufu, and B.~R. Safdi, ``{F-Theorem without
  Supersymmetry},'' {\em JHEP} {\bf 10} (2011) 038,
  \href{http://xxx.lanl.gov/abs/1105.4598}{{\tt 1105.4598}}.

\bibitem{Myers:2010xs}
R.~C. Myers and A.~Sinha, ``{Seeing a c-theorem with holography},'' {\em Phys.
  Rev.} {\bf D82} (2010) 046006, \href{http://xxx.lanl.gov/abs/1006.1263}{{\tt
  1006.1263}}.

\bibitem{Myers:2010tj}
R.~C. Myers and A.~Sinha, ``{Holographic c-theorems in arbitrary dimensions},''
  {\em JHEP} {\bf 01} (2011) 125, \href{http://xxx.lanl.gov/abs/1011.5819}{{\tt
  1011.5819}}.

\bibitem{Casini:2012ei}
H.~Casini and M.~Huerta, ``{On the RG running of the entanglement entropy of a
  circle},'' {\em Phys. Rev.} {\bf D85} (2012) 125016,
  \href{http://xxx.lanl.gov/abs/1202.5650}{{\tt 1202.5650}}.

\bibitem{Pufu:2016zxm}
S.~S. Pufu, ``{The F-Theorem and F-Maximization},'' 2016.
\newblock \href{http://xxx.lanl.gov/abs/1608.02960}{{\tt 1608.02960}}.

\bibitem{Affleck:1991tk}
I.~Affleck and A.~W.~W. Ludwig, ``{Universal noninteger 'ground state
  degeneracy' in critical quantum systems},'' {\em Phys. Rev. Lett.} {\bf 67}
  (1991) 161--164.

\bibitem{Affleck:1992ng}
I.~Affleck and A.~W.~W. Ludwig, ``{Exact conformal-field-theory results on the
  multichannel Kondo effect: Single-fermion Green's function, self-energy, and
  resistivity},'' {\em Phys. Rev.} {\bf B48} (1993), no.~10 7297.

\bibitem{Giombi:2014xxa}
S.~Giombi and I.~R. Klebanov, ``{Interpolating between $a$ and $F$},'' {\em
  JHEP} {\bf 03} (2015) 117, \href{http://xxx.lanl.gov/abs/1409.1937}{{\tt
  1409.1937}}.

\bibitem{Kapustin:2009kz}
A.~Kapustin, B.~Willett, and I.~Yaakov, ``{Exact Results for Wilson Loops in
  Superconformal Chern-Simons Theories with Matter},'' {\em JHEP} {\bf 03}
  (2010) 089, \href{http://xxx.lanl.gov/abs/0909.4559}{{\tt 0909.4559}}.

\bibitem{Herzog:2010hf}
C.~P. Herzog, I.~R. Klebanov, S.~S. Pufu, and T.~Tesileanu, ``{Multi-Matrix
  Models and Tri-Sasaki Einstein Spaces},'' {\em Phys. Rev.} {\bf D83} (2011)
  046001, \href{http://xxx.lanl.gov/abs/1011.5487}{{\tt 1011.5487}}.

\bibitem{Santamaria:2010dm}
R.~C. Santamaria, M.~Marino, and P.~Putrov, ``{Unquenched flavor and tropical
  geometry in strongly coupled Chern-Simons-matter theories},'' {\em JHEP} {\bf
  10} (2011) 139, \href{http://xxx.lanl.gov/abs/1011.6281}{{\tt 1011.6281}}.

\bibitem{Marino:2011eh}
M.~Marino and P.~Putrov, ``{ABJM theory as a Fermi gas},'' {\em J. Stat. Mech.}
  {\bf 1203} (2012) P03001, \href{http://xxx.lanl.gov/abs/1110.4066}{{\tt
  1110.4066}}.

\bibitem{Mezei:2013gqa}
M.~Mezei and S.~S. Pufu, ``{Three-sphere free energy for classical gauge
  groups},'' {\em JHEP} {\bf 02} (2014) 037,
  \href{http://xxx.lanl.gov/abs/1312.0920}{{\tt 1312.0920}}.

\bibitem{Grassi:2014vwa}
A.~Grassi and M.~Marino, ``{M-theoretic matrix models},'' {\em JHEP} {\bf 02}
  (2015) 115, \href{http://xxx.lanl.gov/abs/1403.4276}{{\tt 1403.4276}}.

\bibitem{Drukker:2010nc}
N.~Drukker, M.~Marino, and P.~Putrov, ``{From weak to strong coupling in ABJM
  theory},'' {\em Commun. Math. Phys.} {\bf 306} (2011) 511--563,
  \href{http://xxx.lanl.gov/abs/1007.3837}{{\tt 1007.3837}}.

\bibitem{Witten:2003ya}
E.~Witten, ``{SL(2,Z) action on three-dimensional conformal field theories with
  Abelian symmetry},'' \href{http://xxx.lanl.gov/abs/hep-th/0307041}{{\tt
  hep-th/0307041}}.

\bibitem{Marolf:2006nd}
D.~Marolf and S.~F. Ross, ``{Boundary Conditions and New Dualities: Vector
  Fields in AdS/CFT},'' {\em JHEP} {\bf 11} (2006) 085,
  \href{http://xxx.lanl.gov/abs/hep-th/0606113}{{\tt hep-th/0606113}}.

\bibitem{Beem:2013sza}
C.~Beem, M.~Lemos, P.~Liendo, W.~Peelaers, L.~Rastelli, and B.~C. van Rees,
  ``{Infinite Chiral Symmetry in Four Dimensions},'' {\em Commun. Math. Phys.}
  {\bf 336} (2015), no.~3 1359--1433,
  \href{http://xxx.lanl.gov/abs/1312.5344}{{\tt 1312.5344}}.

\bibitem{Klebanov:1996un}
I.~R. Klebanov and A.~A. Tseytlin, ``{Entropy of near extremal black
  p-branes},'' {\em Nucl. Phys.} {\bf B475} (1996) 164--178,
  \href{http://xxx.lanl.gov/abs/hep-th/9604089}{{\tt hep-th/9604089}}.

\bibitem{Pope:1990kc}
C.~N. Pope, L.~J. Romans, and X.~Shen, ``{A New Higher Spin Algebra and the
  Lone Star Product},'' {\em Phys. Lett.} {\bf B242} (1990) 401--406.

\bibitem{Pope:1989bc}
C.~N. Pope, X.~Shen, and L.~J. Romans, ``{W(INFINITY) AND THE RACAH-WIGNER
  ALGEBRA},''.

\bibitem{Bergshoeff:1989ns}
E.~Bergshoeff, M.~P. Blencowe, and K.~S. Stelle, ``{Area Preserving
  Diffeomorphisms and Higher Spin Algebra},'' {\em Commun. Math. Phys.} {\bf
  128} (1990) 213.

\bibitem{Joung:2014qya}
E.~Joung and K.~Mkrtchyan, ``{Notes on higher-spin algebras: minimal
  representations and structure constants},'' {\em JHEP} {\bf 05} (2014) 103,
  \href{http://xxx.lanl.gov/abs/1401.7977}{{\tt 1401.7977}}.

\bibitem{Dabholkar:2014wpa}
A.~Dabholkar, N.~Drukker, and J.~Gomes, ``{Localization in supergravity and
  quantum $AdS_4/CFT_3$ holography},'' {\em JHEP} {\bf 10} (2014) 90,
  \href{http://xxx.lanl.gov/abs/1406.0505}{{\tt 1406.0505}}.

\bibitem{Dabholkar:2010uh}
A.~Dabholkar, J.~Gomes, and S.~Murthy, ``{Quantum black holes, localization and
  the topological string},'' {\em JHEP} {\bf 06} (2011) 019,
  \href{http://xxx.lanl.gov/abs/1012.0265}{{\tt 1012.0265}}.

\bibitem{Dabholkar:2012nd}
A.~Dabholkar, S.~Murthy, and D.~Zagier, ``{Quantum Black Holes, Wall Crossing,
  and Mock Modular Forms},'' \href{http://xxx.lanl.gov/abs/1208.4074}{{\tt
  1208.4074}}.

\bibitem{Dabholkar:2011ec}
A.~Dabholkar, J.~Gomes, and S.~Murthy, ``{Localization \& Exact Holography},''
  {\em JHEP} {\bf 04} (2013) 062, \href{http://xxx.lanl.gov/abs/1111.1161}{{\tt
  1111.1161}}.

\bibitem{Dabholkar:2014ema}
A.~Dabholkar, J.~Gomes, and S.~Murthy, ``{Nonperturbative black hole entropy
  and Kloosterman sums},'' {\em JHEP} {\bf 03} (2015) 074,
  \href{http://xxx.lanl.gov/abs/1404.0033}{{\tt 1404.0033}}.

\bibitem{Berkovits:1993hx}
N.~Berkovits, ``{A Ten-dimensional superYang-Mills action with off-shell
  supersymmetry},'' {\em Phys. Lett.} {\bf B318} (1993) 104--106,
  \href{http://xxx.lanl.gov/abs/hep-th/9308128}{{\tt hep-th/9308128}}.

\bibitem{Pestun:2007rz}
V.~Pestun, ``{Localization of gauge theory on a four-sphere and supersymmetric
  Wilson loops},'' {\em Commun. Math. Phys.} {\bf 313} (2012) 71--129,
  \href{http://xxx.lanl.gov/abs/0712.2824}{{\tt 0712.2824}}.

\bibitem{Minahan:2015jta}
J.~A. Minahan and M.~Zabzine, ``{Gauge theories with 16 supersymmetries on
  spheres},'' {\em JHEP} {\bf 03} (2015) 155,
  \href{http://xxx.lanl.gov/abs/1502.07154}{{\tt 1502.07154}}.

\bibitem{Freedman:2016yue}
D.~Z. Freedman, K.~Pilch, S.~S. Pufu, and N.~P. Warner, ``{Boundary Terms and
  Three-Point Functions: An AdS/CFT Puzzle Resolved},''
  \href{http://xxx.lanl.gov/abs/1611.01888}{{\tt 1611.01888}}.

\bibitem{Assel:2016pgi}
B.~Assel, D.~Martelli, S.~Murthy, and D.~Yokoyama, ``{Localization of
  supersymmetric field theories on non-compact hyperbolic three-manifolds},''
  {\em JHEP} {\bf 03} (2017) 095,
  \href{http://xxx.lanl.gov/abs/1609.08071}{{\tt 1609.08071}}.

\bibitem{deWit:1982bul}
B.~de~Wit and H.~Nicolai, ``{N=8 Supergravity},'' {\em Nucl. Phys.} {\bf B208}
  (1982) 323.

\bibitem{Cattaneo:1999fm}
A.~S. Cattaneo and G.~Felder, ``{A Path integral approach to the Kontsevich
  quantization formula},'' {\em Commun. Math. Phys.} {\bf 212} (2000) 591--611,
  \href{http://xxx.lanl.gov/abs/math/9902090}{{\tt math/9902090}}.

\bibitem{Elitzur:1989nr}
S.~Elitzur, G.~W. Moore, A.~Schwimmer, and N.~Seiberg, ``{Remarks on the
  Canonical Quantization of the Chern-Simons-Witten Theory},'' {\em Nucl.
  Phys.} {\bf B326} (1989) 108--134.

\bibitem{Banks:1998dd}
T.~Banks, M.~R. Douglas, G.~T. Horowitz, and E.~J. Martinec, ``{AdS dynamics
  from conformal field theory},''
  \href{http://xxx.lanl.gov/abs/hep-th/9808016}{{\tt hep-th/9808016}}.

\bibitem{Harlow:2011ke}
D.~Harlow and D.~Stanford, ``{Operator Dictionaries and Wave Functions in
  AdS/CFT and dS/CFT},'' \href{http://xxx.lanl.gov/abs/1104.2621}{{\tt
  1104.2621}}.

\bibitem{Sen:2011cn}
A.~Sen, ``{State Operator Correspondence and Entanglement in $AdS_2/CFT_1$},''
  {\em Entropy} {\bf 13} (2011) 1305--1323,
  \href{http://xxx.lanl.gov/abs/1101.4254}{{\tt 1101.4254}}.

\bibitem{Maldacena:2001kr}
J.~M. Maldacena, ``{Eternal black holes in anti-de Sitter},'' {\em JHEP} {\bf
  04} (2003) 021, \href{http://xxx.lanl.gov/abs/hep-th/0106112}{{\tt
  hep-th/0106112}}.

\bibitem{Donnelly:2014gva}
W.~Donnelly, ``{Entanglement entropy and nonabelian gauge symmetry},'' {\em
  Class. Quant. Grav.} {\bf 31} (2014), no.~21 214003,
  \href{http://xxx.lanl.gov/abs/1406.7304}{{\tt 1406.7304}}.

\bibitem{Freedman:1998tz}
D.~Z. Freedman, S.~D. Mathur, A.~Matusis, and L.~Rastelli, ``{Correlation
  functions in the CFT(d)/AdS(d+1) correspondence},'' {\em Nucl. Phys.} {\bf
  B546} (1999) 96--118, \href{http://xxx.lanl.gov/abs/hep-th/9804058}{{\tt
  hep-th/9804058}}.

\bibitem{DHoker:1999bve}
E.~D'Hoker, D.~Z. Freedman, S.~D. Mathur, A.~Matusis, and L.~Rastelli,
  ``{Graviton and gauge boson propagators in AdS(d+1)},'' {\em Nucl. Phys.}
  {\bf B562} (1999) 330--352,
  \href{http://xxx.lanl.gov/abs/hep-th/9902042}{{\tt hep-th/9902042}}.

\bibitem{DHoker:1999mqo}
E.~D'Hoker, D.~Z. Freedman, and L.~Rastelli, ``{AdS/CFT four point functions:
  How to succeed at $z$ integrals without really trying},'' {\em Nucl. Phys.}
  {\bf B562} (1999) 395--411,
  \href{http://xxx.lanl.gov/abs/hep-th/9905049}{{\tt hep-th/9905049}}.

\end{thebibliography}\endgroup

\end{document}